\documentclass[journal]{IEEEtran}
\usepackage{amsfonts}
\usepackage{amssymb}
\usepackage{amsthm}
\usepackage{amsmath,amsfonts,amssymb}
\usepackage[dvips]{graphicx}
\usepackage{verbatim}
\usepackage{bm}
\usepackage[ruled,vlined]{algorithm2e}
\usepackage{cite}
\usepackage{color}
\usepackage{multirow}
\usepackage[table,xcdraw]{xcolor}
\usepackage{adjustbox}
\usepackage{multirow}
\usepackage{booktabs}
\usepackage{subfigure}
\usepackage{array}
\newcommand{\PreserveBackslash}[1]{\let\temp=\\#1\let\\=\temp}
\newcolumntype{C}[1]{>{\PreserveBackslash\centering}p{#1}}
\newcolumntype{R}[1]{>{\PreserveBackslash\raggedleft}p{#1}}
\newcolumntype{L}[1]{>{\PreserveBackslash\raggedright}p{#1}}

\usepackage{breakurl}
\newlength{\Oldarrayrulewidth}
\newcommand{\Cline}[2]{%
	\noalign{\global\setlength{\Oldarrayrulewidth}{\arrayrulewidth}}%
	\noalign{\global\setlength{\arrayrulewidth}{#1}}\cline{#2}%
	\noalign{\global\setlength{\arrayrulewidth}{\Oldarrayrulewidth}}}

\newcommand{\figwidth}{8.8}
\IEEEoverridecommandlockouts

\begin{document}
	\title{Antenna Array Enabled Space/Air/Ground Communications and Networking for 6G}
	
\author{Zhenyu Xiao,~\IEEEmembership{Senior Member,~IEEE,}
	Zhu Han,~\IEEEmembership{Fellow,~IEEE,}
	Arumugam Nallanathan,~\IEEEmembership{Fellow,~IEEE,}
Octavia A. Dobre,~\IEEEmembership{Fellow,~IEEE,}
Bruno Clerckx,~\IEEEmembership{Senior Member,~IEEE,}
Jinho Choi,~\IEEEmembership{Senior Member,~IEEE,}
Chong He,~\IEEEmembership{Member,~IEEE,}
	and Wen Tong,~\IEEEmembership{Fellow,~IEEE}
	
	\thanks{The corresponding author is Dr. Zhenyu Xiao with Email xiaozy@buaa.edu.cn.}
	
	\thanks{Zhenyu Xiao is with the School of Electronic and Information Engineering, Beihang University, Beijing 100191, China. (email: xiaozy@buaa.edu.cn).}
	
	\thanks{Zhu Han is with the Department of Electrical and Computer Engineering in the University of Houston, Houston, TX 77004 USA, and also with the Department of Computer Science and Engineering, Kyung Hee University, Seoul, South Korea, 446-701. (email: zhan2@uh.edu).}
	
	\thanks{Arumugam Nallanathan is with the School of Electronic Engineering and Computer Science, Queen Mary University of London, London, U.K. (e-mail: a.nallanathan@qmul.ac.uk).}
	
	\thanks{Octavia A. Dobre is with the Faculty of Engineering and Applied Science, Memorial University, St. John’s,	NL A1C 5S7, Canada. (e-mail:	odobre@mun.ca).}
	
	\thanks{Bruno Clerckx is with the Communications and Signal Processing Group, Department of Electrical and Electronic Engineering, Imperial College London, London SW7 2AZ, U.K. (email: b.clerckx@imperial.ac.uk).}
	
	\thanks{Jinho Choi is with the School of
		Information Technology, Deakin University, Geelong, VIC 3220, Australia.
		(e-mail: jinho.choi@deakin.edu.au).}
	
	\thanks{Chong He is with the Department of Electronic Engineering, Shanghai Jiao Tong University, Shanghai 200240, P. R. China. (email: hechong@sjtu.edu.cn).}
	
	\thanks{Wen Tong is with the Wireless Advanced System and Competency Centre, Huawei Technologies Co., Ltd., Ottawa, ON K2K 3J1, Canada. (email: tongwen@huawei.com).}
}

	\vspace{-10mm}
	\maketitle
	\vspace{-10mm}
	\begin{abstract}
		Antenna arrays have a long history of more than 100 years and have evolved closely with the development of electronic and information technologies, playing an indispensable role in  wireless communications and radar. With the rapid development of electronic  and information technologies, the demand for all-time, all-domain, and full-space network services has exploded, and new communication requirements have been put forward on various space/air/ground platforms. To meet the ever increasing requirements of the future sixth generation (6G) wireless communications, such as high capacity, wide coverage, low latency, and strong robustness, it is promising to employ different types of antenna arrays (e.g., phased arrays, digital arrays, and reconfigurable intelligent surfaces, etc.) with various beamforming technologies (e.g., analog beamforming, digital beamforming, hybrid beamforming, and passive beamforming, etc.) in space/air/ground communication networks, bringing in advantages such as considerable antenna gains,  multiplexing gains, and diversity gains. However, enabling antenna array for space/air/ground communication networks poses specific, distinctive and tricky challenges, which has aroused extensive research attention. 
	This paper aims to overview the field of antenna array enabled space/air/ground communications and networking. The technical potentials and challenges of antenna array enabled space/air/ground communications and networking are presented first. Subsequently, the antenna array structures and designs are discussed. We then discuss various emerging technologies facilitated by antenna arrays to meet the new communication requirements of space/air/ground communication systems. Enabled by these emerging technologies, the distinct characteristics, challenges, and solutions for space communications, airborne communications, and ground communications are reviewed. Finally, we present promising directions for future research in antenna array enabled space/air/ground communications and networking.
	
	\end{abstract}


	\section{Introduction}
Antenna array is a set of multiple connected antennas working cooperatively to transmit or receive radio waves. 
The origination of antenna array can be traced back to more than 100 years ago, when the Nobel Prize winner Ferdinand Braun positioned three monopoles in a triangle and formed a cardioid radiation pattern in the 1900s~\cite{braun1909electr}. At the early stage, antenna arrays were used for radar-related applications. In the 1940s, the requirements of detection distance and accuracy for radars rapidly improved, which led to the accelerating development of high-gain antenna arrays~\cite{haupt2015antenn}. In the mid 1950s, electronic computers became powerful enough  
to achieve rapid electronic beamforming, rather than less flexible mechanical scanning~\cite{fowler1998oldrad}. In the 1960s, the development of semiconductor electronics promoted the miniaturization and integration of array manufacturing. In the 1970s, Bell Labs described the advantages of scanning spot beams for satellites by using adaptive array (also known as smart antennas), including reducing transmit power, and increasing communication capacity~\cite{reudink1977ascann}. The adaptive array technique has greatly improved the quality of satellite communications since it was introduced. The interest in commercial applications of smart antennas owed to the growth of cellular telephone in the 1980s. In the 1990s, the upgrade to digital radio technology in the mobile phone and indoor wireless network created new opportunities for smart antennas. The research on smart antennas led to the development of the multiple-input multiple-output (MIMO) technology used in the fourth generation (4G) wireless communication networks~\cite{yang2015fiftyy}. Nowadays, massive MIMO in the sub-6 GHz and millimeter-wave (mmWave) bands becomes a mainstream technology of the fifth generation (5G) new radio standard and the first 64-antenna massive MIMO base stations (BSs) have been commercially deployed in the sub-6 GHz bands~\cite{sanguinetti2020toward}. In recent years, reconfigurable intelligent surfaces (RISs) have developed rapidly and aroused global attention and interest of both academia and industry, and thus constitute one of the key technologies in future sixth generation (6G) mobile network~\cite{pan2021reconf}. As can be seen, antenna arrays have a long history and develop closely with the advancement of electronic and information technologies, playing an indispensable role in many fields such as radar and satellite/terrestrial wireless communications.


On the other hand, with the rapid development of electronic and information technologies, the demand for all-time, all-domain, and full-space network services has exploded.
Thus, a disruptive 6G wireless system inherently tailored to these requirements will be needed~\cite{saad2020avisio}. One of the accompanying technological trends of 6G is dealing with both ground and aerial users. Therefore, a revolutionary shift from providing terrestrial communication services to support three-dimensional (3D) space ubiquitous communication coverage  is needed.
In this context, non-terrestrial platforms, including satellites, high-altitude platforms (HAPs) and low-altitude platforms (LAPs), are being considered as candidates for deploying wireless communications complementing the terrestrial communication infrastructure~\cite{song2021aerial, jia2021leosat}. 
The standardization efforts of non-terrestrial network (NTN), even before 6G, are ongoing~\cite{3gpp2020studyo, 3gpp2021soluti, 3gpp2019newwid}. 
Moreover, the integration of satellite, airborne, and terrestrial networks, known as space-air-ground integrated network (SAGIN), has become a promising paradigm for the future 6G wireless network~\cite{zhang20196gwirl, saad2020avisio}. 
In this new era, emerging application scenarios put forward new requirements for space/air/ground communication networks regarding spectrum efficiency, data rate, traffic capacity, connectivity density, energy efficiency, latency, and mobility. To name a few, typical scenarios include enhanced ultra-reliable and
low-latency communication (URLLC), long-distance and high-mobility communication, and ultra-low-power communication, as reported in~\cite{zhang20196gwirl}.

As the communication requirement explosively increases, conventional single-antenna transmission faces challenges to meet the insistent demands of high capacity, huge data rate, long distance, low latency, energy efficiency and strong robustness. In contrast, large-scale antenna arrays, including phased arrays, digital arrays, and RISs, are promising to be adopted on space/air/ground platforms for improving the communication qualities, by providing three types of fundamental gains: antenna gain, multiplexing gain, and diversity gain~\cite{lizhong2003divers, mietzner2009multip}.
Firstly, in order to pursue broadband communication, the exploitation of high-frequency bands, such as mmWave frequencies, with rich spectrum resources has become a prevailing trend. However, the high-frequency bands also cause  more severe propagation loss. By steering the radiation energy only to the desired directions, antenna arrays provide considerable antenna gains to compensate propagation loss, thus supporting high-frequency broadband communications. At the same time, the improved signal-to-noise ratio (SNR) at the receiver is also beneficial for supporting long-distance transmission. 
Secondly, by simultaneously transmitting independent information sequences over multiple antennas, antenna arrays can provide substantial spatial multiplexing gains to increase data rate, thus benefiting multiuser high capacity  communications. 
Thirdly, by transmitting and/or receiving redundant signals representing the same information through different paths, antenna arrays can provide diversity gains to combat channel fading, thus enhancing communication reliability. 
In addition, antenna arrays  can facilitate the spectrum reuse, interference mitigation, coverage enhancement, and physical-layer security for space/air/ground communications and networking. Given these promising benefits, antenna array and beamforming technologies have been applied in terrestrial communications such as long term evolution (LTE), 5G and WiFi 6~\cite{3gpp2015studyo, 3gpp2019releas, khorov2019atutor}, satellite communications such as the Starlink project~\cite{shahrzad2018distri}, and airborne communications such as the multifunction advanced data link (MADL) for F-35 aircraft~\cite{steve2015f35lig}.
These applications prove the great potential of antenna array technologies for space/air/ground communications and networking and inspire further exploration.

\begin{table*}[t]
	\caption{{\color{black}summary table of all the sections}}\label{tab:fffff}
	\begin{center}
		\textcolor{black}{
		\begin{tabular}{|c|c|c|l|}
			\toprule[1.5pt]
			& {\color{black}\textbf{Section}}                                                                                    & {\color{black}\textbf{Subsection}}           & \multicolumn{1}{c|}{{\color{black}\textbf{Details \& References}}}                                                                                                      \\ \bottomrule[1.5pt] 
			{\color{black}\textbf{Motivation}}                                                                                             & {\color{black}I. Introduction}                                                                                     &   {\color{black}---}                            & \textcolor{black}{\begin{tabular}[c]{@{}l@{}} The history of antenna arrays are introduced~\cite{braun1909electr,haupt2015antenn,fowler1998oldrad,reudink1977ascann,yang2015fiftyy,sanguinetti2020toward,pan2021reconf}.\\
			The benefits and challenges of antenna array enabled space/air/ground communications\\ and networking are discussed~\cite{ saad2020avisio,song2021aerial, jia2021leosat,3gpp2020studyo, 3gpp2021soluti, 3gpp2019newwid,zhang20196gwirl,lizhong2003divers, mietzner2009multip,3gpp2015studyo, 3gpp2019releas, khorov2019atutor,shahrzad2018distri,steve2015f35lig,xiao2016enabli,xiao2017millim,rappaport2011stateo,gao2016struct, di2014spatia, zhang2020prospe,liu2018spacea, chen2020system, wu2021acompr}.   \end{tabular}}                                                                                                                                                       \\  \hline \Cline{0.8pt}{2-4}
			\multirow{10}{*}{\textbf{\begin{tabular}[c]{@{}c@{}}{\color{black}Hardware}\\ {\color{black}\&}\\ {\color{black}Technologies}\\ {\color{black}Supports}\end{tabular}}}      & \multirow{6}{*}{\begin{tabular}[c]{@{}c@{}}{\color{black}II. Antenna Array}\\ {\color{black}Structure and Design}\end{tabular}}   & {\begin{tabular}[c]{@{}c@{}}Fix-beam\\ Antenna Arrays\end{tabular}}       & {\begin{tabular}[c]{@{}l@{}}Structures of fix-beam antenna arrays, including single beam arrays and multiple beam\\ arrays, are introduced~\cite{Jianping2019}.\end{tabular}}                                                                                                                                                       \\ \cline{3-4} 
			&                                                                                                     & {\color{black}Phased Arrays}                 &   Structures of phased arrays and analog beamforming are introduced~\cite{xiao2018jointp,zhu2020millim}.                                                                                                                                                   \\ \cline{3-4} 
			&                                                                                                     & {\color{black}Digital Arrays}                &   Structures of digital arrays and digital beamforming are introduced~\cite{xiao2021asurve, xiao2016enabli}.                                                                                                                                                       \\ \cline{3-4} 
			&                                                                                                     & {\begin{tabular}[c]{@{}c@{}}Hybrid\\ Antenna Arrays\end{tabular}}        & Structures of hybrid antenna arrays and hybrid beamforming are introduced~\cite{peng2021hybrid, ahmed2018asurve}.                                                                                                                                                       \\ \cline{3-4} 
			&                                                                                                    & {\begin{tabular}[c]{@{}c@{}}Irregular\\ Antenna Arrays\end{tabular}}     &  Structures of irregular antenna arrays are introduced\cite{ma2019patter,rocca2014gabase,bui2018fastan}.                                                                                                                                                     \\ \cline{3-4} 
			&                                                                                                     &{\begin{tabular}[c]{@{}c@{}}Programmable\\ Metasurfaces\end{tabular}}     &  {\begin{tabular}[c]{@{}l@{}}Structures of programmable metasurfaces and passive beamforming are introduced\\~\cite{pan2021reconf, Bai2020,wu2020toward}.\end{tabular}}                                                                                                                                                     \\ \Cline{1pt}{2-4} 
			& \multirow{4}{*}{\begin{tabular}[m]{@{}c@{}}{\color{black}III. Emerging}\\ {\color{black}Communication}\\{\color{black}Technologies}\end{tabular}} & {\begin{tabular}[c]{@{}c@{}}New\\ Beamforming\\ Technologies\end{tabular}}  & \begin{tabular}[c]{@{}l@{}} 
				To form on-demand coverage, new beamforming technologies are developed, including\\
				~$\circ$ Single-RF-Chain Multiple Beams~\cite{wei2019multib, li2020perfor, xiao2018jointp,zhu2019millimTWC, xie2019onthep}
				\\ ~$\circ$ Flexible Beam Coverage~\cite{xiao2016hierar, xiao2018enhanc, kong2018hybrid, zhu2019dbeamf}  
				\\ ~$\circ$ Robust Beamforming~\cite{ayach2014spatia, zhang2019optima, xiu2021reconf, lu2014anover,shu2018lowcom, liu2018roubst, zhu2018robust, xu2015robust, cai2019robust}
			\end{tabular}                                     \\ \cline{3-4} 
			&                                                                                                     & {\begin{tabular}[c]{@{}c@{}}Multi-Antenna\\ Multiple Access\end{tabular}} & \begin{tabular}[c]{@{}l@{}}
					To serve more users, new beam-domain multiple access schemes are studied, including\\
					  ~$\circ$ SDMA~\cite{spencer2004zerofo, stojnic2006ratema, mauricio2018alowco,sun2017agglom} 
				$\circ$ NOMA~\cite{ding2017asurve, zhu2019optima, zhang2020optima, zeng2017capaci,wei2019multib,xiao2018jointp,cui2018optima,zeng2019energy,zhu2019millimAcc, zhu2019millimTWC,khaled2020joints,mao2018energy} 
				$\circ$ RSMA~\cite{Clerckx:2016,Joudeh:2016,naser2020ratesp, mao2017rate, Clerckx:2020, Davoodi:2016,Davoodi:2020,Piovano:2017,Hao:2017,Costa83,Mao:2020,clerckx2021isnoma,mao2019ratesp,Yin:2021,Jaafar:2020,Si:2021}
			\end{tabular}                                                                                               \\ \cline{3-4} 
			&                                                                                                     & {\color{black}RISs}                          & \begin{tabular}[c]{@{}l@{}}
		As an emerging paradigm to achieve high energy efficiency and spectrum efficiency, \\research interests of RIS include\\
			~$\circ$ Passive Relay~\cite{wu2020toward,liu2021risenh, wu2019intell, wu2020jointa, chen2021jointa, wu2020beamfo,rehman2021jointa,abeywickrama2020intell, zuo2020intell,bariah2021largei,hao2021robust, chen2021toward,guo2021learni, li2021robust,hua2021uavass} 
				~$\circ$ Passive Transmitter~\cite{tang2020mimotr,tang2020wirele, zhang2021reconf, guo2020reflec,ma2020passiv,yan2020passi,basar2020reconf, canbilen2020reconf}
			\end{tabular}                                                                              \\ \cline{3-4} 
			&                                                                                                     & {\begin{tabular}[c]{@{}c@{}}Secure\\ Communications\end{tabular}}         & {\begin{tabular}[c]{@{}l@{}}Antenna arrays can improve the physical-layer security via directional transmission\\~\cite{Liang_2008, Bloch11, Bloch21,wang2020energy, wang2021robust, Wyner75TheWiretap, Leung78,Goel08,Khisti_2010,Liao11, Wang14, Li14,zeng2019securi,Choi16,Ng14, Shi15,Lu19,Schraml21,lei2011secure,li2020cooper,fu2020robust, wu2018securi, xiao2021asurve}.\end{tabular}}                                  \\ \Cline{1pt}{2-4}\hline
			\multirow{3}{*}{\textbf{\begin{tabular}[c]{@{}l@{}}{\color{black}Three Layers}\\ {\color{black}$\circ$Scenarios}\\ {\color{black}$\circ$Issues}\\ {\color{black}$\circ$Solutions}\end{tabular}}} & \multirow{3}{*}{\begin{tabular}[c]{@{}c@{}}IV. Space/Air/Ground\\ Communications\end{tabular}}      & {\begin{tabular}[c]{@{}c@{}}Satellite\\ Communications\end{tabular}}      & \textcolor{black}{\begin{tabular}[c]{@{}l@{}}
				The application of antenna arrays poses new challenges to satellite communications,\\ including\\
				~$\circ$ Various Beam Patterns~\cite{Su2019broadb,chang2015broadb,Khidre2013reconf,Khan2016thepur,Zhang2017compac,Yang2019anovel,Qian2014Traffi,Honnaiah2021demand,Zhou2019covera,Wan2016asteer,zhang2019resear} 	~$\circ$ MBA~\cite{moon2019phased,Montero2015cbandm,Lai2016adigit,sow2008beamfo,montesinos2011adapti,zheng2019adapti,Yin:2021,Si:2021}\\ 
				~$\circ$ Beam Management and Handover~\cite{kim2020Beamma,Del1999differ,Riad2001fixedc,Papapetrou2005analyt,Zhao2009Newcha,Yan2008Spotbe,Karapantazis2005Design,boukhatem2003tcraat,Hedjazi1972thehan,Olariu2004OSCARa,Bottcher1994Strate,Zhao1996Combin}
			\end{tabular}}                                                       \\ \cline{3-4} 
			&                                                                                                     & {\begin{tabular}[c]{@{}c@{}}Airborne\\ Communications\end{tabular}}       & \begin{tabular}[c]{@{}l@{}}
 There are varieties of characteristics and challenges in airborne communication, including\\
				~$\circ$ Beam Tracking~\cite{yang2019beamtr, huang20203dbeam, xu2021datadr, zhao2018channe, chiang2021machin, yuan2020learni}    
				~$\circ$ Doppler Effect~\cite{khawaja2019asurve, rangan2014millim, xiao2016enabli,lorca2017onover, ma2020awideb,walter2019analys, jiang2020anovel, ma2020impact,zhang2020dataai}   \\ 
				~$\circ$ Joint Positioning and Beamforming~\cite{yuan2019joint3, zhu2020millim,xiao2020unmann, wei2019multib, yuan2021jointd, xu2020multiu}\\ 
				~$\circ$ Antenna Array Enabled Aerial Ad-hoc Network~\cite{zafar2016flying, cao2018airbor, ramanathan2005adhocn, cai2012neighb, zeng2019aduala, sneha2020freesp, astudillo2017neighb,yang2019networ, chen2017onobli, xu2020improv, gankhuyag2017robust, waheed2019laodli,fawaz2017unmann,shumeye2020routin,leonov2016modeli, khan2020smarti,leonov2016applic,xiao2021asurve,chen2021agamet,wang2018spectr,feng2019spectr,temel2014scalab} \end{tabular} \\ \cline{3-4} 
			&                                                                                                     & {\begin{tabular}[c]{@{}c@{}}Ground\\ Communications\end{tabular}}         & \begin{tabular}[c]{@{}l@{}}
				The potential applications of antenna array in ground communications include\\
				~$\circ$ Cellular Massive MIMO~\cite{rodriguez2014fundam,hosseini2014larges, shin2017coordi, nadeem2019elevat, kim2014fulldi, kuo2016aglanc, li2018interf, li2020ffrbas, dai2016arates,dai2017multiu, thomas2020arates, papazafeiropoulos2017ratesp, dizdar2021ratesp, wu2013approx, chen2017alowco, gao2015lowcom, dai2015lowcom, liu2019adistr, li2017massiv, he2016uplink, maruta2018uplink}  ~$\circ$ Cell-free MIMO~\cite{interdonato2019ubiqui, ngo2017cellfr, zhang2020prospe, ngo2015cellfr, naris2021cellfr, mai2018cellfe, bjornson2020scalab, nayebi2017precod, interdonato2019scalab}   
				\\ ~$\circ$ V2X Communications~\cite{gyawail2021challe, busari2019millim, tassi2017modeli, feng2021beamwi, gao2018dynami, xu2020locati}  \end{tabular}                                                      \\ \hline \Cline{0.8pt}{2-4}
			\textbf{Directions}                                                                                             & V. Future Directions                                                                                & ---                              & Future research directions are highlighted~\cite{wu2020toward, chiang2021machin}.                                                                                                                                                \\ \bottomrule[1.5pt]
		\end{tabular}
	}
	\end{center}
\end{table*}

\begin{table*}[t]
	\caption{Summary of important acronyms}\label{tab:acronym}
	\begin{center}
		\begin{tabular}{p{1.6cm}p{6.4cm}p{0.01cm}p{1.6cm}p{6.4cm}}
			\cline{1-2}\cline{1-2} \cline{4-5}\cline{4-5}
			\textbf{Acronyms} & \textbf{Meaning}                        & \textbf{} & \textbf{Acronyms} & \textbf{Meaning}                                                                                    \\  \cline{1-2} \cline{4-5}
			2-D                         & Two-Dimensional                              &  & LoS                        & Line-of-Sight                                \\
			3-D                         & Three-Dimensional                            &  & LS-MIMO                    & Large-Scale MIMO                             \\
			3GPP                        & Third Generation Partnership Project         &  & LTE                        & Long Term Evolution                          \\
			4G                          & Fourth Generation                            &  & MAC                        & Media Access Control                         \\
			5G                          & Fifth Generation                             &  & MBA                        & Multiple Beam Array                          \\
			6G                          & Sixth Generation                             &  & MEO                        & Medium Earth Orbit                           \\
			A2A                         & Air-to-Air                                   &  & MIMO                       & Multiple-Input Multiple-Output               \\
			A2G                         & Air-to-Ground                                &  & mmWave                     & Millimeter-Wave                              \\
			A2S                         & Air-to-Satellite                             &  & MPC & multi-Path Components                        \\
			AAN                         & Aerial Access Network                        &  & NOMA                       & Non Orthogonal Multiple Access               \\
			ADC                         & Analog-to-Digital Converter                  &  & NTN                        & Non-Terrestrial Network                      \\
			AI                          & Artificial Intelligence                      &  & OMA                        & Orthogonal Multiple Access                   \\
			AoA                         & Angle of Arrival                             &  & PA                         & Power Amplifier                              \\
			AP                          & Access Point                                 &  & PCB                        & Printed Circuit Board                        \\
			AWGN                        & Additive White Gaussian Noise                &  & PIN                        & Positive-Intrinsic-Negative                  \\
			B5G                         & Beyond 5G                                    &  & PSK                        & Phase Shift Keying                           \\
			BDMA                        & Beam Division Multiple Access                &  & QAM                        & Quadrature Amplitude Modulation              \\
			BFN                         & Beamforming Network                          &  & QoS                        & Quality of Service                           \\
			BS                          & Base Station                                 &  & RF  & Radio Frequency                              \\
			CBF                         & Conjugate Beamforming                        &  & RIS                        & Reconfigurable Intelligent Surfaces          \\
			CDMA & Code-Division Multiple Access                &  & RSMA                       & Rate-Splitting Multiple Access               \\
			CoBF                        & Coordinated Beamforming                      &  & SAGIN                      & Space-Air-Ground Integrated Network          \\
			CoMP                        & Coordinated Multipoint                       &  & SIC                        & Successive Interference Cancellation         \\
			CSI                         & Channel State Information                    &  & SINR                       & Signal-to-Interference-plus-Noise Ratio      \\
			CSIT                        & Channel State Information at the Transmitter &  & SNR                        & Signal-to-Noise Ratio                        \\
			DAC                         & Digital-to-Analog Converter                  &  & SPDT                       & Single-Pole Double-Throw                     \\
			DFS                         & Doppler Frequency Shift                      &  & SWAP                       & Size, Weight, and Power                      \\
			DoF                         & Degree of Freedom                            &  & TDD                        & Time Division Duplexing                      \\
			DPC                         & Dirty Paper Coding                           &  & TDMA                       & Time-Division Multiple Access                \\
			Eve                         & Eavesdropper                                 &  & UAV                        & Unmanned Aerial Vehicle                      \\
			FDD                         & Frequency Division Duplexing                 &  & UE                         & User Equipments                              \\
			FDMA                        & Frequency-Division Multiple Access           &  & ULA                        & Uniform Linear Array                         \\
			FD-MIMO                     & Full Dimension MIMO                          &  & UPA                        & Uniform Plane Array                          \\
			GEO                         & Geostationary Orbit                          &  & URLLC                      & Ultra-Reliable and Low-Latency Communication \\
			GNSS                        & Global Navigation Satellite System           &  & V2I                        & Vehicle-to-Infrastructure                    \\
			GPS                         & Global Positioning System                    &  & V2N                        & Vehicle-to-Network                           \\
			HAP                         & High-Altitude Platform                       &  & V2P                        & Vehicle-to-Pedestrian                        \\
			LAP                         & Low-Altitude Platform                        &  & V2V                        & Vehicle-to-Vehicle                           \\
			LEO                         & Low Earth Orbit                              &  & V2X                        & Vehicle-to-Everything                        \\
			LMMSE                       & Linear Minimum Mean Square Error             &  & ZFBF                       & Zero-Forcing Beamforming                     \\
			LNA                         & Low Noise Amplifier                          &  &                            &                                              \\          
			\cline{1-2} \cline{4-5}
		\end{tabular}
	\end{center}
\end{table*}


Despite  various benefits mentioned above, there are also many challenging scientific and technological problems for antenna array enabled space/air/ground communications and networking, which are detailed as follows.
\begin{itemize}
	\item \emph{Antenna Array Design:} Compared to terrestrial infrastructures, the maneuverable platforms in space/air/ground networks usually suffer from stringent constraints on the size, weight, and power (SWAP). The antenna layout, system integration, and power control should be considered in particular. Large-scale antenna arrays require compact circuit implementation, expensive radio frequency (RF) chains, and high power consumption~\cite{xiao2016enabli,xiao2017millim}. 
	Indeed, there is a tradeoff between the spectral efficiency and hardware cost/power consumption for different array structures. Besides, the RF hardware impairments, such as phase noise, non-linear power amplifiers (PAs), I/Q imbalance, and limited analog-to-digital converter (ADC) resolution, become more severe in higher and larger frequency bands~\cite{rappaport2011stateo}. These effects challenge the performance of antenna arrays. Moreover, electromagnetic compatibility needs to be considered when  designing and assembling antenna arrays, to make them compatible with their electromagnetic environments.
	
	\item \emph{Physical Layer:}  The space/air/ground platforms have a common feature of 3D mobility, which makes the communications and networking for these maneuverable platforms different from the conventional terrestrial infrastructures with fixed positions.  Firstly, the fast moving of the space/air/ground platforms, such as satellites, aircraft, and high-speed trains, may result in rapid channel variation in space, time, and frequency domains, which brings in a more stringent time constraint for channel state information (CSI) acquisition. The employment of large-scale antenna array may also result in prohibitively high pilot overhead, which challenges fast and accurate acquisition of CSI~\cite{gao2016struct}.  Besides, due to the high-mobility and jittering of space/air/ground platforms, the narrow beams are prone to misalignment, which may cause serious loss of antenna gain, decline of SNR, or even interruption of connection. What’s more, subject to the power consumption and hardware cost, fully-digital arrays could not be adopted in most of the space/air/ground platforms. Instead, phased arrays and programmable metasurfaces are usually exploited, which results in more constraints such as constant amplitudes of the beamforming vector, limited knowledge of CIS, etc.
Finally, the beam pointing of antenna array is natively coupled with the 3D position of maneuverable platforms, which complicates the optimal design of the communication systems.

	\item \emph{Multiple Access Control Layer:} The adoption of antenna arrays and beamforming technologies accomplishes considerable antenna gains, multiplexing gains, and diversity gains. However, in multi-user scenario, the narrow beam may limit the number of accessible users as well as other quality of service (QoS). Therefore, new multiple access schemes are required to overcome the contradiction between narrow beam and users' QoS. In the high-dynamic antenna array enabled space/air/ground communication scenarios, multiple access schemes need to improve the performance by joint beam/time/frequency domain optimizations. 

	\item \emph{Network Layer:} 
	The high mobility of the space/air/ground platforms results in the rapid change of the channel states and network topologies, and thus real-time beam management is required. Mechanisms that were designed for conventional   terrestrial communication networks  need to be redesigned for the antenna array enabled space/air/ground communication networks. For instance, the high-dynamic feature of space/air/ground platforms and the directional transmission feature of antenna array may challenge protocol design such as  neighbor discovery, routing, handover, and resource management.

\end{itemize}

In a word, it is time to address the above technical challenges to enable antenna arrays for space/air/ground communications and networking. There are several overview papers related to MIMO or multi-antenna technologies~\cite{mietzner2009multip, di2014spatia, zhang2020prospe}. However, these papers above focus more on terrestrial cellular networks. On the other hand, there exist several review papers related to satellite communications, air communications, or integrated communication systems~\cite{liu2018spacea, chen2020system, wu2021acompr}. However, these papers do not highlight the potentials, challenges, and solutions for antenna array aided communications and networking for space/air/ground platforms. 
Different from the above works, this paper aims to review the contributions and progress in antenna array enabled space/air/ground communications and networking, which has not yet been completely explored.	
First, we start with discussing the antenna array structures and designs, looking at the strengths and weaknesses of adopting different types of antenna arrays, such as fix-beam antenna arrays, phased arrays, digital arrays, hybrid antenna arrays, programmable metasurfaces, and irregular antenna arrays, on space/air/ground communication systems.
Second, we discuss various emerging technologies facilitated by antenna arrays to meet the new communication requirements of ubiquitous, flexible, and robust coverage, massive connectivity, and secure communications. These include new beamforming technologies, new beam-domain multiple access schemes, RISs, and physical-layer security enhancement. Third, 
substantial new characteristics and challenges in the aspect of antenna array enabled space/air/ground communication systems are covered. Specifically, we discuss unique features of satellite communications in beam pattern, beam coverage, and beam management, address issues of airborne communications in beam tracking, Doppler effect, joint positioning and beamforming, and aerial ad-hoc network, and consider scenarios of ground communications in cellular massive MIMO, cell-free MIMO, and vehicle-to-everything (V2X) communications.


{\color{black}Table~\ref{tab:fffff} provides an overview of the main content as well as the references of this paper.}
Specifically, antenna array structures and design are discussed in Section II, where different types of antenna array and their features are introduced. 
Antenna arrays facilitate various emerging technologies, such as new beamforming, multi-antenna multiple access, physical-layer security, which are covered in Section III. In Section IV, the potential paradigm shifts for enabling antenna arrays to space/air/ground communications and networking are discussed. Future research directions are highlighted in Section V, and the paper is concluded in Section VI. For ease of reading, the acronyms employed in this paper are summarized in Table~\ref{tab:acronym}.

\section{Antenna Array Structures and Design}
Because of the high path loss between space/air and ground, antenna arrays with a large aperture are usually exploited to provide high gain, narrow beam, low sidelobe level, \emph{et al.} According to structures, antenna arrays can be classified as fix-beam antenna arrays, phased arrays, digital arrays,  hybrid antenna arrays, irregular antenna arrays, and programmable metasurfaces. {\color{black}Based on these, various beamforming technologies, such as analog beamforming, digital beamforming, hybrid beamforming, and passive beamforming, are developed. These architectures for antenna arrays as well as beamforming technologies are introduced below.
} 

\subsection{Fix-beam Antenna Arrays}
\begin{figure}[t]
	\centering
	\includegraphics[width=8.89cm]{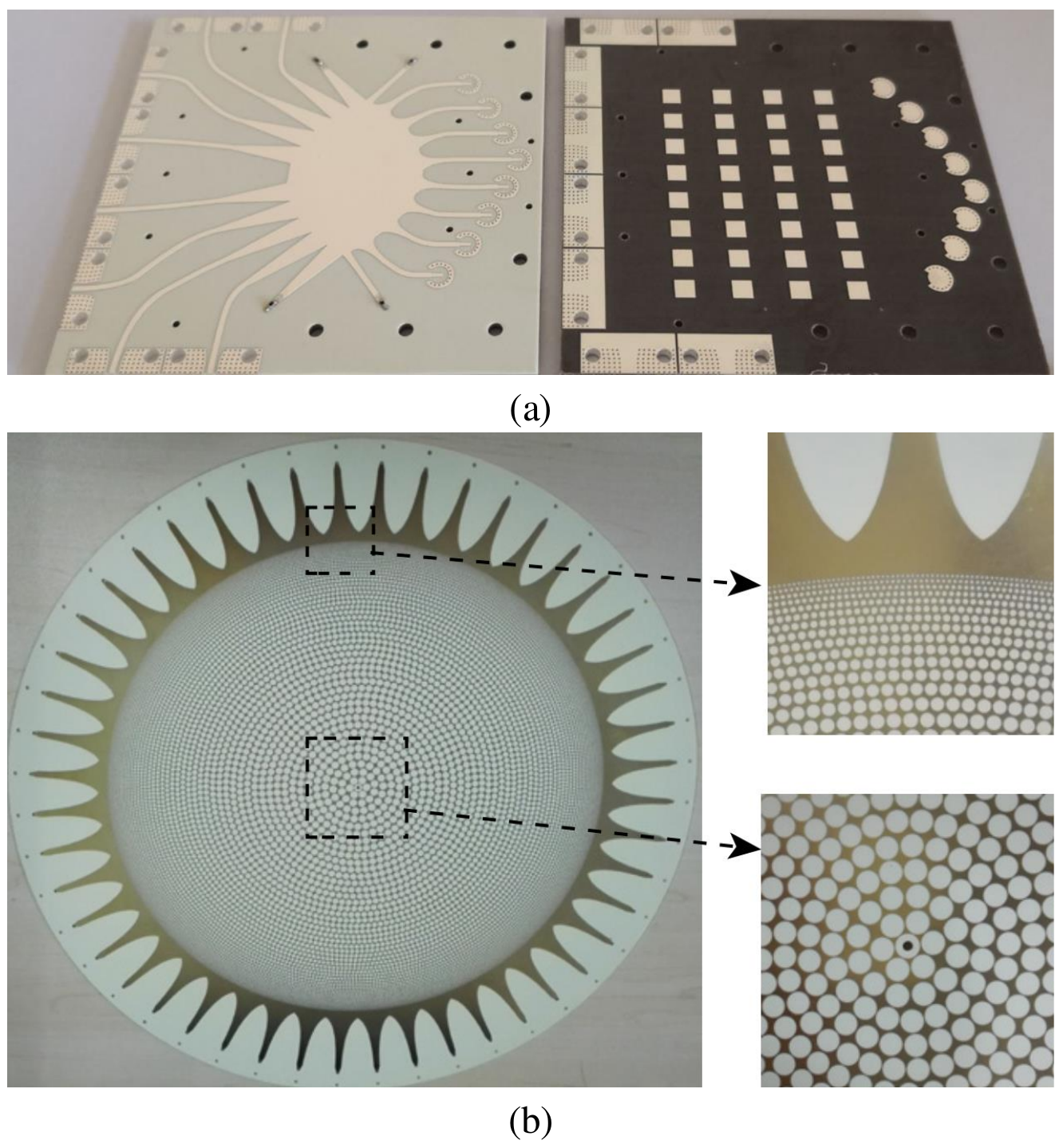}
	\caption{Two kinds of microwave lenses. (a) Rotman lens, (b) Luneberg lens.}
	\label{fig-Lens}
\end{figure}

Fix-beam antenna arrays include single beam arrays and multiple beam arrays (MBAs). The single beam arrays usually consist of radiation elements and relatively simple feeding network, which are cheap to be exploited to track satellites and airplanes on an electronic-controlled turntable. MBAs are usually installed on the satellite to serve users in different regions with multiple narrow beams generated by the Butler matrix, lens, reflector antennas with multiple feedings. The lens-based MBAs mainly include the Rotman lens and Luneberg lens~\cite{Jianping2019}. The Rotman lens has a two-dimensional (2D) true-time delay. By moving the position of the feed, the phase differences between the adjacent output port of the lens will be changed, thus the radiation beam points to different angles. Fig.~\ref{fig-Lens}(a) shows a Ka-band Rotman lens antenna designed to cover an angular range of $ \pm 56^\circ $ with 1.5 dB gain drop. The planar Luneburg lens is the dielectric gradient index lens as shown in Fig.~\ref{fig-Lens}(b). It has infinite focus points, which makes it a promising candidate for wide-scanning antennas. The full-angle beam scanning can be realized by introducing microstrip port that can work in both transmitting and receiving mode.

\subsection{Phased Arrays}\label{PhasedArrays}
\begin{figure}[t]
	\centering
	\includegraphics[width=\linewidth]{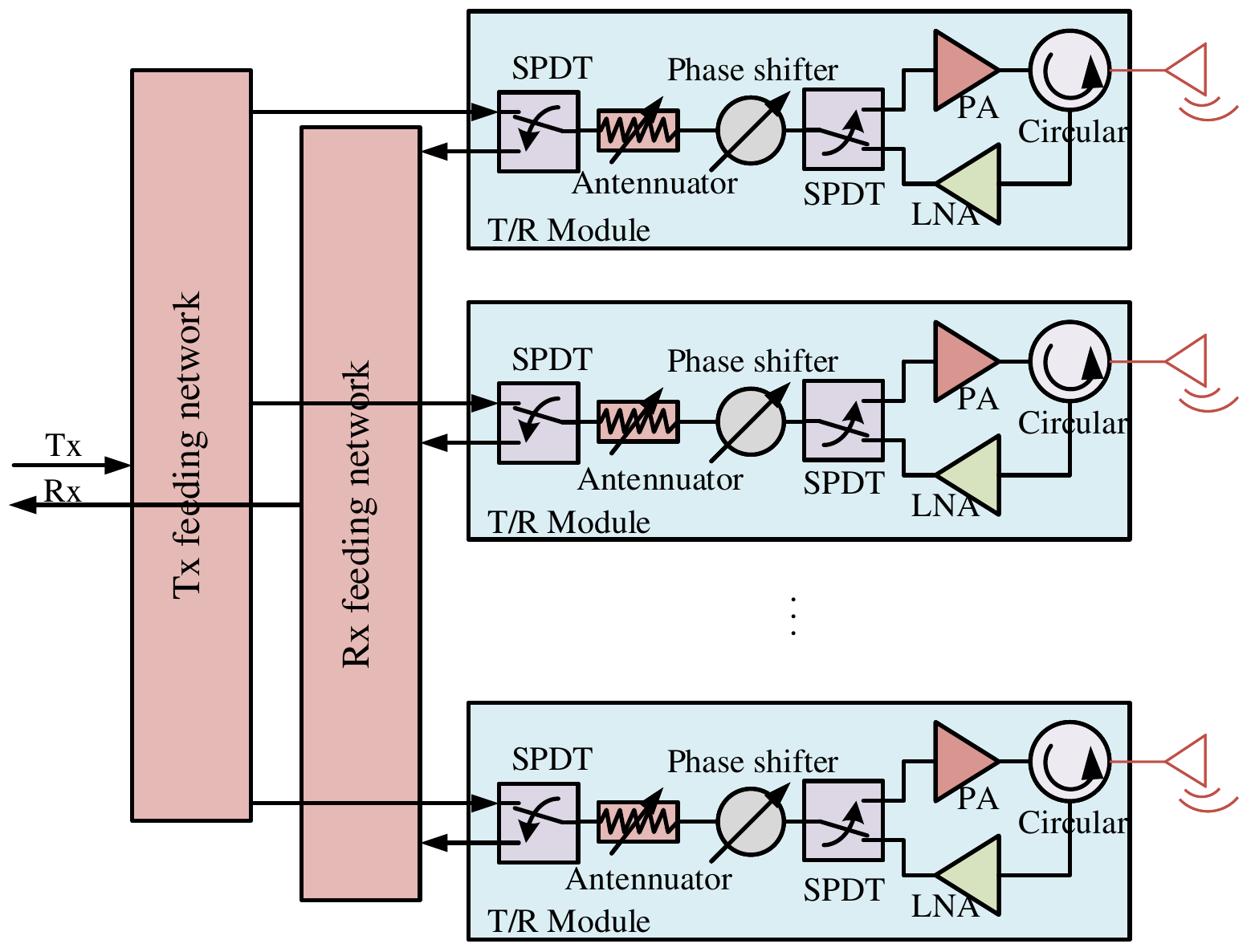}
	\caption{Structure of phased array.}
	\label{fig-structure of phased array}
\end{figure}

Phased arrays are extensively exploited in radars, wireless communications and electronic reconnaissance, whose structure is shown in Fig.~\ref{fig-structure of phased array}. T/R modules are the core elements of the phased array, which are placed between antennas and the feeding network. T/R modules consist of circulars, PAs, low noise amplifiers (LNAs), single-pole double-throw (SPDT) switches, phase shifters and attenuators. The circulars and SPDT switches make it possible to reuse phase shifters and attenuators when transmitting and receiving are separate in time. The phase shifters and attenuators control and switch the beam pointing fast, and therefore the satellites with phased arrays can serve multiple wireless users simultaneously. As beamforming is realized in the RF domain, the beam synthesis of phased arrays is called \emph{analog beamforming}, or single-RF-chain beamforming. 

{\color{black}
We consider an end-to-end communication scenario, where the transmitter sends signal $s$ to the receiver with power $P$. The transmitter, equipped with $N_\mathrm{t}$ antennas, can perform transmit beamforming (also known as \emph{precoding}) with the beamforming
vector $\mathbf{f}\in \mathbb{C}^{N_{\mathrm{t}} \times 1}$. While the receiver, equipped with $N_\mathrm{r}$ antennas, can perform receive beamforming (also known as \emph{combining}) with the beamforming vector $\mathbf{w}\in \mathbb{C}^{N_{\mathrm{r}} \times 1}$. Then, the received signal at the receiver is given by
\begin{align}\label{eq:ABF}
y=\mathbf{w}^{\mathrm{H}}  \mathbf{H} \mathbf{f} \sqrt{P} s+ \mathbf{w}^{\mathrm{H}}\mathbf{n},
\end{align}
where $\mathbf{H}\in\mathbb{C}^{N_{\mathrm{r}} \times N_{\mathrm{t}}}$ is the channel matrix between the transmitter and the receiver. 
$\mathbf{n} \sim \mathcal{CN}(\mathbf{0},\sigma^2 \mathbf{I}_{N_\mathrm{r}})$ is the white Gaussian noise at the receiver, and $\sigma^2$ is the average power of the noise at each antenna branch.
Obviously, beamforming vectors $\mathbf{w}$ and $\mathbf{f}$ can be optimized to improve communication performance, such as increasing SNR and reducing power consumption. Note that in general, all the PAs have the same scaling factor. Therefore, the analog beamforming vectors have constant-modulus (CM) elements~\cite{xiao2018jointp,zhu2020millim}, which are
\begin{align} \label{CM-constraint}
	\left \{
	\begin{aligned}
	&|[\mathbf{w}]_n|=1/\sqrt{N_\mathrm{r}}, n=1,2,...,N_\mathrm{r}, \\
&|[\mathbf{f}]_n|=1/\sqrt{N_\mathrm{t}}, n=1,2,...,N_\mathrm{t}. 
	\end{aligned}
	\right.
\end{align}
}

\begin{figure}[t]
	\centering
	\includegraphics[width=0.45\textwidth]{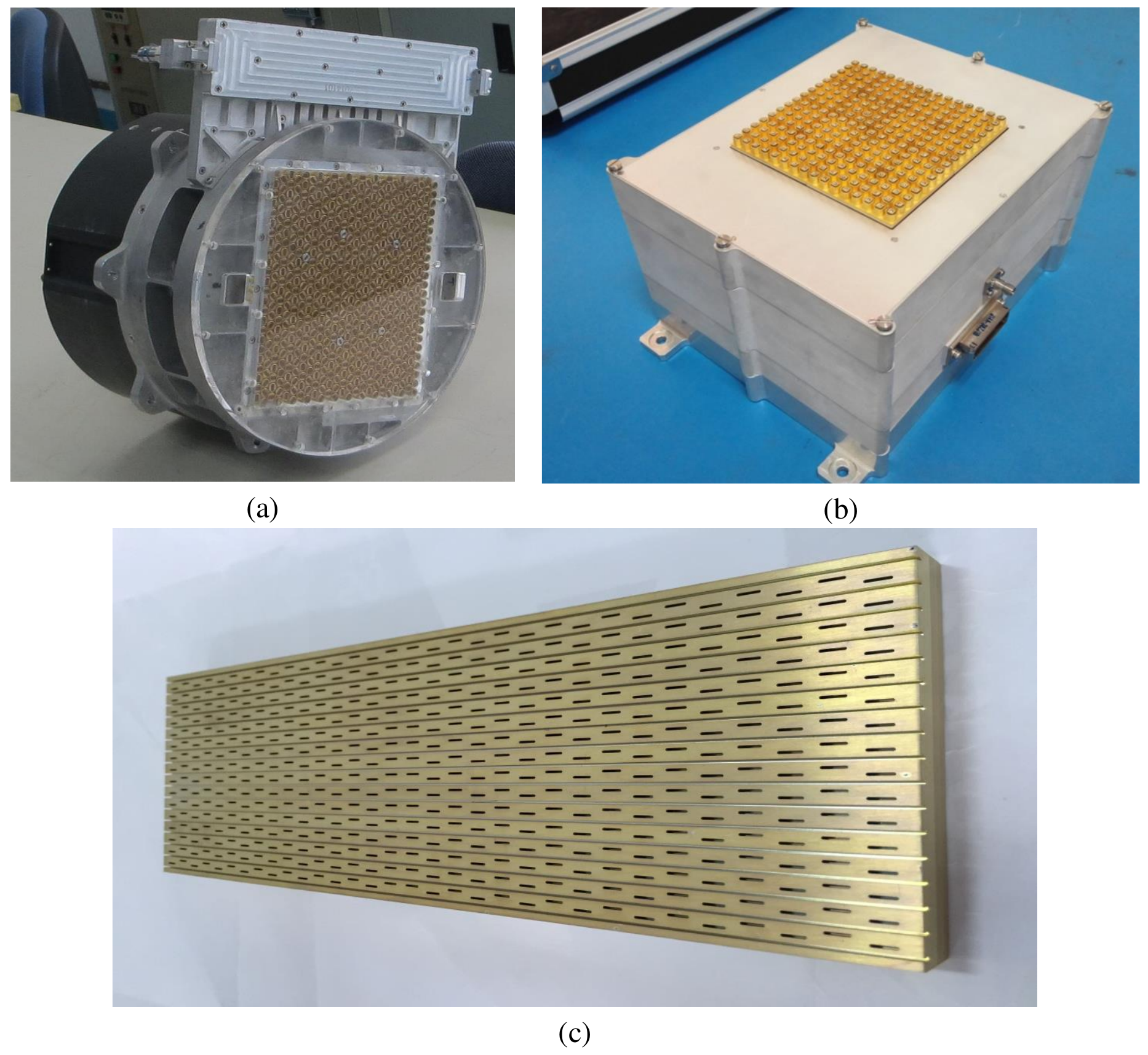}
	\caption{Phased arrays on satellite applications. (a) brick-structure phased array, (b) tile-structure phased array, (c) high efficiency phased array.}
	\label{fig-phased arrays on satellite applications}
\end{figure}

Earlier phased arrays are with a brick structure, and their T/R modules are separate from radiation elements. Fig.~\ref{fig-phased arrays on satellite applications}(a) shows a brick-structure phased array in the Ka band with 576 elements, whose beam scanning scope is $ \pm 60^\circ $. With the development of the integrated circuit technology, the tile structure is more popular for recent phased arrays, whose T/R modules are integrated with radiation elements, and multiple channels are designed together to decrease the thickness and power. A tile-structure phased array in the Ka band is shown in Fig.~\ref{fig-phased arrays on satellite applications}(b), which has a weight less than 2.4 kg and power less than 20 W. For satellite applications, the efficiency of arrays is a key point, then high efficiency radiation element and feeding network are required. Fig.~\ref{fig-phased arrays on satellite applications}(c) shows a 2D Ka-band phased array with high efficiency, consisting of waveguide slot antennas and feeding networks. 

\subsection{Digital Arrays}
\begin{figure}[t]
	\centering
	\includegraphics[width=8.89cm]{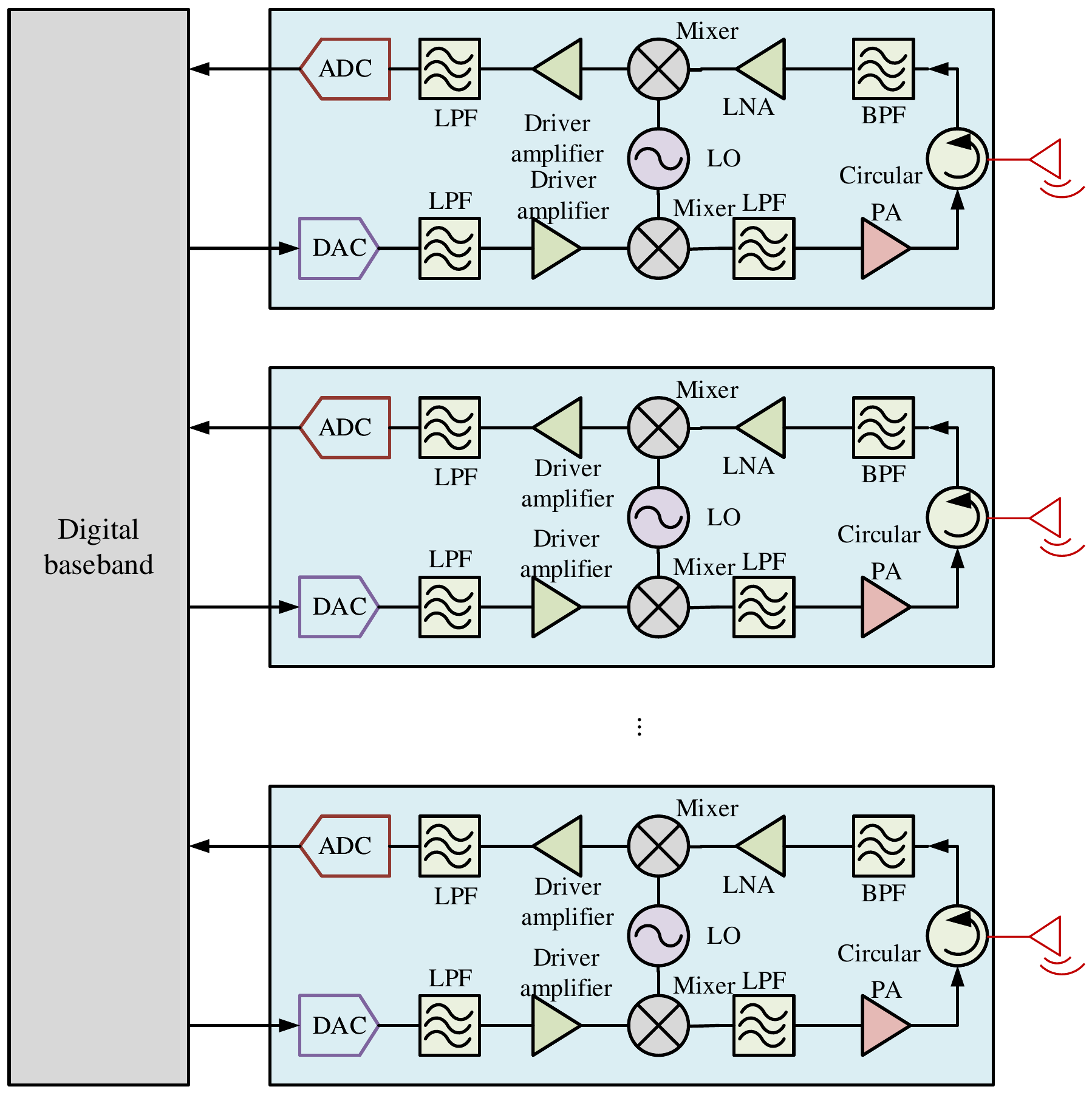}
	\caption{Structure of digital array.}
	\label{fig-Structure of digital array}
\end{figure}

Compared to phased arrays, digital arrays have better dynamic range and are easy to generate multiple beams simultaneously, which can serve users in different areas at the same time~\cite{xiao2021asurve, xiao2016enabli}. The structure of a digital array is shown in Fig.~\ref{fig-Structure of digital array}. For each channel, ADC and digital-to-analog converter (DAC) are required to receive and transmit baseband signals, and \emph{digital beamforming} is implemented in the baseband by digital weighting. 
\textcolor{black}{
Since each antenna is connected to an independent RF chain, multi-stream transmission can be supported. Supposing that a signal vector $\mathbf{s} \in \mathbb{C}^{N_{\mathrm{s}}\times 1}$ is transmitted from the transmitter to the receiver with power $P$, then the received signal vector at the receiver is given by
\begin{align}\label{eq:DBF}
	\mathbf{y}=\mathbf{W}^{\mathrm{H}}  \mathbf{H} \mathbf{F} \mathbf{s}+ \mathbf{W}^{\mathrm{H}}\mathbf{n},
\end{align}
where $\mathbf{W}\in \mathbb{C}^{N_{\mathrm{r}} \times N_{\mathrm{s}}}$, $\mathbf{F}\in \mathbb{C}^{N_{\mathrm{t}} \times N_{\mathrm{s}}}$ are the beamforming matrices at the receiver and the transmitter, respectively. $\mathbf{F}$ is supposed to satisfy the power constraint of 
\begin{align}\label{eq:DBF}
	\|\mathbf{F}\|_\mathrm{F}^2=P.
\end{align}
Since the signal processing is performed in the digital domain, there are sufficient degrees of freedom (DoFs) to implement efficient beamforming algorithms, to achieve satisfactory communication performance.
}

However, the digital array is usually expensive and requires more power than the phased array. Currently, it is mainly exploited in shipborne integrated electronic systems. Some digital arrays are also used in narrow band satellite communications now. However, with the development of technology, more and more digital phased array will be fabricated in space/air and ground communications. 

\subsection{Hybrid Antenna Arrays}
To balance the performance and cost, hybrid antenna arrays with analog-digital \emph{hybrid beamforming} are proposed and fabricated for coverage in satellite communications~\cite{peng2021hybrid}. In 5G wireless communications, the hybrid antenna arrays are also exploited in massive MIMO systems to decrease the cost with hybrid precoding and combining~\cite{ahmed2018asurve}. 
{\color{black}
For a hybrid antenna array, a large number of antennas are connected with small amount of RF chains via phase shifters, and thus multi-stream transmission are enabled with high gains and relatively low cost.  Let $N_\mathrm{t}^{\mathrm{RF}}$ and $N_\mathrm{r}^{\mathrm{RF}}$ denote the number of RF chains at the transmitter and receiver respectively, where we usually have $N_\mathrm{t}^{\mathrm{RF}} \ll N_\mathrm{t}$ and $N_\mathrm{r}^{\mathrm{RF}} \ll N_\mathrm{r}$. At the transmitter, signals are first processed by low-dimensional digital beamforming, and then processed by high-dimensional analog beamforming. While at the receiver, the order is reversed. The received signal vector at the receiver is given by
\begin{align}\label{eq:HBF}
	\mathbf{y}=\mathbf{W}_\mathrm{BB}^\mathrm{H} \mathbf{W}_\mathrm{RF}^\mathrm{H} \mathbf{H} \mathbf{F}_\mathrm{RF} \mathbf{F}_\mathrm{BB} \mathbf{s}+ \mathbf{W}_\mathrm{BB}^\mathrm{H} \mathbf{W}_\mathrm{RF}^\mathrm{H}\mathbf{n},
\end{align}
where $\mathbf{W}_\mathrm{BB}\in \mathbb{C}^{N_{\mathrm{r}}^\mathrm{RF} \times N_{\mathrm{s}}}$ and $\mathbf{W}_\mathrm{RF} \in \mathbb{C}^{N_{\mathrm{r}} \times N_{\mathrm{r}}^\mathrm{RF}}$ are the digital beamforming matrix and analog beamforming matrix at the receiver, respectively. While $\mathbf{F}_\mathrm{BB}\in \mathbb{C}^{N_{\mathrm{t}}^\mathrm{RF} \times N_{\mathrm{s}}}$ and $\mathbf{F}_\mathrm{RF} \in \mathbb{C}^{N_{\mathrm{t}} \times N_{\mathrm{t}}^\mathrm{RF}}$ are the digital beamforming matrix and analog beamforming matrix at the transmitter, respectively. In particular, analog beamforming matrices $\mathbf{W}_\mathrm{RF}$ and $\mathbf{F}_\mathrm{RF}$ are supposed to satisfy CM constraint of 
\begin{align}
		\left \{
	\begin{aligned}
		&|[\mathbf{W}_\mathrm{RF}]_{i,j}|=1/\sqrt{N_\mathrm{r}}, i=1,2,...,N_\mathrm{r}, j=1,2,...,N_\mathrm{r}^\mathrm{RF}, \\
&|[\mathbf{F}_\mathrm{RF}]_{i,j}|=1/\sqrt{N_\mathrm{t}}, i=1,2,...,N_\mathrm{t}, j=1,2,...,N_\mathrm{t}^\mathrm{RF}. 
	\end{aligned}
	\right.
\end{align}
Besides, transmit hybrid beamforming is supposed to satisfy power constraint of
\begin{align}\label{eq:HBF}
 \|\mathbf{F}_\mathrm{RF} \mathbf{F}_\mathrm{BB}\|_\mathrm{F}^2=P
\end{align}
}


\subsection{Irregular Antenna Arrays}
\begin{figure}[t]
	\centering
	\includegraphics[width=8.89cm]{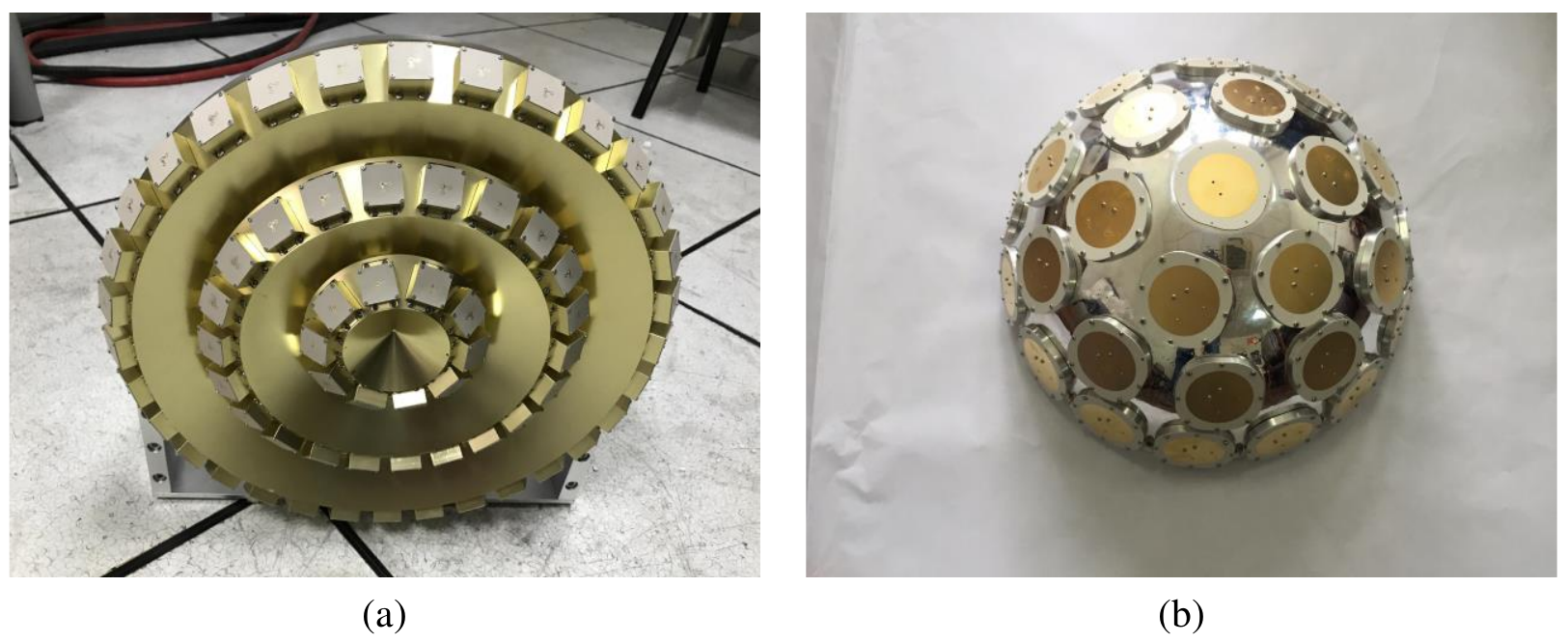}
	\caption{Irregular antenna arrays for space/air/ground communications. (a) triple round phased arrays in X-band, (b) hemisphere antenna array in S-band.}
	\label{fig-irregular antenna arrays}
\end{figure}

For wide coverage, lots of irregular antenna arrays were designed and fabricated for spaceborne and airborne communications. Compared to conventional antenna arrays, they are usually divided in several independent parts to cover as much space as possible~\cite{ma2019patter,rocca2014gabase,bui2018fastan}. For example, as shown in Fig.~\ref{fig-irregular antenna arrays}(a), triple round phased arrays in the X band were designed together to acquire beam coverage up to $ \pm 80^\circ $. In Fig.~\ref{fig-irregular antenna arrays}(b), multiple antennas were integrated on a hemisphere to track satellites in different direction by switching. The irregular antenna array provided a low-cost solution to the S-band satellite-ground digital datalink. Other irregular sparse antenna arrays, such as nested and co-prime arrays, are usually exploited in direction finding for wireless communications to provide high resolution.

\subsection{Programmable Metasurfaces}
In recent years, programmable metasurfaces were developed to substitute the phased array partially because of the simple structure, low power consumption, and low cost. Programmable metasurfaces have potential in various platforms including satellites and airplanes. They can also be exploited in the user end as high gain antennas. Programmable metasurfaces empowered RISs have aroused global attention and interest of both academia and industry, and thus constitute one of the key technologies in future 6G mobile network~\cite{pan2021reconf}.

\begin{figure}[t]
	\centering
	\includegraphics[width=\linewidth]{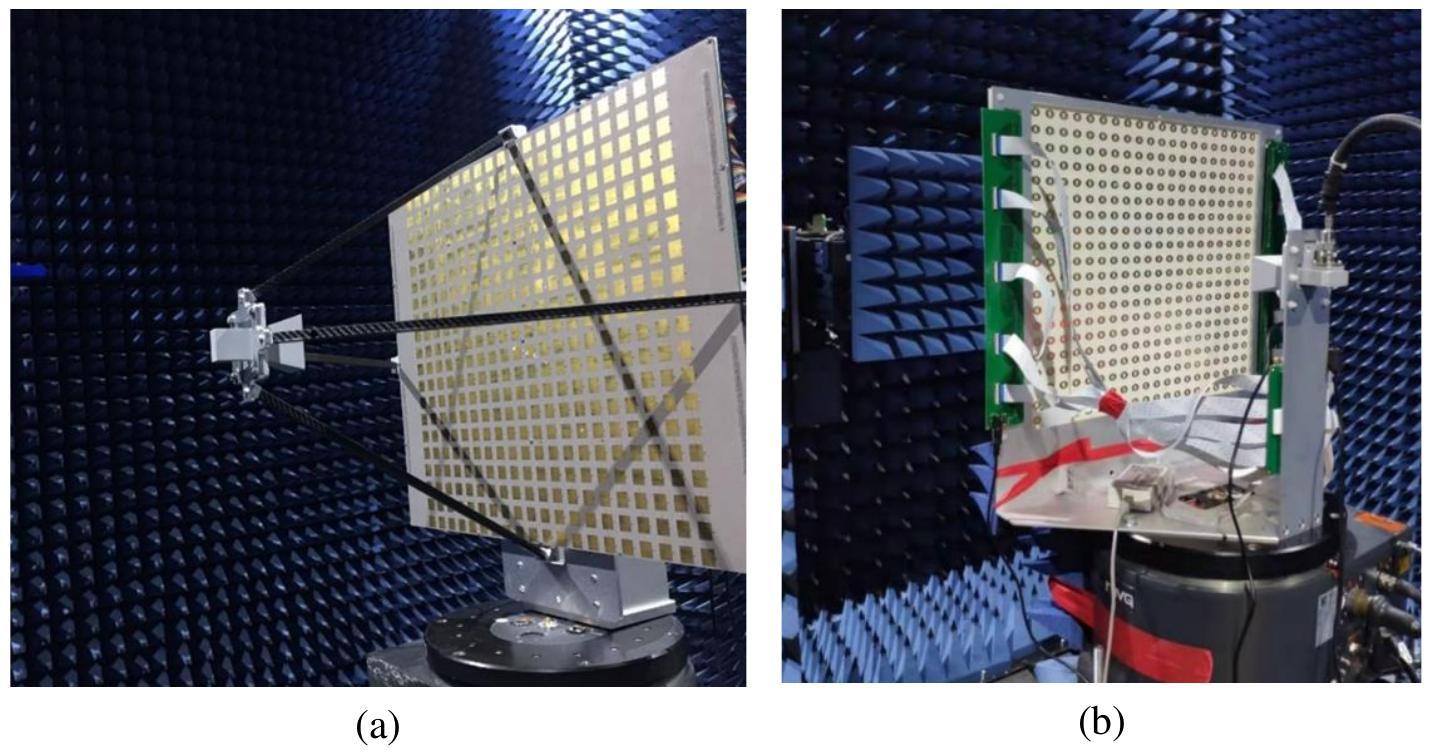}
	\caption{Two kinds of programmable metasurfaces. (a) reflective type, (b) transmissive type.}
	\label{fig-two kind of programmable metasurfaces}
\end{figure}

The beam of a programmable metasurface is controlled by the periodic structure of the metasurface, which is also know as \emph{passive beamforming}.  The phase state of each element is usually controlled with 0 and 180 degrees, while sometimes finer degree control may also be implemented. 
The binary phase states are easily implemented by PIN diodes on the surface.  Programmable metasurfaces use feeding antennas such as horn antennas to transmit and receive electromagnetic waves. The feeding antennas can be placed on both sides of the metasurface, as shown in Fig.~\ref{fig-two kind of programmable metasurfaces}. If the electromagnetic wave is reflected to the feeding antenna from the surface, the programmable metasurface is reflective. In contrast, if the electromagnetic wave penetrates the metasurface to the feeding antenna, the programmable metasurface is transmissive. Two kinds of metasurfaces are shown in Fig.~\ref{fig-two kind of programmable metasurfaces}(a) and Fig.~\ref{fig-two kind of programmable metasurfaces}(b), which work in C and X bands respectively~\cite{Bai2020}. For the reflective metasurface, it is fabricated in a two-layer printed circuit board (PCB), while the transmissive metasurface requires a four-layer PCB. Both metasurfaces adjust  beams with diodes, requiring no phase shifter and attenuator. Then compared to phased arrays, the programmable metasurface is much cheaper. Nevertheless, it may have disadvantages of a higher sidelobe level and lower efficiency.

{\color{black}
Taking the reflective metasurface as an example, the signal model is introduced as follows. An antenna of the transmitter sends a signal $s$, which  arrives at the $n$-th element of the metasurface through channel $h_n$. Then, the phase state of the reflected signal $y_n$ can be changed by manipulating the reflection coefficient $e^{j\theta_n}$ of the element, i.e., $y_n=e^{j\theta_n} h_n s$. Therefore, the reflected signal vector $\mathbf{y}=[y_1,y_2,...,y_N]^\mathrm{T}$ is given by
\begin{align}
	\mathbf{y}=\mathbf{\Theta}\mathbf{h}s,
\end{align}
where $\mathbf{\Theta}=\mathrm{diag}\{e^{j\theta_1},...,e^{j\theta_N}\}$ is the reflection coefficient matrix. $\mathbf{h}=[h_1,h_2,...,h_N]^\mathrm{T}$ is the channel vector between the transmit antenna and RIS. $N$ is the total number of reflecting elements. As can be seen, the reflection coefficient matrix is similar to an intermediate beamforming matrix, which can steer the reflected signal towards the desired receiver in a passive manner. Therefore, this process is known as passive beamforming~\cite{wu2020toward}.

}

{\color{black}
\subsection{Summary and Discussion}
In this section, varieties of antenna arrays were introduced. Fix-beam antenna arrays are cheap and have simple structures. However, beamforming is not supported and thus beam control is not flexible enough. 
Digital arrays require a dedicated RF chain for each antenna and have sufficient DoFs to implement efficient beamforming, which however require high hardware cost and power consumption. 
In contrast, phased arrays require only one RF chain connecting to all antennas, which are more energy and cost efficient at the sacrifice of the DoFs for beamforming. Hybrid antenna arrays require a small number of RF chains connecting to a large number of antennas via phase shifters or switches. This structure can strike a balance between communication performance and hardware cost, and thus becomes a promising paradigm in antenna array enabled space/air/ground communication systems with limited payload, power, and computation capacity. 
As a revolutionary technique for future 6G systems, programmable metasurfaces are able to change phase states of signals through integrating massive low-cost passive reflecting elements, which provide a low-complexity and cost-effective manner to manipulate the wireless propagation environment.
}
	
\section{Antenna Array Enabled Emerging Communication Technologies}\label{sec:emerging_technologies}
The conventional antenna array based beamforming techniques usually form a single narrow beam steering or tracking to one single target/user, which may not meet the new communication requirements of ubiquitous, flexible, and robust coverage. Hence, new beamforming techniques need to be developed to form on-demand coverage. In addition, to serve more users, new beam-domain multiple access schemes need to be studied. {\color{black}Moreover, because of the potential to simultaneously achieve high energy efficiency and spectrum efficiency, RIS has drawn significant attention.} Finally, antenna array may also be used to reduce information leakage and improve the physical-layer security via directional transmission. These new techniques are introduced below.

\subsection{New Beamforming Technologies}

Traditional beamforming techniques may not be suitable for future space/air/ground communication networks that require massive access and high dynamic due to the limited number of RF chains and high cost. Therefore, some new beamforming techniques have attracted wide concern and study.

\subsubsection {Single-RF-Chain Multiple Beams}

To balance hardware cost and beam gain in antenna array enabled communication systems, hybrid beamforming structure is a suitable paradigm, in which a few RF chains are able to be connected to a large antenna array. In general, the number of served users is no more than the number of RF chains because one RF chain can only shape one data stream. As the number of users increases, covering all the users becomes challenging due to the limited number of RF chains. Hence, single-RF chain multi-beam techniques are necessary to be developed, where one RF chain can generate multiple beams along different directions to serve multiple users.

\begin{figure}[t]
	\centering
	\subfigure[Illustration of the sub-array structure.]{\includegraphics[width=0.29\textwidth]{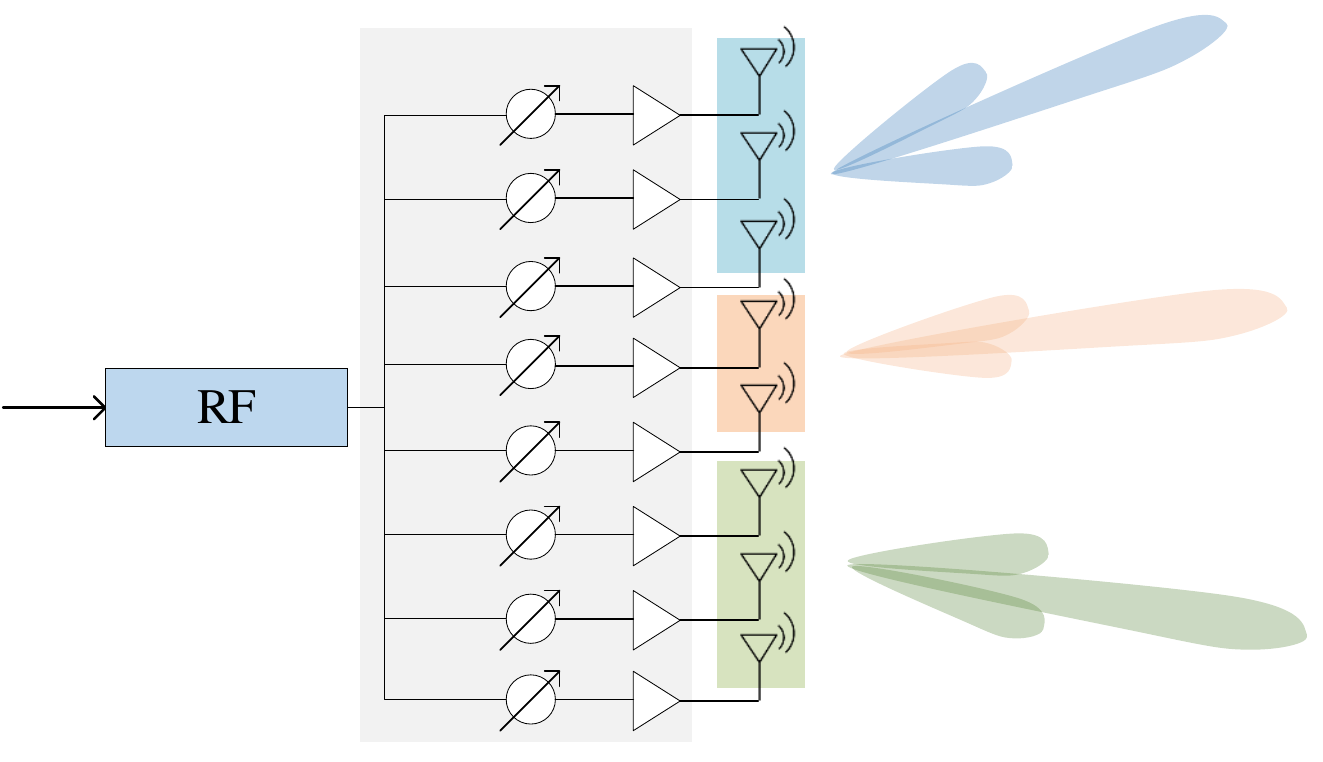}}
	\subfigure[{\color{black}Beam gain (dB).} ]{\includegraphics[width=0.19\textwidth]{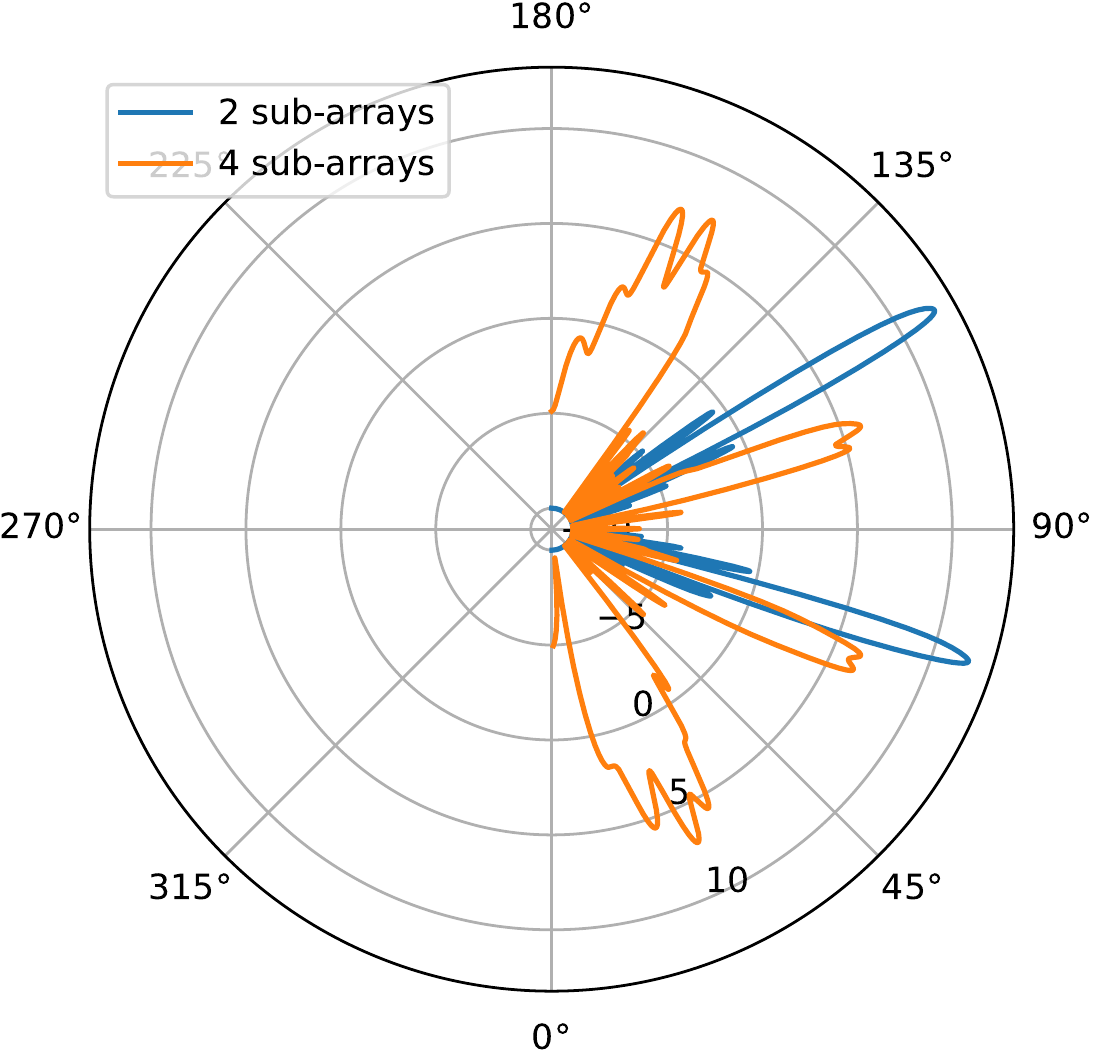}}
	\caption{Sub-array techniques for realizing multiple beams.}
	\label{sub-array technique}
\end{figure}

One possible method for single-RF-chain multi-beam is to divide the antenna array into several sub-arrays and each sub-array can generate a beam pointing to a specific direction as shown in Fig.~\ref{sub-array technique} (a). A certain number of adjacent antennas are separated in a group to form multiple sub-arrays that can generate analog beams steering to desired users~\cite{wei2019multib, li2020perfor}. 
{\color{black} We present the simulation results of generating multiple beams via sub-array techniques. First, a phased array with 64 elements is uniformly divided into  two sub-arrays and four sub-arrays, respectively. Then, analog beamforming is performed to generate  multiple beams  pointing  to different directions. As shown in  Fig.~\ref{sub-array technique} (b),
with different numbers of sub-arrays, the number of beams can be adjusted flexibly. Besides, as the number of sub-arrays increases, the number of antennas in each sub-array decreases, which results in a loss of the beam gain. Therefore, the tradeoff between the number of beams and beam gains should be considered carefully. }

Another way to synthesize multiple beams with a single-RF-chain antenna array is the optimization approach. In particular, an optimization problem may be formulated to maximize the beam gains along multiple desired directions. Meanwhile, the beam gains along other directions are constrained to a low level. 
{\color{black}
For an array with $N_\mathrm{t}$ elements, we denote the steering vector of the $i$-th desired beam and the analog beamforming vector as $\mathbf{a}_i$ and $\mathbf{f}$, respectively. Then, the array gain of the $i$-th beam is given by $G_i=|\mathbf{a}_i^\mathrm{H}\mathbf{f}|^2$. The optimization problem can be formulated as~\cite{xiao2018jointp,zhu2019millimTWC}
\begin{align}
	\max\limits_{\mathbf{f}, \alpha}~~ & \alpha  \label{prob_multibeam}   \\
	\mbox{s.t.}~~ 
	&|\mathbf{a}_i^\mathrm{H}\mathbf{f}|^2 \geq \alpha, i=1,2,...,I,  \tag{\ref{prob_multibeam}a}  \\
	&|[\mathbf{f}]_n|=1/\sqrt{N_\mathrm{t}}, n=1,2,...,N_\mathrm{t}, \tag{\ref{prob_multibeam}b}
\end{align}
where $I$ is the total number of desired beams. $\alpha$ is a slack variable to improve the beam gain along the desired  directions. 
}
It is shown in~\cite{xiao2018jointp,zhu2019millimTWC} that with the optimization approach multiple beams can also be well shaped.

\begin{figure}[t]
	\centering
	\includegraphics[width=0.43\textwidth]{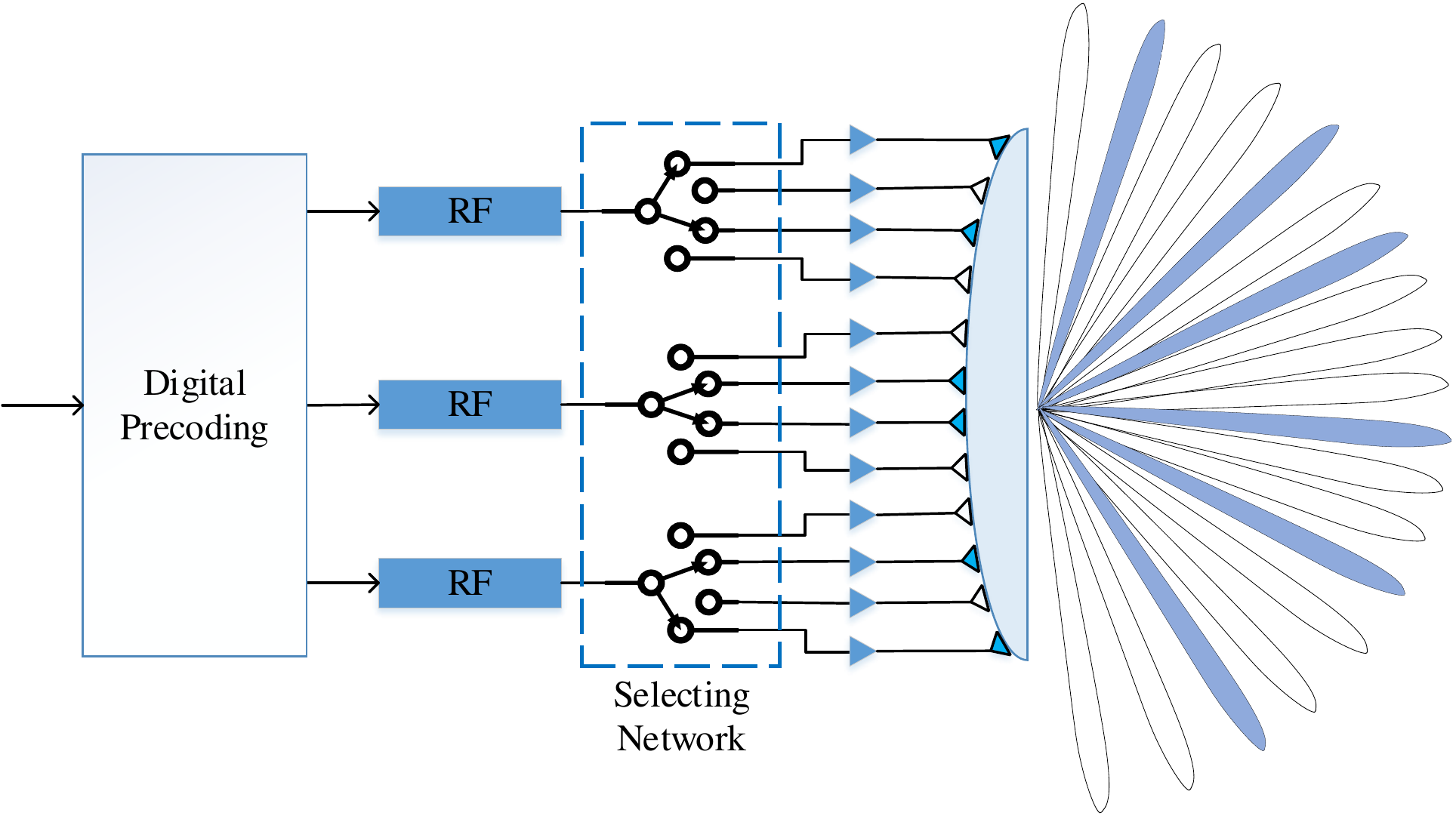}
	\caption{Architecture of the lens antenna array.}
	\label{lens array technique}
\end{figure}

In addition to the sub-array and optimization techniques, the RF chain can also be connected to a lens antenna array, which is a new path division multiplexing paradigm as shown in Fig.~\ref{lens array technique}, using power splitter/mixer and switch to generate multiple beams~\cite{xie2019onthep}. The RF chain can select different antennas to connect in order to generate beams steering to different directions after lens refraction, which greatly reduces beamforming complexity and hardware cost.


\subsubsection {Flexible Beam Coverage}

For conventional analog beamforming, a steering vector is usually selected as the analog beamforming vector to generate a narrow beam pointing to a specific direction. However, in some communication scenarios, a narrow beam is unable to meet the coverage requirement for limited coverage area. Moreover, it is worth noting that the target region may have random shapes and sizes. Therefore, a wide beam with larger coverage area and even a flexible beam that can cover a region with arbitrary shape and size are required.

\begin{figure}[t]
	\centering
	\includegraphics[width=0.45\textwidth]{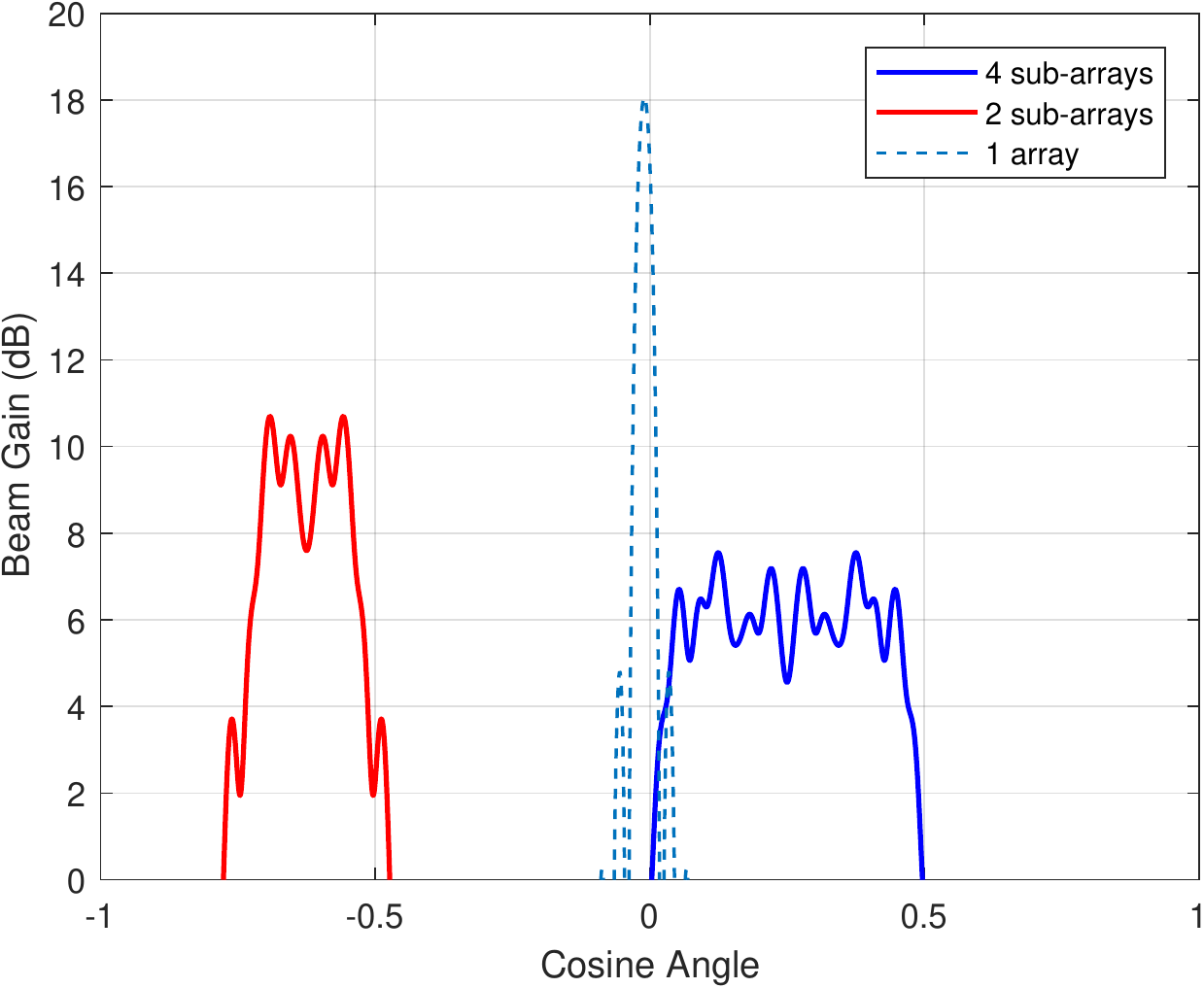}
	\caption{{\color{black}Sub-array scheme to realize flexible beam coverage.}}
	\label{flexibleBeam}
\end{figure}

To fully cover the target region, a wide beam that can cover a wide range of angles is needed, and the width of the beam should be adaptively adjusted according to the range of the target area. An analog phased array is more likely to adjust beam width in the angle domain because the beamwidth is approximately inversely proportional to the array size. For a specific angle range to cover, our previous work has proposed a sub-array scheme to divide the large array into several virtual sub-arrays and the beams generated by these sub-arrays are steered to evenly-spaced angled within the beam coverage~\cite{xiao2016hierar, xiao2018enhanc}. The combination of multiple wide beams can further expand the coverage range. 
\textcolor{black}{
Fig.~\ref{flexibleBeam} shows the sub-array scheme for realizing flexible beam coverage using a 64-element phased array~\cite{xiao2018enhanc}. As can be seen, two sub-arrays are used to generate beams to cover the wide area with cosine angle ranging from -0.75 to -0.5. To cover a cosine angle space ranging from 0 to 0.5, four sub-arrays are needed.}
In addition to the sub-array scheme, optimization scheme is also an effective method to design wide beam. For example, in order to form wide beams to ensure the coverage of broadcast control signals for massive MIMO systems, an alternating optimization algorithm was proposed in~\cite{kong2018hybrid} to optimize the hybrid precoder, which required the coverage probability for each user to be larger than a pre-determined threshold, and the transmit power to be as small as possible.

For certain special user distributions, a wide beam with circular covered area may not be appropriate because of unnecessary power cost in areas without users. Flexible beam coverage that can cover an area with arbitrary shape and size is more practical. A 3D beamforming approach with a uniform plane array (UPA) to realize flexible coverage is proposed for unmanned aerial vehicle (UAV)-enabled mmWave communications in our previous work~\cite{zhu2019dbeamf}. The large array is divided into multiple sub-arrays to generate wide beam and steer to different sub-areas. First, the position coordinates of the target area are transformed to a special-angle coordinates. Then  wide beams are used to cover the minimum rectangular envelop in angle domain that cover the whole desired area. Besides, phase rotation for sub-arrays is designed to reduce the fluctuation between different sub-arrays. The beamforming gain can be mainly concentrated in the target coverage area. For more complex scenarios, such as the shape and size of target region change dynamically, low-complexity beam pattern design is needed.

\subsubsection {Robust Beamforming}

Many existing works on beamforming are based on known user positions and perfect CSI at BSs~\cite{ayach2014spatia, zhang2019optima, xiu2021reconf}. In fact, it is hard for BSs to obtain accurate user position because of mobility and Global Positioning System (GPS) positioning error. Complex propagation environment, finite length of pilot signals and limited feedback bandwidth also bring great challenges for BSs to obtain perfect CSI. Besides, conventional training overhead for CSI estimation grows proportionally with the BS antenna size, which is quite large for large-antenna-array systems~\cite{lu2014anover,shu2018lowcom}. Therefore, in order to guarantee system performance and the quality of service, robust beamforming under imperfect CSI is needed. 

A commonly used model of channel estimation error is the bounded CSI error model, assuming that the estimation error of the CSI is within a specific range~\cite{liu2018roubst, zhu2018robust, xu2015robust}. The norm of these channel estimation errors is assumed to be bounded by a threshold and corresponding robust beamforming is designed for different objective functions.
{\color{black}
In the following, we give an example of maximizing the achievable rate with CSI errors, for the end-to-end communication scenario described in Section~\ref{PhasedArrays}. Instead of knowing the perfect CSI $\mathbf{H}$, we can only acquire the imperfect CSI $\mathbf{\tilde{H}}$ with CSI error $\boldsymbol{\Delta}$, i.e., $\mathbf{H}=\mathbf{\tilde{H}}+\boldsymbol{\Delta}$.
Therefore, the beamforming optimization problem of maximizing the achievable rate in the worst case can be formulated as follows
\begin{align}
	\max\limits_{\mathbf{f}, \mathbf{w}}~~ & \min\limits_{\boldsymbol{\Delta}} \log_2\left(1+\frac{\left|\mathbf{w}^\mathrm{H} (\mathbf{\tilde{H}}+\boldsymbol{\Delta}) \mathbf{f}\right|^{2} P}{\sigma^{2}}\right) \label{prob_robust}  \\
	\mbox{s.t.}~~ 
	&|[\mathbf{w}]_n|=1/\sqrt{N_\mathrm{r}}, n=1,2,...,N_\mathrm{r}, \tag{\ref{prob_robust}a}  \label{CM1} \\
&|[\mathbf{f}]_n|=1/\sqrt{N_\mathrm{t}}, n=1,2,...,N_\mathrm{t}, \tag{\ref{prob_robust}b} \label{CM2} \\
	&\left\|\boldsymbol{\Delta}\right\|_{2} \leq \epsilon, \tag{\ref{prob_robust}c} \label{CSI-error}
\end{align}
where $\sigma^2$ is the power of white Gaussian noise at the receiver. (\ref{CM1}) and (\ref{CM2}) are the CM constraints of $\mathbf{w}$ and $\mathbf{f}$, respectively. (\ref{CSI-error}) is the bounded CSI error model, which assumes that the CSI error $\boldsymbol{\Delta}$ is confined within an origin-centered hyper-spherical region with radius $\epsilon$.} Note that the robust beamforming problems are usually non-convex and are difficult to be solved by existing toolboxes. A feasible approach is converting the initial problem into several convex sub-problems that can be solved by convex optimization tools~\cite{liu2018roubst, xu2015robust}.


Another channel estimation error model is the probabilistic CSI error model, where the channel estimation errors are assumed to be statistically independent of the estimated channel matrix and characterized by a matrix-variate complex circular Gaussian distribution. With this model, a penalty dual decomposition based algorithm can be adopted to jointly optimize the digital and analog beamforming at BSs in order to maximize the system's worst-case sum rate~\cite{cai2019robust}.

\subsection{Multi-Antenna Multiple Access}


In the multi-antenna scenario, in addition to orthogonal multiple access (OMA), such as time-division multiple access (TDMA), frequency-division multiple access (FDMA), and code-division multiple access (CDMA), the utilization of antenna arrays opens the door to the spatial domain and multi-antenna processing.
The unique advantages of antenna array, like offering a high antenna gain by controlling the beam direction, increasing data rate by transmitting independent information simultaneously, and improving the communication reliability by transmitting and receiving redundant signals, inspire several new multiple access strategies.
Three advanced multi access technologies are proposed and discussed below.

\subsubsection{SDMA} Apart from OMA, space-division multiple access (SDMA), which is also known as beam division multiple access (BDMA), is the most common multiple access approach in multi-antenna system~\cite{spencer2004zerofo, stojnic2006ratema}. SDMA makes use of linear precoders/beamformers at the transmitter to separate users in the spatial domain, which is an effective way to increase the capacity and quality of wireless communications. By using SDMA, the BS can generate different beams and allocate them to cover users in different positions. Therefore, users covered by different beams can transmit their signals in parallel in the same time-frequency resource block.


SDMA can be used with various antenna array structures. With fully digital structures, SDMA can simultaneously support a number of users at most as high as the number of antennas at the BS. 
{\color{black}
With out of generality, we take  downlink two-user communication scenario as an example, where a BS equipped with an $N_\mathrm{t}$-element digital array serves two users with a single antenna. The signal vector is denoted by 
$
	\mathbf{s}=[s_1,s_2]^\mathrm{T}.
$
The signals are respectively weighted by the beamforming vectors $\mathbf{f}_1\in \mathbb{C}^{N_\mathrm{t}\times 1}$ and $\mathbf{f}_2\in \mathbb{C}^{N_\mathrm{t}\times 1}$ as follows
\begin{align}
	\mathbf{x}=\mathbf{F}\mathbf{s}=\mathbf{f}_1 s_1+\mathbf{f}_2 s_2,
\end{align}
where $\mathbf{F}\triangleq [\mathbf{f}_1, \mathbf{f}_2]$. 
Thereafter, the signal received at user $k$ is 
\begin{align}\label{y_SDMA}
	y_k&=\mathbf{h}_k^\mathrm{H}\mathbf{x}+n_k, k=1,2,
\end{align}	
where $\mathbf{h}_k \in \mathbb{C}^{N_\mathrm{t}\times 1}$ is the channel response vector between user $k$ and the BS. 
$n_k$ denotes the Gaussian white noise at user $k$ with power $\sigma_k^2$.
In SDMA, each user $k$ only decodes its desired signal $s_k$ by treating the interference from other streams as noise. Therefore, the achievable rate $r_k$ of user $k$ is given by
 \begin{align}
 		r_{k}=\log _{2}\left(1+\frac{\left|\mathbf{h}_k^\mathrm{H}\mathbf{f}_k \right|^{2} }{  \left|\mathbf{h}_k^\mathrm{H}\mathbf{f}_j \right|^{2} +\sigma_k^{2}}\right), (k,j)=\{(1,2), (2,1)\}.
 \end{align}
}
Popular designs of precoders include zero-forcing beamforming (ZFBF), though further enhancements are possible by other regularized ZFBF and optimized precoders. Such an architecture is very popular in 4G and 5G multi-user MIMO, massive MIMO, and coordinated multipoint (CoMP). 

However, the performance of SDMA depends on the number of RF chains. If the number of users is larger, all users cannot be ensured to be covered by a beam generated by an independent RF chain, and thus the inter-beam interference becomes inescapable. Therefore, the key issue is to design a proper group method, allowing users in different groups to access the BS simultaneously without interfering each other. A possible way is to group users according to the number of RF chains. Since users with highly correlated channels covered by different beam will lead to high interference, such interference can be cancelled by assigning those users into the same group~\cite{mauricio2018alowco,sun2017agglom}. By doing so, users with low-correlation channels are divided into different groups and can be served by the BS at the same time-frequency-code domain by performing SDMA. For users in the same group, each RF chain can shape one or more beams to fully cover them according to the users distribution. Then, the users in the same group can be served by performing OMA strategies.

\subsubsection{NOMA}\label{NOMA} 
Non orthogonal multiple access (NOMA) is a promising technology to support multiple access. Different from OMA whose performance is limited by the orthogonal resources, NOMA strategy allows multiple users to access in the same time-frequency-code domain, and distinguish them in the power domain. Specifically, the transmitter superimposes the signals in the resource block, where the power level is decided according to effective channel gain of each user, where higher power is allocated to the signal of user with lower channel gain. The receiver uses successive interference cancellation (SIC) technology to decode the signals in a successive way. 

{\color{black}
Let us consider the above downlink two-user communication scenario as an example. Contrary to SDMA, superposition coding is used in NOMA and one of the two users is required to decode the interfering signal before decoding the desired signal~\cite{mao2018energy}. 
There are two decoding orders~\cite{xiao2018jointp}, which are \emph{decoding $s_1$ first} and \emph{decoding $s_2$ first}. 
For the former, user 1 directly decodes $s_1$ by treating the signal component of $s_2$ as noise. While user 2 first decodes $s_1$ and subtracts it from the received signal $y_2$, and then decodes $s_2$ without the interference from $s_1$. Therefore, the achievable rates of the two users are given by
\begin{align}
\left\{\begin{array}{l}
	r_{1}=\log _{2}\left(1+\frac{\left|\mathbf{h}_{1}^{\mathrm{H}} \mathbf{f}_1\right|^{2} }{\left|\mathbf{h}_{1}^{\mathrm{H}} \mathbf{f}_2\right|^{2} +\sigma_1^{2}}\right), \\
	r_{2}=\log _{2}\left(1+\frac{\left|\mathbf{h}_{2}^{\mathrm{H}} \mathbf{f}_2\right|^{2} }{\sigma_2^{2}}\right).
\end{array}\right.
\end{align}}
According to the analysis in~\cite{ding2017asurve, zhu2019optima, zhang2020optima}, for downlink NOMA, signals of users with worse channel gain will be decoded first. While for uplink NOMA, signals of users with the higher channel gain and lower data rate requirements are decoded first. Even though these works~\cite{ding2017asurve, zhu2019optima, zhang2020optima} focus on single-antenna NOMA systems, the above conclusions on decoding order could be used for multi-antenna NOMA (though no guarantee on optimality).

The high propagation attenuation of high frequency signals, such as mmWave signals, enlarges the channel difference of users,  while NOMA may achieve higher achievable-rate gains compared to OMA for users with significantly different channels.
Moreover, due to the characters of directional propagation, users in the same beam can take full advantage of the array gain by applying NOMA. Several comparisons are made in~\cite{zeng2017capaci,wei2019multib,xiao2018jointp,cui2018optima,zeng2019energy} and it is found that NOMA can be designed to achieve a better performance than OMA in the aspects of spectral efficiency, sum-rate and the number of served users, due to exploiting the additional freedom of beamforming besides power allocation.
Therefore, instead of OMA, NOMA can be combined with SDMA and utilized to transmit signals of users in the same group~\cite{zhu2019millimAcc}. It was shown that the joint SDMA and NOMA scheme outperforms the SDMA scheme and the joint SDMA and OMA scheme in terms of achievable sum-rate~\cite{zhu2019millimTWC,khaled2020joints}.

\subsubsection{RSMA} To further improve the spectral and energy efficiency, multiple access may utilize multi-antenna rate-splitting multiple access (RSMA) techniques~\cite{Clerckx:2016,Joudeh:2016,naser2020ratesp}. RSMA relies on linear precoded rate-splitting at the transmitter and SIC at the receivers~\cite{mao2017rate}. In contrast to SDMA and NOMA where each message to transmit is directly encoded into a corresponding stream, in RSMA, messages are split into common and private parts such a part of the message of a given user is decoded by all users.

{\color{black}Let us still  consider the above downlink two-user communication scenario as an example. At the BS, the messages intended to both users are split into two parts, a common part and a private part. The common parts of the messages of both users are then encoded into the common stream $s_\mathrm{c}$, while the private parts of the messages are independently encoded into the private streams $s_1$ and $s_2$. Therefore, three streams $s_\mathrm{c}, s_1,s_2$ are created, and the transmit signal vector is 
$
	 \mathbf{s}=[s_\mathrm{c}, s_1, s_2]^\mathrm{T}.
$
After that, linear precoding can be performed with digital beamforming matrix $\mathbf{F}=[\mathbf{f}_\mathrm{c}, \mathbf{f}_1, \mathbf{f}_2]$, such that the signal vector is precoded as 
\begin{align}
\mathbf{x}=\mathbf{F}\mathbf{s}= \mathbf{f}_{\mathrm{c}} s_{\mathrm{c}}+\mathbf{f}_{1} s_{1}+\mathbf{f}_{2} s_{2},
\end{align}
where the common stream $s_\mathrm{c}$ is precoded by $\mathbf{f}_\mathrm{c}$ in a multicast manner so as to deliver it to both users, while the private 
streams $s_1$ and $s_2$ are mapped to the transmit antenna array through legacy multiuser-precoders $\{\mathbf{f}_{k}\}$ (e.g. ZFBF). The formula of the received signal at user $k$ is the same with (\ref{y_SDMA}).  At each user $k$, the common stream $s_c$ is first decoded  by treating the interference from the private streams as noise. 
Therefore, the achievable rate of the common stream  is given by
 \begin{align}
	r_{\mathrm{c},k}=\log _{2}\left(1+\frac{\left|\mathbf{h}_k^\mathrm{H}\mathbf{f}_\mathrm{c} \right|^{2} }{  \left|\mathbf{h}_k^\mathrm{H}\mathbf{f}_1 \right|^{2}+\left|\mathbf{h}_k^\mathrm{H}\mathbf{f}_2 \right|^{2} +\sigma_k^{2}}\right), k=1,2.
\end{align}
Note that in order to ensure successful decoding of the common stream $s_\mathrm{c}$ at both users, the common rate shall not exceed $r_\mathrm{c}=\min (r_{\mathrm{c},1},r_{\mathrm{c},2})$.
Using SIC, the common stream is subtracted from the received signal, such that user $k$ can decode its private stream $s_k$ without the interference from $s_\mathrm{c}$. Therefore, the achievable rate of the private stream is given by
 \begin{align}
	r_{k}=\log _{2}\left(1+\frac{\left|\mathbf{h}_k^\mathrm{H}\mathbf{f}_\mathrm{k} \right|^{2} }{  \left|\mathbf{h}_k^\mathrm{H}\mathbf{f}_j \right|^{2} +\sigma_k^{2}}\right), \forall (k,j)\in \{(1,2), (2,1)\}.
\end{align}

To illustrate the superiority of RSMA over various baselines, Fig.~\ref{fig_RSMA} displays the max min fair (MMF) rate (schemes are designed to maximize the minimum rate among all users) performance of various schemes for a scenario of 6 users and 4 transmit antennas under perfect channel state information at the transmitter (CSIT)~\cite{clerckx2021isnoma}. The conventional multi-user linear precoding MU-LP is the most common SDMA schemes and relies on optimized precoders. Two multi-antenna NOMA schemes are also provided and optimized, one optimized for a single group (G=1) of users (hence one user decodes the messages of 5 other users), and the other optimized for 3 groups (G=3) of 2 users (hence in each group, there are two users, and one of the users decodes the messages of the other users). We note that RSMA, implemented using the so-called 1-layer RS scheme~\cite{Clerckx:2016, Joudeh:2016, mao2017rate}, provides significant performance gains over all other baselines and only requires one SIC at the receivers. In contrast NOMA scheme with G=3 requires 5 SIC layers at the receivers and achieves significantly lower performance. This demonstrates how RSMA can achieve significant performance gain with relatively low receiver complexity thanks to the ability to partially decode interference and partially treat it as noise~\cite{clerckx2021isnoma}.
\begin{figure}[t]
	\centering
	\includegraphics[width=0.45\textwidth]{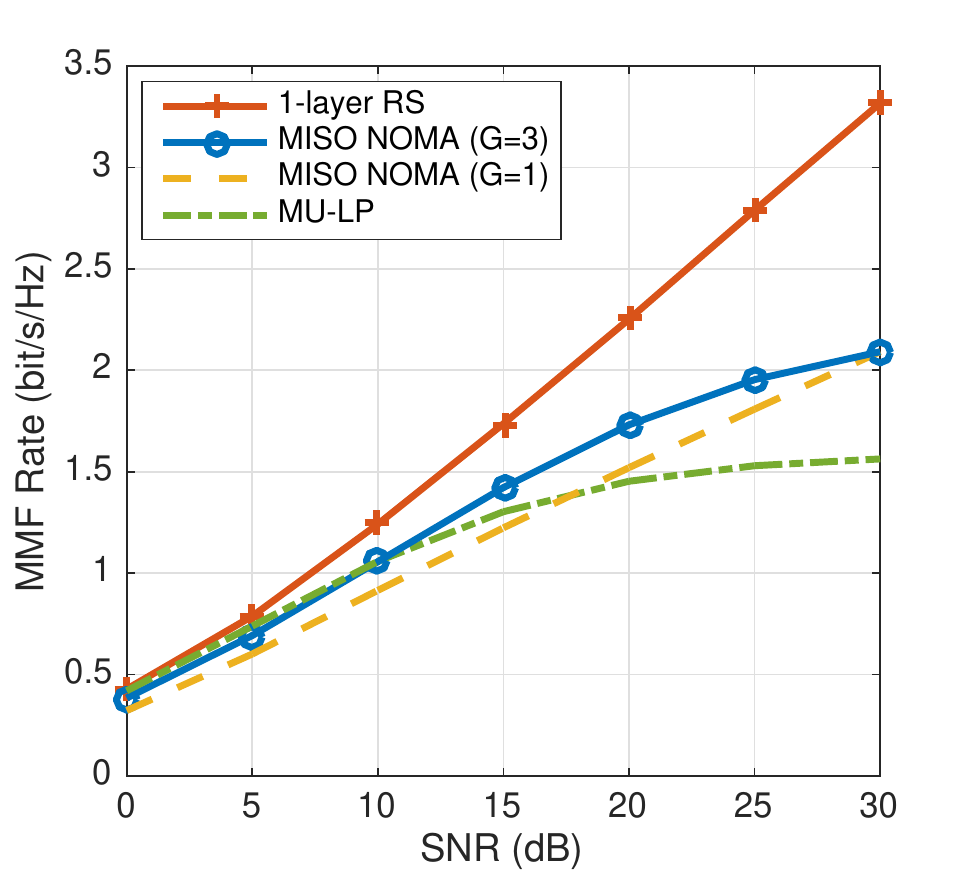}
	\caption{{\color{black}the MMF rate  performance of various multiple access schemes.}}
	\label{fig_RSMA}
\end{figure}
}
RSMA can adjust the amount of interference that is decoded at each user by adjusting the common stream (power and content). Since SDMA treats inter-beam interference as noise and NOMA fully decodes the interference, the two schemes (along with OMA and other schemes as multicasting) can be seen as special cases of RSMA~\cite{mao2017rate, Clerckx:2020}. This shows how powerful RSMA can be at unifying unrelated multiple access techniques into a single framework seemingly.

Compared with classical approaches developed under the assumption of perfect CSIT,  RSMA is information theoretically optimal in terms of achieving the largest achievable multiplexing gains in the presence of imperfect CSIT~\cite{Davoodi:2016,Davoodi:2020,Joudeh:2016,Piovano:2017,Hao:2017}. In other words, RSMA is inherently robust to imperfect CSIT. This optimality provides a firm theoretical ground and further motivates the design of multiple access and robust interference management techniques based on the rate-splitting principle. In particular, it changes our perspective on how to exploit imperfect CSIT. For instance, it was shown that though Dirty Paper Coding (DPC)~\cite{Costa83} is capacity achieving in multi-antenna broadcast channel with perfect CSIT, RSMA outperforms DPC in the presence of imperfect CSIT despite relying on relatively simpler linear precoding~\cite{Mao:2020}. Nevertheless, a capacity-achieving strategy in the presence of imperfect CSIT remains unknown.

Some works including~\cite{mao2017rate, Clerckx:2020,clerckx2021isnoma,mao2019ratesp} and references therein have also compared RSMA, SDMA, and NOMA and showed the superiority of RSMA over SDMA and NOMA in terms of spectral and energy efficiency, robustness to imperfect CSIT, capability in supporting a larger number of users, flexibility to user deployments (in terms of channel alignment/orthogonality and channel strength disparity among users) and network load (underloaded or overloaded), etc. Superiority has also been studied and demonstrated in satellite and aerial networks~\cite{Yin:2021,Jaafar:2020,Si:2021}.

\subsection{RISs}



Thanks to lots of key technologies, such as massive MIMO, ultra-dense network, and mmWave communication, the target of 5G has been largely accomplished. However, the prohibitive hardware cost and complexity, and increasing energy consumption have become by-products and are remaining unsolved. 
As a result, RIS, as an emerging paradigm to simultaneously achieve high energy efficiency and spectrum efficiency, has drawn significant attention in both academia and industry~\cite{wu2020toward,gong2020toward, mahmoud2021intell,lei2021reconf}. Specifically, a RIS is a planar surface composed by an array of passive programmable reflecting elements, each of which can independently induce different reflection amplitude and/or phase shift on the incoming signal. Thus, RIS is able to manipulate electromagnetic waves to reconfigure them toward their desired directions, which is usually known as \emph{passive beamforming}. RISs are more energy and cost efficient, since they only  reflect the incoming signals passively, without the need of baseband signal processing modules and RF chains. Besides, RISs can achieve higher spectrum efficiency, because they can provide powerful passive beamforming gains and naturally operate in full-duplex mode without self-interference or introducing thermal noise. Therefore, there are growing interests on the use of RISs for realizing future 6G wireless networks, mainly including \emph{passive relay} and \emph{passive transmitter}, as introduced below.

\subsubsection{Passive Relay}
\begin{figure}[t]
	\begin{center}
		\includegraphics[width=\linewidth]{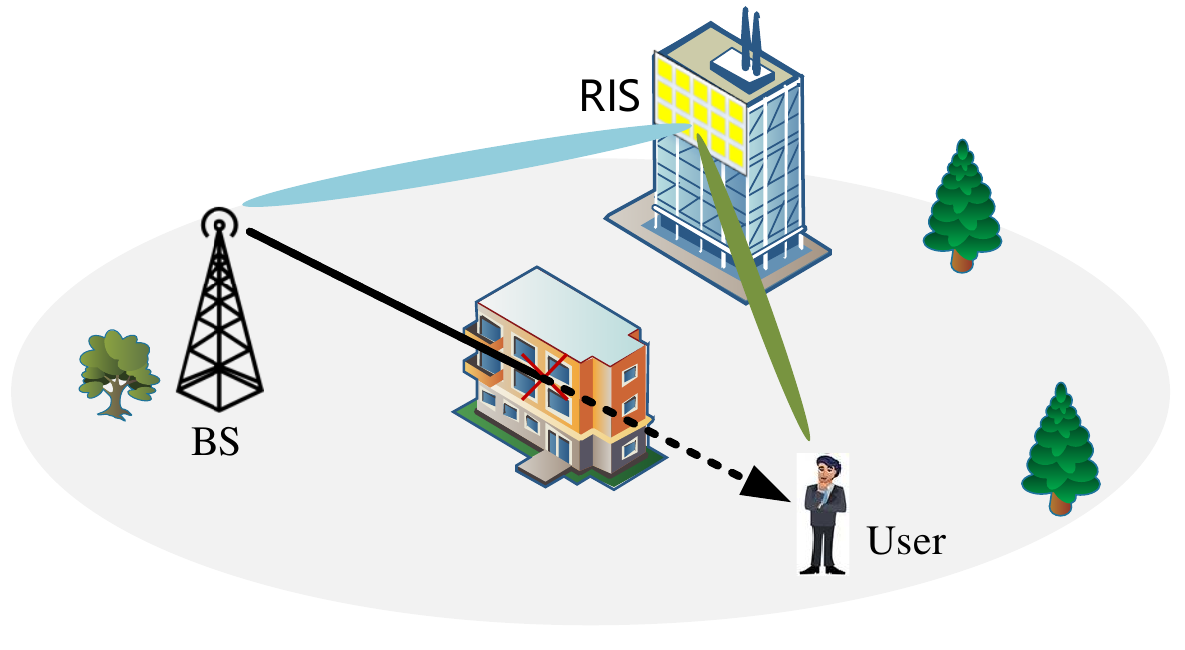}
		\caption{RIS for coverage enhancement.}
		\label{fig:RIS_coverage_enhancement}
	\end{center}
\end{figure}

As a promising technology which can manipulate electromagnetic waves, RISs can be deployed as passive relays for communication coverage enhancement and extension by intelligently changing the propagation environment between the transmitter and the receiver. As shown in Fig.~\ref{fig:RIS_coverage_enhancement}, the heavily blocked BS-User direct communication link is replaced by two clear line-of-sight (LoS) links, namely, the BS-RIS link and the RIS-user link, where RIS essentially acts as a passive relay. 
{\color{black}Suppose that the RIS has $M$ reflecting elements, whose reflection coefficient matrix is $\Theta=\mathrm{diag}\{e^{j\theta_1},...,e^{j\theta_M}\}$. The BS is equipped with an $N_\mathrm{t}$-element phased array, and thus beamforming can be performed. The user is equipped with a single antenna. Suppose that the direct BS-user link is heavily blocked by buildings and the received signal is negligible. Therefore, the signal arrives at the user through the BS-RIS channel $\mathbf{G}\in \mathbb{C}^{M\times N_\mathrm{t} }$ and RIS-user channel  $\mathbf{h}^\mathrm{H} \in \mathbb{C}^{1 \times M}$ is given by
\begin{align}
	y=(\mathbf{h}^\mathrm{H} \Theta \mathbf{G} )\mathbf{f} \sqrt{P} s+ n,
\end{align}	
where $s$ denotes the transmitted signal and $P$ is the transmit power. $\mathbf{f}\in \mathbb{C}^{N_\mathrm{t}\times 1}$ is the beamforming vector of BS. $n$ denotes the Gaussian white noise at the user.
In this way, the user located in a dead zone of the BS can be successfully covered with the aid of RIS. Besides, the coverage performance can be optimized via jointly designing passive beamforming $\Theta$ and transmit beamforming $\mathbf{f}$.}

To achieve satisfactory performance, the deployment of RISs in a hybrid wireless network including both active transmitters and passive RISs is a crucial problem. Generally speaking, the deployment of RISs should consider the link conditions with both transmitters and receivers, spatial user density, inter-cell interference issue, and so on~\cite{wu2020toward,liu2021risenh}. 
Besides, passive beamforming design is essential to steer the reflected signals toward the desired directions. To optimize the network performance, passive beamforming of RIS is necessarily jointly designed with active beamforming of transmitters/receivers~\cite{wu2019intell, wu2020jointa, chen2021jointa}. It is preferable to consider practical hardware constraints such as discrete amplitude and phase-shift levels~\cite{wu2020beamfo,rehman2021jointa,abeywickrama2020intell} in the beamforming design, to ensure practical communication performance. Moreover, the combination of RIS with other technologies such as NOMA~\cite{liu2021risenh, zuo2020intell,bariah2021largei} and terahertz communications~\cite{hao2021robust, chen2021toward}, and the application of RIS in various platforms such as UAV~\cite{guo2021learni, li2021robust,hua2021uavass} would open up new forward directions for the future 6G communication networks.

\subsubsection{Passive Transmitter}
In addition to be deployed in the far-field region of the transmitter as a passive relay, RIS can also be utilized as a passive transmitter fed by a nearby RF signal generator. Specifically, the RF signal generator is responsible for feeding an unmodulated carrier signal to RIS. While the RIS  modulates and delivers information symbols by exploiting the carrier signal through carefully control reflection coefficients of each reconfigurable reflecting element.

\begin{figure}[t]
	\begin{center}
		\includegraphics[width=0.85\linewidth]{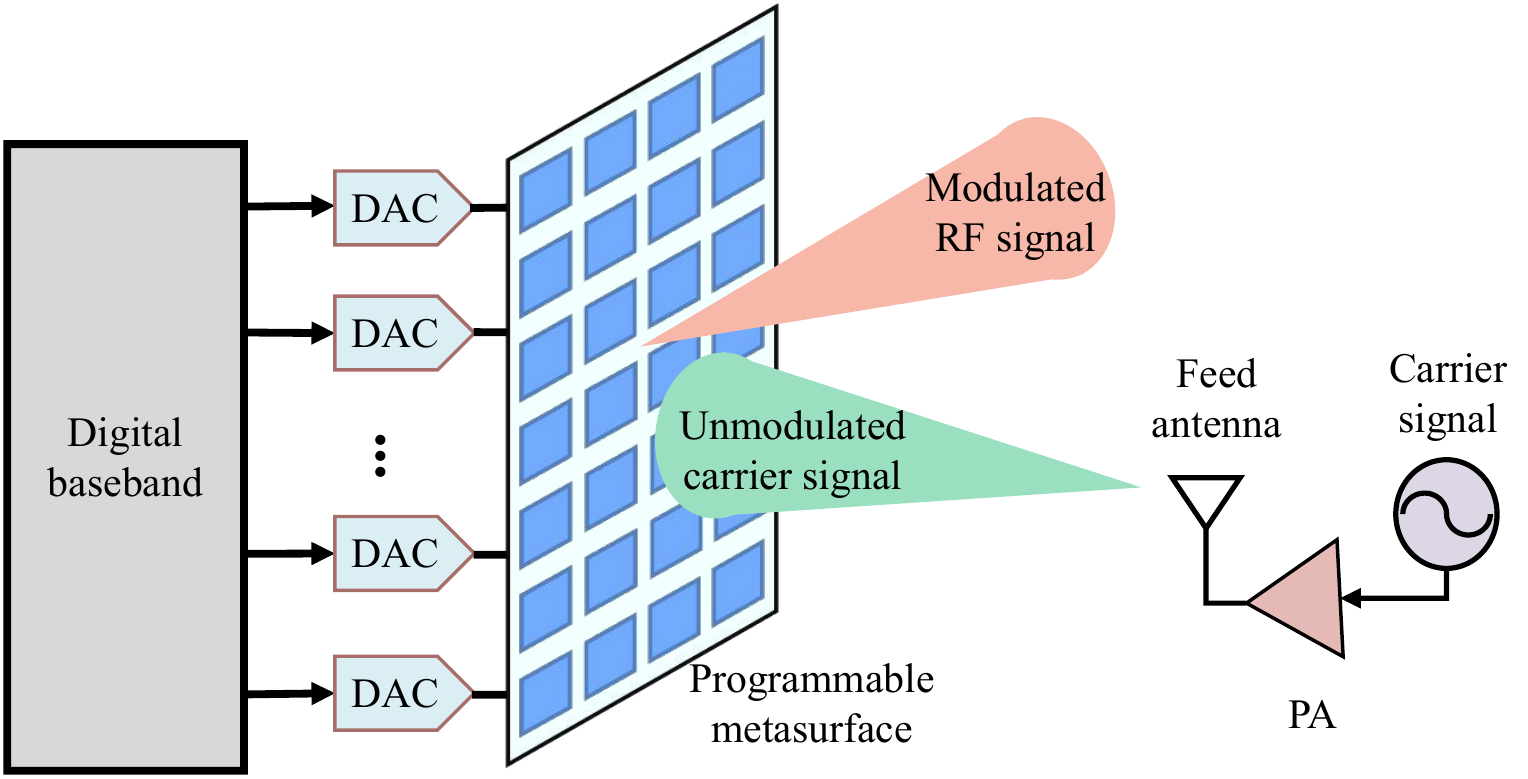}
		\caption{RIS-based wireless transmitter.}
		\label{fig:RIS_transmitter1}
	\end{center}
\end{figure}

Fig.~\ref{fig:RIS_transmitter1} shows the architecture of a RIS-based transmitter proposed in~\cite{tang2020mimotr}. Different from conventional  transmitter requires multiple RF chains where each RF chain needs DACs, mixers, PAs, and filters, the proposed RIS-based transmitter is RF-free and requires only one narrow band PA to manage the transmit power of the air-fed carrier signal. Compared with conventional architecture, this RF-free architecture greatly reduces the hardware complexity, cost, and power consumption. By mapping the control signals generated by the digital baseband to the RIS, phase shift keying (PSK) modulation~\cite{tang2020wirele,tang2020mimotr} and quadrature amplitude modulation (QAM)~\cite{zhang2021reconf} can be achieved by manipulating different phase/amplitude of the reflected RF signals.
Simultaneous transmission of multi-channel RF signal is supported by independently controlling phase/amplitude  through a dedicated DAC for each reconfigurable element. Thus, advanced signal processing methods such as beam steering and space-time modulation for MIMO communicaitons are enabled. 

\begin{figure}[t]
	\begin{center}
		\includegraphics[width=0.85\linewidth]{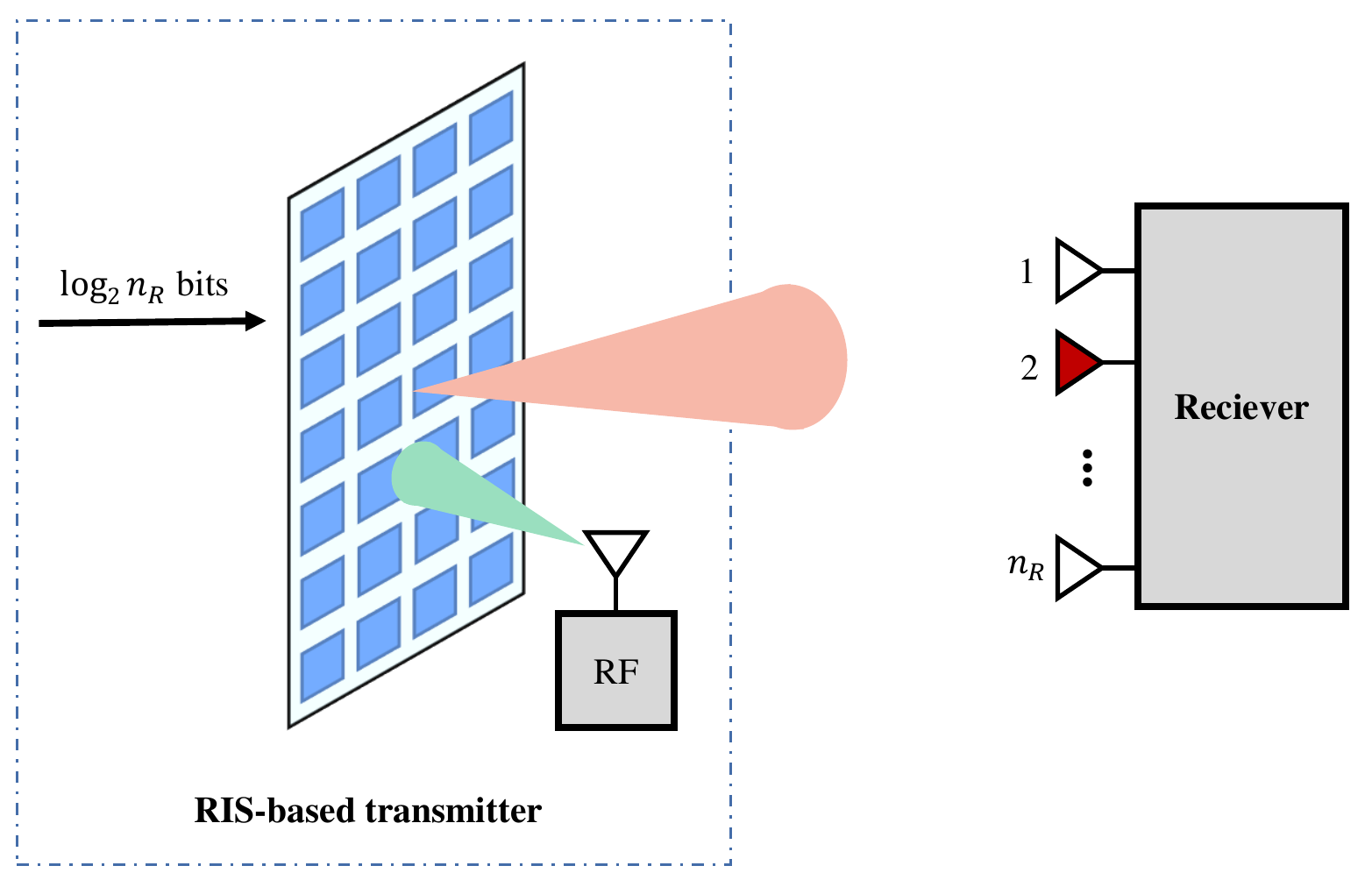}
		\caption{RIS-based index modulation system.}
		\label{fig:RIS_transmitter2}
	\end{center}
\end{figure}

In addition to conventional  modulation such as PSK and  QAM, RIS-based transmitter can also realize various reflection pattern-related modulation schemes~\cite{guo2020reflec,ma2020passiv,yan2020passi,basar2020reconf, canbilen2020reconf}. Essentially, these schemes map information bits to different phase shift matrices of the RIS, while the receiver can demodulate the information through detecting which phase shift matrix is used. For example, Fig.~\ref{fig:RIS_transmitter2} shows a RIS-based index modulation system proposed in~\cite{basar2020reconf}, which maps the information to the index of receive antenna.
Specifically, the receiver equipped with $N_\mathrm{R}$ antennas lies in the far-filed of the RIS-based transmitter. At the transmitting end, the incoming $\log_2 N_\mathrm{R}$ information bits specify the index $m$ of a receive antenna. Then the phase shifts of the RIS are adjusted accordingly to maximize the received SNR at the $m$-th receive antenna. Thereafter, the unmodulated carrier signal generated by the RF signal generator is modulated through the RIS and reflected to the receiver. At the receiving end, the information can be demodulated by detecting the instantaneous received SNR at each receive antenna. 

In summary, RIS-based transmitter opens up a new paradigm to achieve cost-effective and energy-efficient information modulation for the future 6G communication, which is worth for further research efforts.

\subsection{Secure Communications}

Antenna arrays can reduce information leakage and improve  physical-layer security~\cite{Liang_2008, Bloch11, Bloch21,wang2020energy, wang2021robust} thanks to the spatial selectivity. Physical-layer security means security that can be guaranteed in the physical layer by using the difference in channel characteristics. 
Thus, physical-layer security differs from computational security, which is at the heart of traditional cryptographic algorithms.

\begin{figure}[t]
	\centering
	\includegraphics[width=6.52cm]{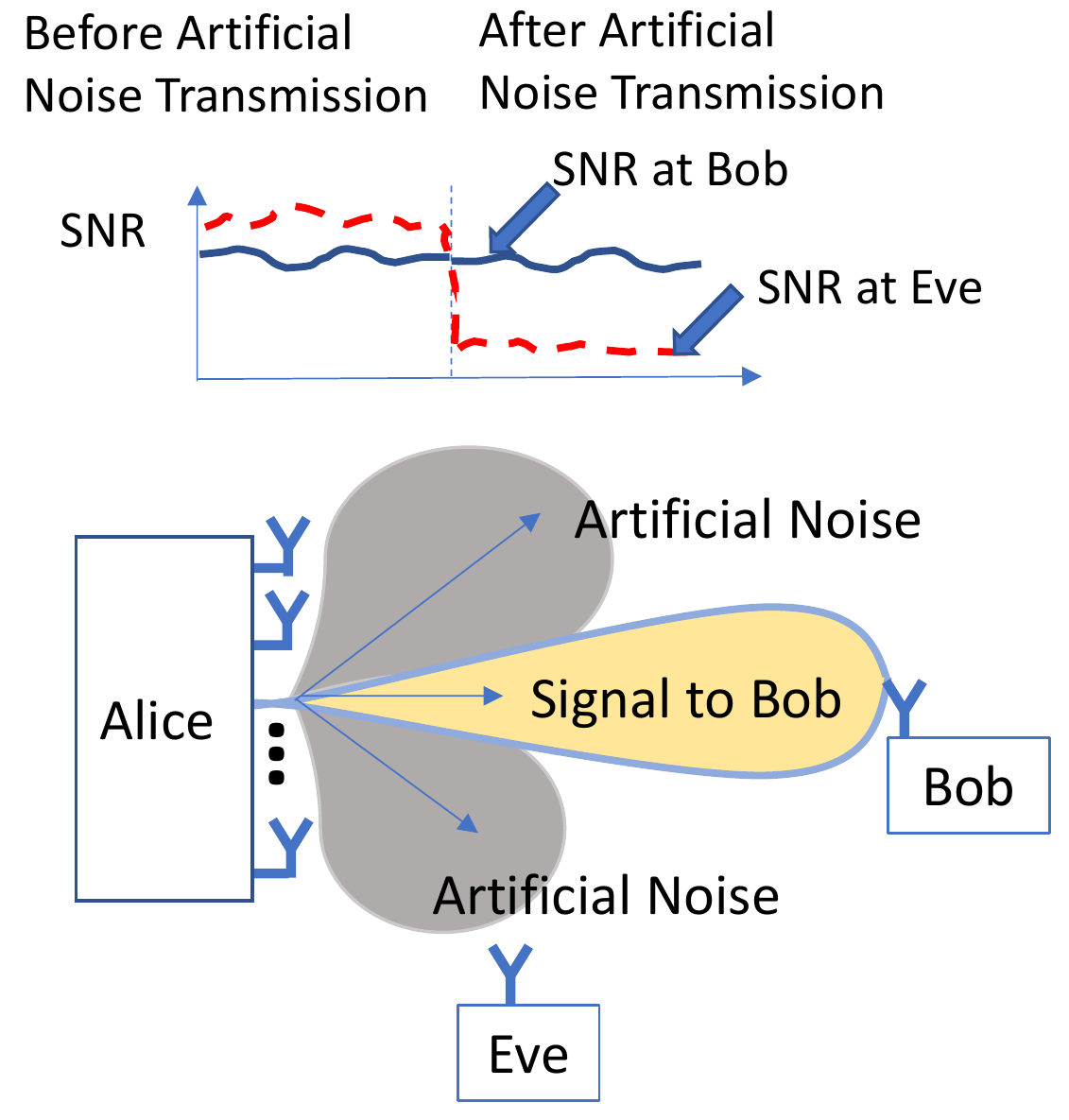}
	\caption{Secure beamforming from a legitimate transceiver (Alice) to a legitimate receiver (Bob) in the presence of an eavesdropper (Eve).}
	\label{Fig:S1}
\end{figure}

In order to exploit
the difference in channel characteristics for secure communications, a channel model, called the wire-tap channel, is considered in~\cite{Wyner75TheWiretap} with three parties: Alice, Bob, and Eve as illustrated in Fig.~\ref{Fig:S1}.  Here, a transmitter (Alice) wishes to convey a message to a legitimate receiver (Bob) while keeping it secret from an eavesdropper (Eve).
In a nutshell, for additive white Gaussian noise (AWGN) channels, if the SNR $\gamma_\mathrm{Bob}$ of the channel between Alice and Bob, which is called the legitimate channel, is higher than that $\gamma_\mathrm{Eve}$ of the channel between Alice and Eve, which is called the wiretap channel, secure communication is possible from Alice to Bob~\cite{Leung78}.
{\color{black}
Suppose that Alice and Bob are equipped with $N_\mathrm{t}$-element and $N_\mathrm{r}$-element phased arrays, respectively. Thus, precoding and combining can be performed. The legitimate channel is denoted by $\mathbf{H}$. We suppose that Eve is equipped with  a single antenna and the wiretap channel is denoted by $\mathbf{h}_\mathrm{Eve}$. Then, the received SNRs of Bob and Eve are respectively given by 
\begin{align}
	\gamma_\mathrm{Bob} &= \frac{|\mathbf{w}^{\mathrm{H}}  \mathbf{H} \mathbf{f}|^2P}{\sigma_1^2}\label{SNR-Bob}, \\
	\gamma_\mathrm{Eve} &= \frac{| \mathbf{h}_\mathrm{Eve} \mathbf{f}|^2P}{\sigma_2^2}\label{SNR-Eve}, 
\end{align}
where $P$ is the transmit power of Alice. $\mathbf{w}$ and $\mathbf{f}$ are the beamforming vectors of Bob and Alice, respectively. $\sigma_1^2$ and $\sigma_2^2$ are the powers of  Gaussian white noise at Bob and Eve, respectively. 
Based on (\ref{SNR-Bob}) and (\ref{SNR-Eve}), \emph{secrecy throughput}, which is defined as the effective average transmission rate of the confidential message~\cite{ wu2018securi, xiao2021asurve}, can be derived as
\begin{align}
	S_\mathrm{Bob} = [\log_2 (1+\gamma_\mathrm{Bob})-\log_2 (1+\gamma_\mathrm{Eve})]^+,
\end{align}
with $[x]^+ \triangleq\max(x,0)$.
}

{\color{black}Obviously, beamforming vectors can be optimized to increase the SNR of the legitimate channel as well as decreasing the SNR of the wiretap channel,  thus improving communication security.} In~\cite{Goel08}, in wireless communications, the use of antenna array is proposed to generate artificial noise that can degrade the SNR of the wiretap channel. As shown in Fig.~\ref{Fig:S1}, the SNR of the wiretap channel can be higher than that of the main channel. However, once artificial noise is transmitted to other directions other than that of Bob, the SNR of the wiretap channel can be lower than that of the main channel.
 The resulting approach is referred to as random masked beamforming, and in~\cite{Khisti_2010}, its secrecy rate is analyzed. 
A salient feature of random masked beamforming is that it can generate artificial jamming signals without knowing instantaneous CSI of Eve, i.e., an eavesdropper.
With known  statistical properties of the eavesdropper channel, it is possible to obtain the ergodic secrecy rate from a transmitter to a legitimate receiver. For guaranteed performances, various beamforming optimization problems are considered~\cite{Liao11, Wang14, Li14,zeng2019securi} to decide the transmit beam as well as covariance matrix for artificial noise vector.
In general, the ergodic (or long-term average) secrecy rate is to be maximized in most 
formulations. However, for slow fading channels, the ergodic secrecy rate may not be useful because codewords are not sufficiently long to experience varying degrees of fading.
As in~\cite{Choi16}, instantaneous secrecy rate needs to be taken into account to formulate beamforming 
optimization problems. 

In~\cite{Ng14, Shi15}, with antenna arrays, beamforming is used to facilitate the dual use of artificial noise and energy signals to provide secure communication and facilitate efficient wireless energy transfer.
Furthermore, as shown in~\cite{Lu19,Schraml21,lei2011secure}, robust beamforming can also be considered for secure communications in satellite systems. Finally, secrecy can be considered as part of secure multiple access designs. Secure RSMA was shown to outperform secure NOMA and secure SDMA~\cite{li2020cooper,fu2020robust}.

{\color{black}
\subsection{Summary and Discussion}
In this section, varieties of emerging communication technologies enabled by antenna array were discussed. For the future space/air/ground communication systems which  require large-scale coverage, a single narrow beam may not be able to efficiently cover all users at the same time. Therefore, it is necessary to develop single-RF-chain multiple beam and flexible beam coverage technologies. Sub-array scheme and mathematical optimization method are popular approaches to generating multiple beams and wide flexible beams, and coping with imperfect CSI. However, the computational complexity must be taken seriously when designing beamforming approach, especially for the  space/air platforms with high real-time requirements and limited computing resources. 

Enabled by antenna arrays, the multiple access technologies that exploit the spatial characteristics are promising for future communication systems. SDMA separates users in the spatial domain via beamforming. However, if two neighboring users have highly correlated channels, the interference between the two beams will become severe, which can seriously affect the communication performance.  NOMA distinguishes signals via different power levels and employs SIC for decoding, and thus can separate users within the same beam. In contrast, RSMA splits messages into common and private parts to further improve the spectral and energy efficiency. It is interesting to explore the performance of different multiple access technologies for different space/air/ground platforms in various scenarios. For example, A2G communications with 3D beamforming offer more refined angle resolutions in both azimuth and elevation dimensions~\cite{zeng2019access}, and thus facilitate the use of SDMA. 

With the ability to manipulate electromagnetic waves passively, RIS is regarded as a revolutionary technique for enhacing the spectrum and energy efficiency of wireless systems. It is capable of mitigating the challenging blockage and coverage issues and  carrying additional information. Besides, it is interesting to investigate the potential integration of RISs with other wireless technologies, including NOMA, mmWave, and cognitive radio systems~\cite{pan2021reconf}. 

By utilizing the spatial selectivity, physical-layer security can be enhanced in antenna array enabled systems. In addition to physical layer security techniques, such as artificial noise injection and cooperative jamming, the distinct characteristics of space/air/ground platforms can be utilized. For example, aerial platforms can
adjust their 3D position to avoid the jamming area, or transmit artificial noise to eavesdroppers more efficiently.

}

\section{Antenna Array Enabled Space/Air/Ground Communications}\label{sec:space_air_ground}

With these emerging technologies described in Section~\ref{sec:emerging_technologies}, the antenna array enabled space/air/ground communication network is becoming a promising paradigm for next generation communication network. In addition to ground wireless communications, the satellites can provide globally seamless communication coverage, while the aircraft can achieve on-demand deployment and wide-area communication coverage in emergencies. Meanwhile, the application of antenna arrays and the mobility of space/air/ground platforms poses substantial new characteristics and challenges to the antenna array enabled space/air/ground communication systems. These issues are introduced in this section.

\subsection{Satellite Communications}
Satellite communications usually mean the communications between a satellite platform and a ground station or different satellite platforms. Satellites can operate in a geostationary Earth orbit (GEO) constellation, a medium Earth orbit (MEO) constellation, and a low Earth orbit (LEO) constellation, according to the orbital height. Compared to terrestrial networks and airborne networks, satellite communication networks have a much larger coverage area. However, long-distance communication between satellites and ground leads to much larger link loss and transmission delay. Satellite communication networks enabled by antenna arrays can make up for the above shortcomings and obtain more flexible beam coverage to meet the needs of users to access the network anytime and anywhere. Meanwhile, the particular characters of satellite altitude, frequency and movement bring several unique features to the satellite communication networks in beam coverage, beamforming, beam management and handover, as described below.
\subsubsection{Various Beam Patterns}
In satellite communications, a variety of service scenarios may require different coverage schemes, thus calling for various beam patterns. Generally speaking, broad coverage  requirements are usually accomplished by \emph{wide beams}, which include global beams, hemispherical beams and regional beams. However, wider beams are usually accompanied by smaller antenna gains. Therefore, wide beams are more suitable for transmitting/receiving user control signals or broadcasting communications.
On the other hand, \emph{spot beams} are proposed to improve antenna gains and promote multiplexing gains. The more concentrated beams can reduce transmit power, and increase communication capacity, but with smaller coverage area. Therefore, spot beams are more suitable for providing high-speed data services. Besides, to balance the stable transmission requirements of control signals and high-speed requirements of data signals, a \emph{hybrid wide-spot beam} is proposed in~\cite{Su2019broadb}, which is essentially the combination of wide beams and spot beams.

For \emph{wide beam}, one of the main technologies that provides such kind of beam pattern is reconfigurable antennas.
According to their electrical performance, reconfigurable antennas can be divided into three main categories: reconfigurable frequency, reconfigurable pattern, and reconfigurable polarization. 
In~\cite{chang2015broadb}, a type of antenna with a frequency bandwidth from 1.15 GHz to 1.6 GHz  was designed for wide-bandwidth beam global navigation satellite system
(GNSS). By adjusting the effective aperture of the antenna, the radiation pattern of the antenna can be reconstructed, thereby achieving wide beam coverage.
In~\cite{Khidre2013reconf}, a beamwidth reconfigurable microstrip patch antenna of H-plane pattern was designed to achieve wide beam coverage, where the beam width can be continuously adjusted from $50^{\circ}$ to $112^{\circ}$. However, a single wide-beam antenna usually results in the loss of gain as the antenna beam width increases, thereby reducing the quality of service for users. To solve this problem, a left-bias pattern and a right-bias pattern were combined through pattern reconfigurable technology~\cite{Khan2016thepur,Zhang2017compac,Yang2019anovel}, where the wide beam coverage area of the reconfigurable pattern antenna is the union of the coverage provided by the left-bias pattern and coverage of right-bias pattern.

For \emph{spot beam}, it is necessary to flexibly adjust the center point of the  beam to ensure that the communication target is within the coverage area, due to the limited coverage area of the spot beam and the mobility of both satellites and users. 
In different traffic scenarios, the distribution of business volume is not uniform, for example, metropolis regions and emergency communications during disasters. Therefore, traffic-based dynamic coverage schemes are needed to adjust the size of a spot beam and resource allocation~\cite{Qian2014Traffi}. To support the non-uniform distribution of users and varying traffic requirements, adaptive multi-beam pattern and footprint planning were developed~\cite{Honnaiah2021demand}, where spot beams with flexible sizes and positions were designed according to user spatial clustering to improve the flexibility of satellite communication systems. In~\cite{Zhou2019covera}, a coverage metric was proposed to measure the average coverage level of satellite constellations of different orbital altitudes for backhaul. Among spot beams, TDMA spot-beam communication process was further formulated as a discrete-time queuing problem to calculate the quantity of accessed equipments in a unit area. In addition, the relationship between the equipment density, maximum tolerable delay, and satellite constellation coverage level was derived. A steerable spot-beam reflector antenna was explored in~\cite{Wan2016asteer}, where the steerable spot beam can be quickly repositioned to provide flexible coverage by rotating the reflector around its apex (referred to as vertex rotation). In~\cite{zhang2019resear}, an effective optimization method of multiple-feed per beam antenna based on genetic algorithm was proposed to improve the coverage performance of spot beams, where the orthogonality constraint introduced by the lossless beamforming network (BFN) was taken into account.

The main idea of the \emph{hybrid wide-spot beam} is to provide a wide beam and multiple spot beams at the same time. 
The wide beam, with fixed direction and coverage, is utilized to cover the whole service area
	for the transmission of control signals such as mobility management, session management, bearer establishment and mapping. 
	On the other hand, the spot beams are always steered to the users for the high-speed transmission of user data. 
	In order to enable efficient modulation and coding techniques for data transmission, spot beams usually require much higher power consumption than that of the wide beams. Note that spot beams are more flexible for planning the system capacity and resource configuration according to the needs of users, due to the steerable beams.
	In summary, with the hybrid wide-spot beam strategy, the structure of the satellite access network is actually reconstructed, that is, the separation of the control plane and the user plane is realized.

\subsubsection{MBA}
Under the circumstance of exponentially increasing communication demands, designing a satellite system with high throughput is becoming a hot-spot in both academia and industry~\cite{moon2019phased}. However, the limited resources available for satellites make it challenging to fulfill the requirements. Multiple beam array (MBA) and the corresponding multi-beam forming techniques are promising solutions~\cite{Montero2015cbandm}. MBA is an antenna that uses the same aperture to generate multiple beams with different directions simultaneously. By achieving polarization isolation and space isolation effectively, MBA can realize spectrum multiplexing thus increase communication throughput. Moreover, a global or regional beam coverage can be split into several small cells and covered by independent spot beams. 
In this way, the ground terminal may use a small aperture antenna to realize high-speed data transmission.
To avoid interference, different beams work in different frequency bands or adopt different polarization modes. By proper beamforming schemes, the multi-beam forming can help to achieve high gains in the target areas, while leaking low gains outside the serving areas. Therefore, the transmit power can be reduced. 

MBA can be reflector-based architectures, phased array architectures, and lens-based architectures~\cite{Lai2016adigit}. Reflector antennas and lens-based antennas leverage optical elements such as reflectors and lens to reach higher gains. Therefore, they are applied in MEO/GEO satellites to serve of remote transmission. On the other hand, the phased array architecture is more suitable for LEO satellites with high-flexibility requirements, by means of beamforming. The multi-beam forming in phased array MBA includes analog beamforming and digital beamforming. Globalstar leveraged analog beamforming in its MBA with the BFN composed of power dividers. Iridium utilized the BFN composed of Butler matrix. Once the BFN is determined, the beam shape, the intersection level and beam direction of adjacent beams are fixed and difficult to change. Notably, if the number of beams increases, the BFN of analog beamforming will be complex to realize. In addition, the fixed scheme is difficult to be adaptive. Thus, digital beamforming is attracting more interests. The RF signals received by multiple antenna array elements are respectively converted to baseband through multiple channels, and beamforming is realized through the digital signal processor. Supported by digital architecture, the adaptive beamforming can be  applied in satellite MBAs. The possibility of using digital BFN to design satellite antenna systems with adaptive beamforming was discussed in~\cite{sow2008beamfo}. Aiming to reduce the complexity of beamforming design for antennas with large number of emitters, a low complexity algorithm was proposed in~\cite{montesinos2011adapti}. The authors in~\cite{zheng2019adapti} presented an adaptive beamforming method based on user locations. The locations could be provided by users, whose terminals were equipped with the navigation subsystem. 

It is worth noting that no matter the analog or digital beamforming, after dividing cells, the shape of beam for the cell needs to be decided. Therefore, it is necessary to find the appropriate amplitude and phase weighting values for each element of the array. This problem can be formulated as the optimization problem. With proper algorithms, the required beam pattern can be obtained. Multi-beam forming can also be combined with RSMA and on-board processing to boost performance and better manage interference between users compared to SDMA and NOMA~\cite{Yin:2021,Si:2021}.



\subsubsection{Beam Management and Handover}
Satellite systems provide a wide range of communication service coverage. LEO satellite has the characteristics of low orbit height and short electromagnetic wave round-trip time, which can effectively solve the delay problem for satellite communication. However, the rapid movement of LEO satellite may cause frequent handover of user calls, which challenges the beam management technology for LEO systems~\cite{kim2020Beamma}. The beam management mainly consists of beam handover and beam scheduling. Beam handover is also called cellular handover or intra-satellite handover, which refers to the handover of links between adjacent beams within the coverage area of the same satellite. 

Beam handover technologies mainly include the non-priority handover, queuing priority handover and reserved channel strategies. The non-priority handover strategy employs a fixed channel allocation method to allocate a fixed number of channels to each cell and each type of service. Although this strategy is simple, it can not adapt to the dynamic changes of the network traffic, which reduces the efficiency of resource utilization for the system. It is generally used in combination with other strategies~\cite{Del1999differ}. The queuing priority handover strategy~\cite{Riad2001fixedc,Papapetrou2005analyt,Zhao2009Newcha,Yan2008Spotbe,Karapantazis2005Design,boukhatem2003tcraat} is based on queuing technology to distinguish the priority of various types of calls or requests and determine the network resource allocation accordingly. When the satellite receives a new call or handover request, while there is no channel available for the next beam, the request will be placed in a special queue for waiting. If the channel is idle at a specific time, then the channel can be scheduled by the next beam, otherwise,  the channel will be forcibly interrupted. Calls or requests in the same queue are allocated according to the first-in-first-out principle. Different priorities can also be set for different queues. The queues with higher priority get more network resources. The reserved channel strategy uses the concept of a protected channel, which is set up in each cell specifically for handover services~\cite{Hedjazi1972thehan}. The key issue of the reserved channel strategy is to set a reasonable threshold so that the reserved channel resources conform to the actual situation of the network, so as to avoid a waste of network resources or affect the effectiveness of the strategy. In addition to being a fixed value, the threshold of the reserved channel can be dynamically adjusted according to the network status, which may improve the network resource usage~\cite{kim2020Beamma}. The existing methods of dynamically adjusting the threshold are presented as follows. Adjustment strategies based on forecast, probability models or state quantities were used to predict different types of requests, and dynamically adjust the threshold of the reserved channel according to the prediction results~\cite{Olariu2004OSCARa}. In~\cite{Bottcher1994Strate}, an adaptive dynamic channel allocation strategy was proposed to reduce the overall handover blocking probability. An opportunistic call admission protocol was proposed to avoid the cost of researching resources for users in a series of beams along the predicted user trajectory in~\cite{Zhao1996Combin}. In time-based adjustment strategy, the reserved channel of the next beam can be adjusted according to the time the user stayed in the current beam or the expected channel usage time~\cite{Bottcher1994Strate}. In~\cite{Olariu2004OSCARa}, a time-based channel reservation algorithm was proposed to ensure the probability of zero handover failure. The LEO satellite communication network usually adopts multiple earth orbit satellites with limited coverage to form a specific constellation. To form a communication link, the user needs to connect to one of the serving satellites. Due to the fast moving characteristics of LEO satellites, inter-satellite handover occurs frequently. Once handover occurs, it involves the problem of beam scheduling. The user terminal always selects the maximum instantaneous elevation when handover happens in~\cite{kim2020Beamma}. In~\cite{Zhao1996Combin}, an adjustment strategy based on QoS was proposed to dynamically adjust reserved channels, and overcome the low bandwidth utilization rate problem in the reserved channel mechanism. Moreover, user satisfaction was utilized to measure system QoS. 
An inter-satellite handover algorithm based on the position and signal strength of the active user terminal in~\cite{Olariu2004OSCARa} was proposed to maximize the user throughput. By measuring the transmission delay and Doppler shift of user terminal, the network can estimate and measure the position of user terminal during the call process, so as to reserve channel resources for the user.

In practical applications, the non-priority handover strategy, queuing priority strategy, and channel reservation strategy can be selected according to the actual situation, or multiple strategies can be utilized simultaneously. In general, the queuing priority strategy has a better performance in terms of the blocking rate and the drop rate for LEO satellite communication networks~\cite{Hedjazi1972thehan}. Nevertheless, the comparative analysis of these strategies and other QoS indicators and system capacity needs to be further studied.

\subsection{Airborne Communications}
\begin{figure*}[t]
	\begin{center}
		\includegraphics[width=0.8\linewidth]{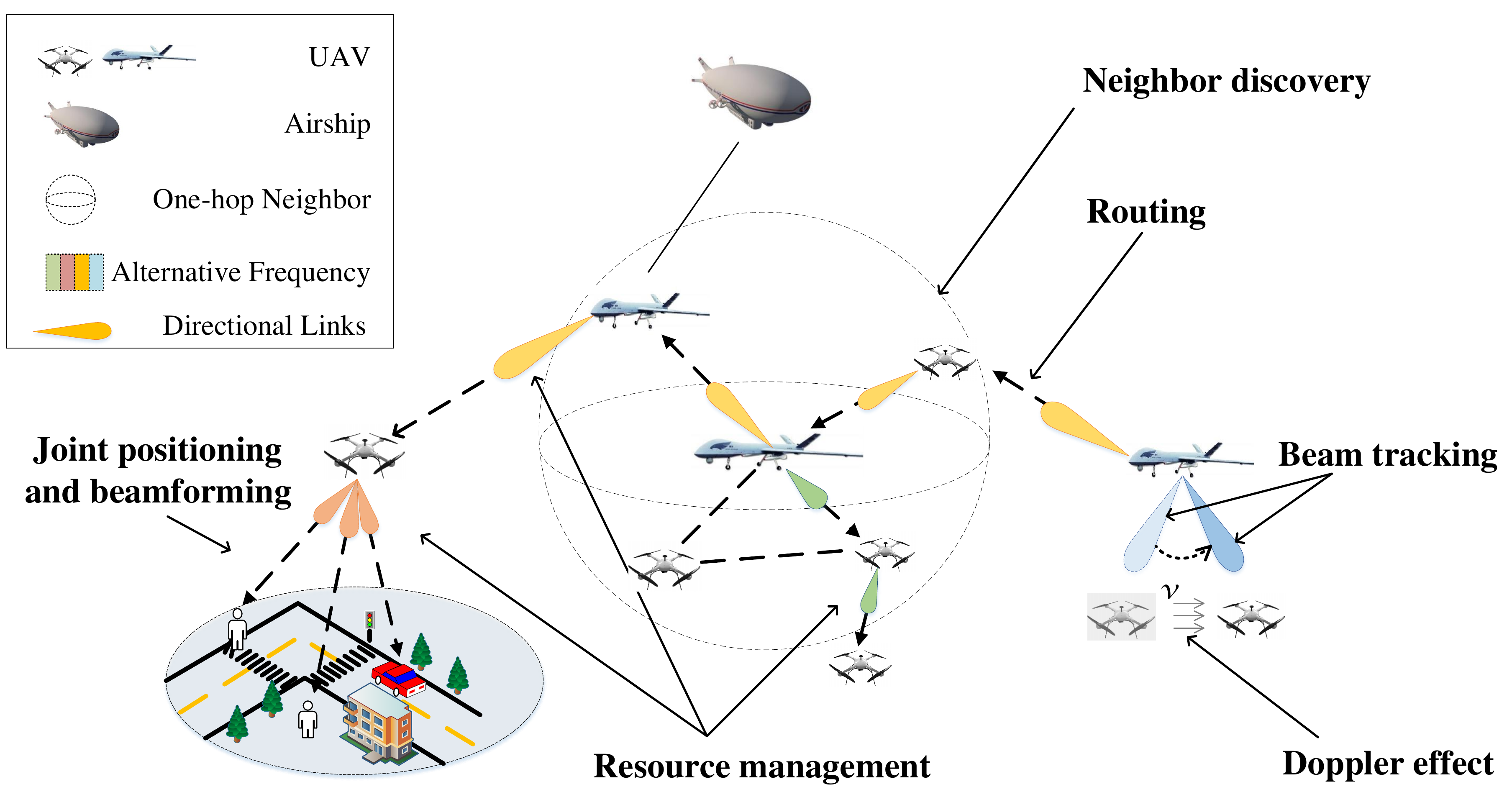}
		\caption{Illustration of the typical scenarios for antenna array enabled airborne communications and networking.}
		\label{fig:airborne_scenario}
	\end{center}
\end{figure*}

Airborne communication systems utilize various aircraft equipped with transceivers and sensors, to build communication access platforms~\cite{cao2018airbor}. These aircraft mainly include UAVs, airships, and balloons, making up the LAPs and HAPs. Compared with ground communication systems, airborne communication systems can be flexibly deployed in a cost-effective manner, irrespective of terrain. Compared with satellite communication systems, airborne communication systems have much shorter range LoS links, resulting in lower latency and less propagation loss. Therefore, airborne communication is a key part of space/air/ground communications. Enabled by antenna array, the system can obtain new benefits. For instance, antenna array provides considerable beam gains to compensate propagation loss through directional transmission, which improves the channel quality. Besides, the directional transmission is beneficial to the reuse of spectrum resource in the spatial domain.  At the same time, antenna array enabled airborne communication systems have varieties of distinct characteristics and challenges in both communications and networking. Fig.~\ref{fig:airborne_scenario} illustrates the typical scenarios for the airborne systems, where varieties of aircraft form an aerial ad-hoc network and accomplish missions collaboratively. From the perspective of communications, we mainly focus on beam tracking, Doppler effects,  and joint positioning and beamforming. From the perspective of networking, directional neighbor discovery, routing, and resource management are addressed.

\subsubsection{Beam Tracking}
Compared with terrestrial systems, 3D mobility with very high dynamic is one of the typical characters for airborne systems, especially for large-scale UAVs. Due to the mobility, the directional transmission mechanism enabled by antenna array brings serious beam misalignment problem, which leads to the deterioration of communication performance or even the interruption of connection. To maintain the beam alignment, typical beam tracking scheme for the conventional terrestrial systems is to train the beam direction periodically through transmitting pilot signals. However, the high mobility as well as platform constraints (e.g., SWAP) of airborne system may result in unacceptable burden of pilot transmission, challenging the training-based beam tracking scheme from being applied to that.  Therefore, beam tracking is practically important while quite a challenging problem for antenna array enabled airborne communication system. A distinct property for air-to-ground (A2G)/air-to-air (A2A)/air-to-satellite (A2S) communications is that their channels are dominated by LoS paths. By exploiting this property, mobility prediction-based beam tracking schemes are more efficient when LoS paths exist between airborne platforms and other platforms. The angular velocity estimation, and angular domain information, i.e., elevation angle and azimuth angle can be utilized to save the pilot overhead, rapidly establish and reliably maintain communication links for A2G communications~\cite{yang2019beamtr, huang20203dbeam} and A2A communications~\cite{xu2021datadr}. Airborne sensors which provide movement state information, like GPS and micro inertial measurement unit, can assist the coarse beam alignment~\cite{zhao2018channe}. Besides, machine learning-based schemes, such as Q-learning~\cite{chiang2021machin}, long short-term memory recurrent neural network~\cite{yuan2020learni} were exploited to assist beam tracking, for their ability to predict beam alignment based on sequential beam tracking experience. 
In summary, the high dynamic of aircraft makes the beam tracking a challenging problem for antenna array enabled airborne communication systems. Both efficiency and overhead need to be balanced when designing solutions. Available information and new techniques that help predict the mobility of airborne platforms can be exploited to assist beam tracking and help reduce the overhead. 

\subsubsection{Doppler Effect}
For airborne communication systems, an inevitable problem is the Doppler effect, which can introduce carrier frequency offset and inter-carrier interference. It is known that the Doppler shift $f_D$ of a received signal is a function of carrier frequency $f_c$, relative velocity $v$, angle of arrival (AoA) $\theta$, and angle of the relative velocity $\theta_v$, expressed as $f_D=(v/c)f_c (\theta-\theta_v)$, where $c$ is the speed of light. If multi-path components (MPCs) arrive at the receiver with different AoA, e.g., with large angular dispersion, the resulting different Doppler shifts will produce spectral broadening, called Doppler spread~\cite{khawaja2019asurve}. 
In high dynamic airborne systems, one might initially think that Doppler spread would be high and cause catastrophic effects on communications. In fact, airborne platforms operate in high altitudes, the channels are mainly dominated by LoS paths. MPCs are expected to have very similar Doppler shift with relatively small angular spread. This is especially true for high carrier frequency, such as mmWave frequency~\cite{rangan2014millim}. Besides, directional transmission enabled by antenna arrays will further reduce the number of MPCs and in turn reduce the angular spread~\cite{xiao2016enabli,lorca2017onover}. Large Doppler shift with small Doppler spread can be well mitigated by frequency synchronization. 
Doppler power spectrum is an important statistical property to characterize the Doppler spread, which expresses the power spectral density of the received signals as a function of the Doppler shift~\cite{lorca2017onover}. As a result, Doppler power spectrum were derived and analyzed in many studies on wideband non-stationary A2A/A2G channel model~\cite{ma2020awideb,walter2019analys, jiang2020anovel, ma2020impact}. It was found that the UAV rotations significantly affect channel correlations~\cite{ma2020impact}.
To reduce the Doppler effect, it is necessary to perform Doppler frequency shift (DFS) estimation and compensation. The idea of coarse estimation plus fine estimation may be favorable to achieve a fast and accurate DFS estimation and compensation~\cite{zhang2020dataai}. 
In a word,  the high mobility of airborne platforms, the use of higher carrier frequency and the directional transmission make the Doppler effect of airborne communications different from that of conventional terrestrial communications. It is worth the research effort to model this property and compensate the effect for multiple airborne communication scenarios. 

\subsubsection{Joint Positioning and Beamforming}
With 3D mobility, airborne platforms can design their positions or trajectories according to the mission to improve communication performance. Enabled by antenna array, beamforming can be designed not only to improve the received signal power but also to mitigate mutual interference. Therefore, there is a great deal of freedom for antenna array enabled airborne communication systems for jointly positioning (also trajectory) and beamforming design. However, the joint design is challenging. Different from the positioning of an aircraft with single antenna, positioning and directional beamforming are coupled for antenna array enabled airborne systems. 
The channel state among the transmitter and receiver may change according to the aircraft's position and posture. Because of the coupling variables, the optimization problem is non-convex and difficult to solve. 
Moreover, the positioning design of multiple airborne platforms is more tricky, since interference between different terminals needs to be properly considered. 
To solve the challenging joint positioning and beamforming problem, a feasible solution is the iterative algorithm, where beamforming and trajectory are alternately optimized~\cite{yuan2019joint3}. Specifically, in each iteration, the trajectory is optimized by fixing beamforming direction, and then beamforming is optimized with fixed trajectory.
Alternatively, the ideal beam pattern was introduced and the joint optimization problems were solved in two steps~\cite{zhu2020millim,xiao2020unmann}.
The ideal beam pattern states that the summation of the beam gains in different directions is approximately equal to the number of antennas~\cite{wei2019multib}. After substituting the ideal beam gain, a more tractable joint deployment and beam gain allocation problem is obtained, followed by approaching the ideal beam pattern through multi-beam forming techniques. 
In addition, by applying a modified cosine antenna pattern approximation of uniform linear array (ULA), the UAV trajectory and directional beamforming can be jointly optimized in a single convex optimization problem~\cite{yuan2021jointd}. 
Besides, to ensure a robust joint trajectory and transmit beamforming design, practical considerations such as UAV jittering, user location uncertainty, wind speed uncertainty, and no-fly zones are necessary~\cite{xu2020multiu}.
In summary, to give full play to the unique advantages of the antenna array enabled airborne communication systems, flexible positioning needs to be simultaneously designed with effective beamforming. As an appealing and challenging research direction, it is worth the effort for exploiting both optimization strategy and practical communication scenario.





\subsubsection{Antenna Array Enabled Aerial Ad-hoc Network}

Aerial ad-hoc networks refer to multi-aircraft systems organized in an ad-hoc fashion, aiming to accomplish complex missions cooperatively. Compared to single aircraft aerial system, aerial ad-hoc networks are more flexible, reliable and survivable through redundancy. Therefore, aerial ad-hoc networks have broad military, civilian, and commercial applications such as remote sensing, traffic monitoring, border surveillance, and relay networks~\cite{zafar2016flying}. At the same time, aerial ad-hoc networks have distinct characteristics such as a high level of network heterogeneity, highly dynamic, frequently changed network topologies, weakly connected communication links, and vulnerable to jamming and eavesdropping~\cite{cao2018airbor}. 
Directional communication enabled by antenna arrays provides significant performance gain for aerial ad-hoc networks. By focusing electromagnetic energy only in the intended direction, antenna arrays can enlarge transmission distance for a given power level, which improve network connectivity. On the other hand,  directional beams increase spatial reuse, which allows more simultaneous transmissions and enhances anti-jamming/eavesdropping abilities, thus providing higher network capacity and security~\cite{ramanathan2005adhocn}. Nevertheless, these benefits are accompanied by certain unique challenges. Mechanisms that were designed for terrestrial ad-hoc networks or with omnidirectional communications need to be redesigned for the antenna array enabled aerial ad-hoc networks. We provide an overview on the important issues and potential solutions, mainly about neighbor discovery, routing, and resource management.

\emph{Neighbor discovery}, also known as \emph{routing discovery}, refers to the process of finding one-hop neighbors, which is a crucial initial step for establishing connections among the nodes~\cite{cai2012neighb}. For omnidirectional antenna enabled networks, simple broadcast can reach all neighbors. The problem is more challenging for antenna array enabled networks, since nodes need to determine \emph{when} and \emph{where} to point their directional beams simultaneously to discover each other. A natural approach to contour the challenge is to use omnidirectional antenna in the neighbor discovery process~\cite{zeng2019aduala, sneha2020freesp}. For example, a dual-antenna collaborative communication strategy was proposed in~\cite{zeng2019aduala} for aerial ad-hoc networks, where neighbor discovery is based on low-frequency heartbeat location information piggybacked on control frames enabled by omnidirectional antenna. The main drawback of this approach is that an additional omnidirectional antenna is required. Following the similar idea, an antenna array can work in a quasi-omnidirectional manner by omnidirectional beamforming to perform neighbor discovery~\cite{astudillo2017neighb,yang2019networ}. However, wider beam means lower beamforming gain, thus causing shorter discovery range. Without synchronization and any available information, \emph{probabilistic} approach can be performed, where each node randomly chooses a direction to steer its beam. Obviously, this approach lacks performance guarantee in terms of discovery delay~\cite{chen2017onobli}. With time synchronization among nodes, e.g., with satellite positioning system as common clock source, \emph{deterministic} approach can be developed, where the beam of each node is steered based on a predefined sequence. For example, the antenna scans its beam clockwise to perform neighbor discovery in~\cite{ramanathan2005adhocn, astudillo2017neighb}. In this case, neighbors can be discovered within one cycle with a high probability. With partial prior information available, such as the location of other nodes piggybacked through routing updates~\cite{ramanathan2005adhocn} or the location/motion prediction~\cite{zeng2019aduala}, neighbor discovery may be performed more rapidly  and achieve fast convergence, known as \emph{informed discovery}~\cite{ramanathan2005adhocn}. 


After neighbor discovery, an aerial ad-hoc network requires mechanisms for discovering routes and forwarding packets along these routes. \emph{Routing} plays the role, and has a major impact on network throughput and packet delay. Compared to conventional ad-hoc networks, the 3D high mobility of aircraft brings intermittent connections and frequent topology changes for aerial ad-hoc networks, which need to be emphatically considered during routing design. The routing schemes for aerial ad-hoc network can be categorized into \emph{topology-based}~\cite{xu2020improv, gankhuyag2017robust, waheed2019laodli}, \emph{geographic/location-based}~\cite{fawaz2017unmann,shumeye2020routin}, and \emph{bio-inspired}~\cite{leonov2016modeli, khan2020smarti,leonov2016applic}. 
\emph{Topology-based} routing requires to obtain the routing path before data transmission begins, which has high transmission efficiency but may cause high overhead for routing discovery and maintenance. \emph{Geographic} routing utilizes geographic positions of the aircraft for routing decisions, which requires hardware installations of aircraft. \emph{Bio-inspired} routing is inspired by collective behavior of biological systems, such as the honey collection in a bee colony, or food finding in an ant colony. Since there is no significant difference from routing in omnidirectional aerial ad-hoc networks. Routing schemes for antenna array enabled aerial ad-hoc networks can draw lessons from that designed for omnidirectional aerial ad-hoc networks~\cite{shumeye2020routin,xiao2021asurve}.

To encourage the quality of communication in a network, there is a need for a framework to dynamically manage various resources including time domain, frequency domain, power domain, space domain, and so on~\cite{chen2021agamet}. Therefore,  \emph{resource management} plays a key role in aerial ad-hoc networks. Typically, resource management includes spectrum management, task assignment, interference management, power control, and so on. The goal for spectrum management is to improve spectrum utilization as well as to reduce mutual interference, ensuring efficient and robust wireless communication for a network~\cite{wang2018spectr}. 
Control-data separation architecture may achieve both stable and high-rate communication for aerial ad-hoc networks~\cite{zeng2019aduala}. Specifically, lower frequency was utilized for one ommnidirectional antenna enabled control channel, ensuring stable control frames transmission. Higher frequencies were utilized for directional antenna enabled data channels, enabling broadband data transmission. 
Oppositely, control-data sharing the same bandwidth may achieve higher spectrum utilization but also a potential interference problem~\cite{feng2019spectr}. 
Due to the directional transmission characteristic and platform restriction, particular attention should be paid to resources in space domain and power domain for antenna array enabled aerial ad-hoc networks. Benefiting from antenna array, narrow beams increase spatial reuse, and thus enable more simultaneous transmissions and decrease mutual inferences~\cite{ramanathan2005adhocn,temel2014scalab}. Efficient beam management scheme is necessary to guarantee network performance. At the same time, beam misalignment problem should be addressed considering the high mobility of aircraft. Besides, the onboard energy of aircraft, especially for small UAVs, are usually limited. Thus, energy-efficient operations such as transmit power control, load balancing and node sleep are essential for aerial ad-hoc networks~\cite{feng2019spectr}.

In summary, the high-dynamics of aircraft and directional transmission bring unique challenges on antenna array enabled airborne ad-hoc networks. Prior information and geographic positions are helpful, and can be exploited to facilitate the process of neighbor discovery and routing. Bio-inspired routing scheme is a promising routing solution worth exploring. Besides, directional transmission and platform constraints bring more considerations regarding resource management in space domain and power domain.

\subsection{Ground Communications}

Massive antenna array technology has been widely used in 5G communication systems nowadays, such as beamforming technology based on antenna arrays \cite{lota2017gunifo, noh2020fastbe, hu2021anorth}. Beyond 5G (B5G) and 6G communications need to address more challenges on high data rate, low latency, massive connectivity, seamless coverage and high-speed mobility. Antenna array will be one of the key technologies to support ground communications by providing high beamforming gain and multiplexing of users. We will introduce the applications of antenna array in ground communications in details as follows.



\subsubsection{Cellular Massive MIMO}



In order to deal with massive connectivity and provide better service for users, cellular networks with smaller cells compared to 4G are widely used in 5G communication networks. The combination of mmWave and large-scale antenna array brings new solutions for high throughput. Nevertheless, it brings new challenges at the same time. To overcome high path loss and blockage of mmWave signals, an effective approach is that dividing smaller cells to provide better user QoS by getting the transmitters and receivers closer. The small cells, which are defined as low-power wireless access points (APs) operated in licensed spectrum, can provide improved cellular coverage, capacity and applications for homes, enterprises and other connectivity~\cite{rodriguez2014fundam}, compensate mmWave pass loss and contribute to seamless coverage.

Although multi-cell systems can provide better performance for users, they may suffer severe inter-cell interference caused by frequency reuse, especially for cell-edge users. Inference management and elimination is one of the most significant challenges for multi-cell transmissions, which needs the cooperation among BSs in different cells. The coordinated beamforming (CoBF) designed for massive MIMO multi-cell networks, where BSs are equipped with a large antenna array, has attracted great concern to achieve interference suppression.  There are two important downlink multi-cell interference mitigation techniques, i.e. large-scale MIMO (LS-MIMO) and network MIMO~\cite{hosseini2014larges}. In a LS-MIMO system, BSs equipped with multiple antennas not only serve their scheduled users, but also null out interference caused to other users within cooperating cluster using ZFBF. In a network MIMO system, BSs eliminate interference through data and CSI exchange over the backhaul links and joint transmission using ZFBF. It was proved that LS-MIMO can be the preferred approach for multi-cell interference mitigation in wireless networks. To improve the throughput of cell-edge users, two interference alignments, termed interfering channel alignment based CoBF and interference alignment based CoBF, can be used~\cite{shin2017coordi}. Two BSs jointly optimize their beamforming to improve the data rates of cell-edge users without data sharing between two cells.

\begin{figure}[t]
	\centering
	\includegraphics[width=6.52cm,height=5.14cm]{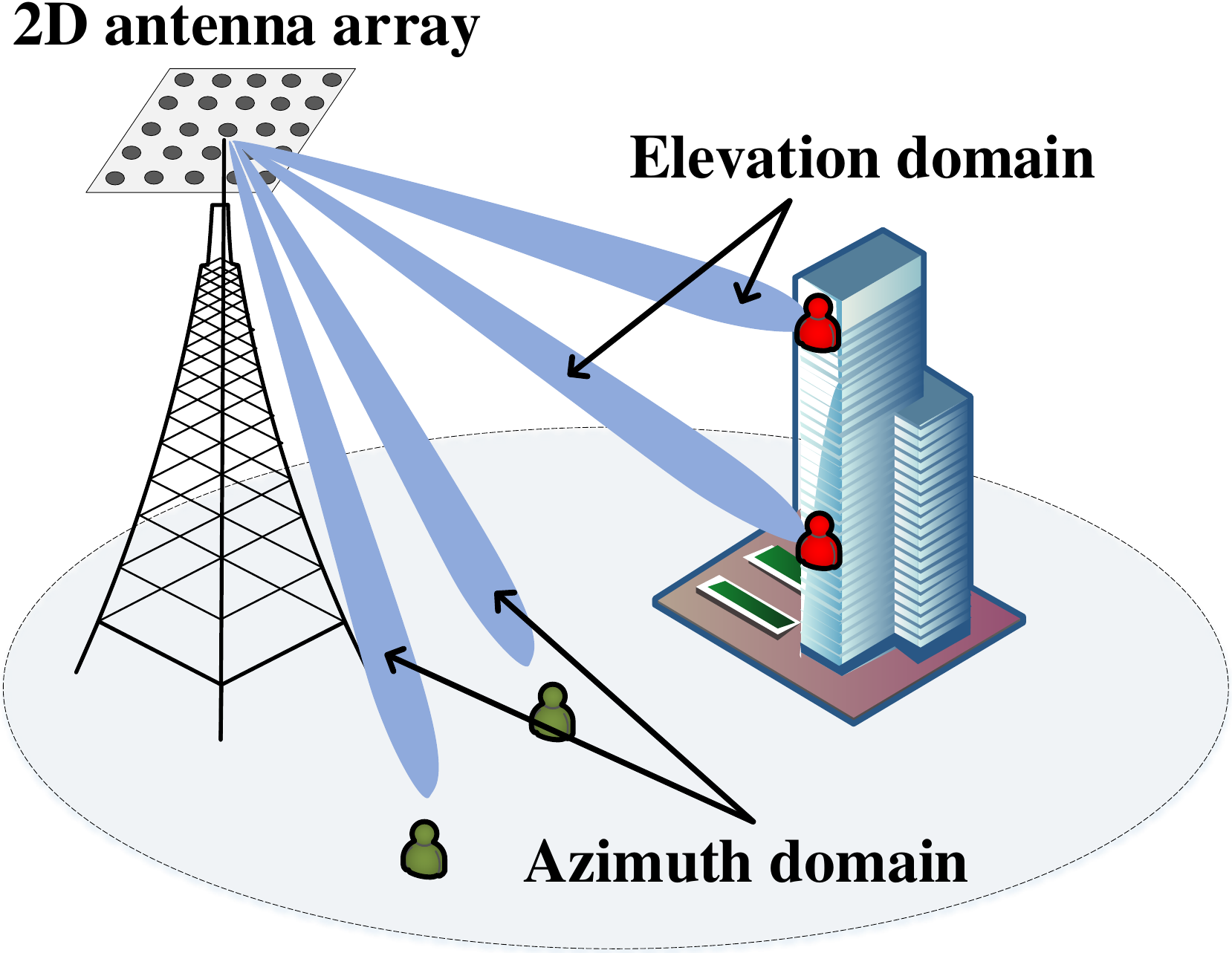}
	\caption{FD-MIMO.}
	\label{fig:FD_MIMO}
\end{figure}

As the number of antennas increases, one of the immediate problems is that the spatial limitations at the top of BS tower limit the use of massive linear antenna array. For example, the length of 64 half-wave antennas in linear array paradigm will reach 4 meters at the carrier frequency of 2.4 GHz. Hence, it is crucial to limit massive antenna array in a smaller form factor. To overcome this problem, full dimension MIMO (FD-MIMO) that utilizes a large number of antennas placed in a 2D antenna array at BSs has attracted substantial research attention from both wireless industry and academia in the past few years~\cite{nadeem2019elevat}. It is defined in 3GPP and is considered as a critical technology for 5G cellular systems to improve network capacity as it allows cellular systems to support a large number of users by using multi-user MIMO technology. It allows the extension of spatial separation to elevation domain as well as traditional azimuth domain as shown in Fig.~\ref{fig:FD_MIMO}, which can reduce the form factor of antenna array at the same time~\cite{kim2014fulldi}. Both azimuth and elevation angles of the downlink beams can be steered dynamically~\cite{nadeem2019elevat, kuo2016aglanc}, which exploit full DoFs. Benefiting from the additional DoF of FD-MIMO, flexible 3D beamforming can be employed in BSs to achieve effective interference coordination in cellular networks~\cite{li2018interf, li2020ffrbas}. Nevertheless, both the works in~\cite{li2018interf} and~\cite{li2020ffrbas} utilize statistical CSI in order to reduce the feedback overhead of channel estimation. How to obtain instantaneous CSI is a great challenge for large-scale antenna arrays. Fortunately, the use of RSMA in large-scale antenna array and massive MIMO systems has been shown to boost the performance over conventional massive MIMO in the presence of imperfect CSI due to frequency division duplexing (FDD) quantization~\cite{dai2016arates,dai2017multiu}, 
time division duplexing (TDD) pilot contamination~\cite{thomas2020arates}, phase noise and hardware impairments~\cite{papazafeiropoulos2017ratesp}, or due to mobility and latency~\cite{dizdar2021ratesp}.


Although massive MIMO further improves spectral efficient and link reliability, it comes at the cost of significantly increased computational complexity compared to small-scale MIMO systems. In particular, uplink signal detection becomes inefficient and has high complexity because of the large increase of dimensions caused by massive antennas. Conventional optimal method, such as maximum-likelihood detection, is not suitable anymore for high complexity. Massive MIMO systems at BSs requires novel detection algorithms that fit for high-dimensional problems with low complexity~\cite{wu2013approx, chen2017alowco}. There has been several reduced-complexity  linear minimum mean square error (LMMSE)-based detectors proposed, but still require high hardware complexity and power consumption as the number of transmit antenna increases~\cite{gao2015lowcom} or the number of users increases~\cite{dai2015lowcom}. An iterative data detection algorithm based on the coordinate descent method can be used to further reduce complexity~\cite{chen2017alowco}, which is able to achieve the same or even higher bit error rate performance compared with the classical LMMSE algorithm. To reduce the signal processing pressure at BSs, distributed algorithm, where BS antennas are divided into different clusters and each cluster has independent computing hardware is an effective method~\cite{liu2019adistr}.

Except for data detection, interference management is quite important for uplink transmission in multi-cell MIMO systems, where a large number of small cells result in severe uplink interference for pilot reuse in channel estimation. The simulation results in~\cite{li2017massiv} shown that higher level of pilot reuse results in lower achievable sum spectral efficiency and  an uplink detector was developed to suppress both intra-cell and inter-cell interference based on MMSE. In~\cite{he2016uplink}, the authors improved uplink performance in massive MIMO macrocells through uplink power control and cell range extension in a two-tier massive heterogeneous cellular network, which ingrates both cellular network and massive MIMO. In multi-cell systems, the cooperation of BSs that regard BSs as distributed antennas is common and effective method to achieve interference elimination, but requires a large amount of CSI between BSs and users among cooperating cells. Interference suppression approach that doesn't require cell cooperation is convenient and the novel semi-blind uplink interference suppression method for multi-cell multi-user massive MIMO systems in~\cite{maruta2018uplink} is confirmed to be the most effective solution evolving spectral use for future wireless networks.

However, most proposed data detection and interference management methods depend on perfect CSI at BSs, which is impractical. Hence, how to design low-complexity data detector at BSs to reduce power consumption and design effective pilot to suppress interference under imperfect CSI is quite important in uplink networks and need further study.


\subsubsection{Cell-free MIMO}

5G cellular communication networks can provide much higher peak data rates and traffic throughput and lower latency compared to previous cellular technologies. However, this outstanding performance can only be achieved by the users nearby BSs. For severe inter-cell interference, the experience performance of edge-users can be much worse. In a conventional cellular network, each user is connected to the BS in one of the cells and the BSs have multiple active users to serve at a certain time, which causes inter-cell interference inevitably~\cite{interdonato2019ubiqui}. All the service antennas are located in a compact area and have low backhaul requirements.

\begin{figure}[t]
	\centering
	\subfigure[Conventional cellular network.]{\includegraphics[width=0.22\textwidth]{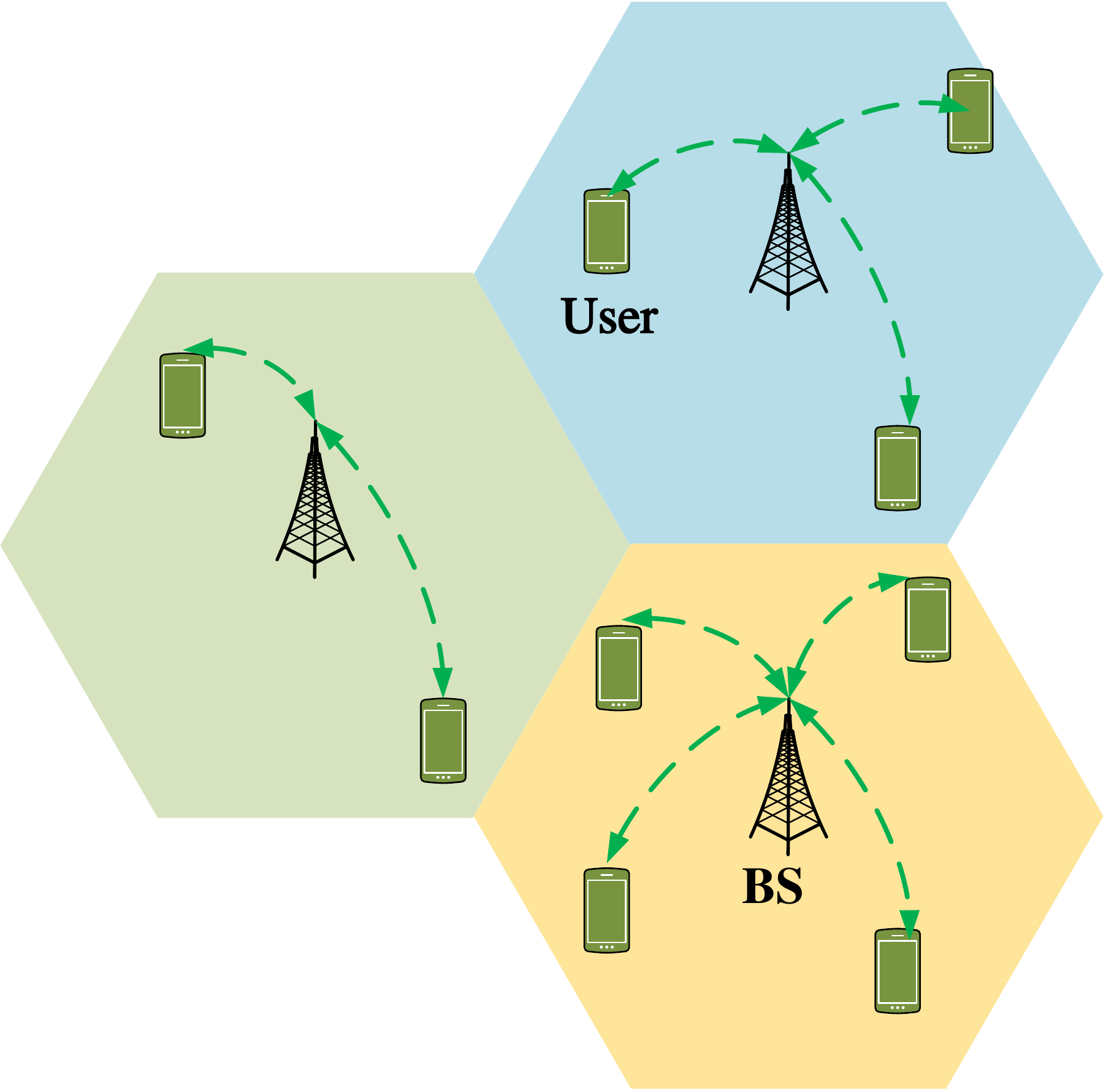}}
	\subfigure[Cell-free network.]{\includegraphics[width=0.26\textwidth]{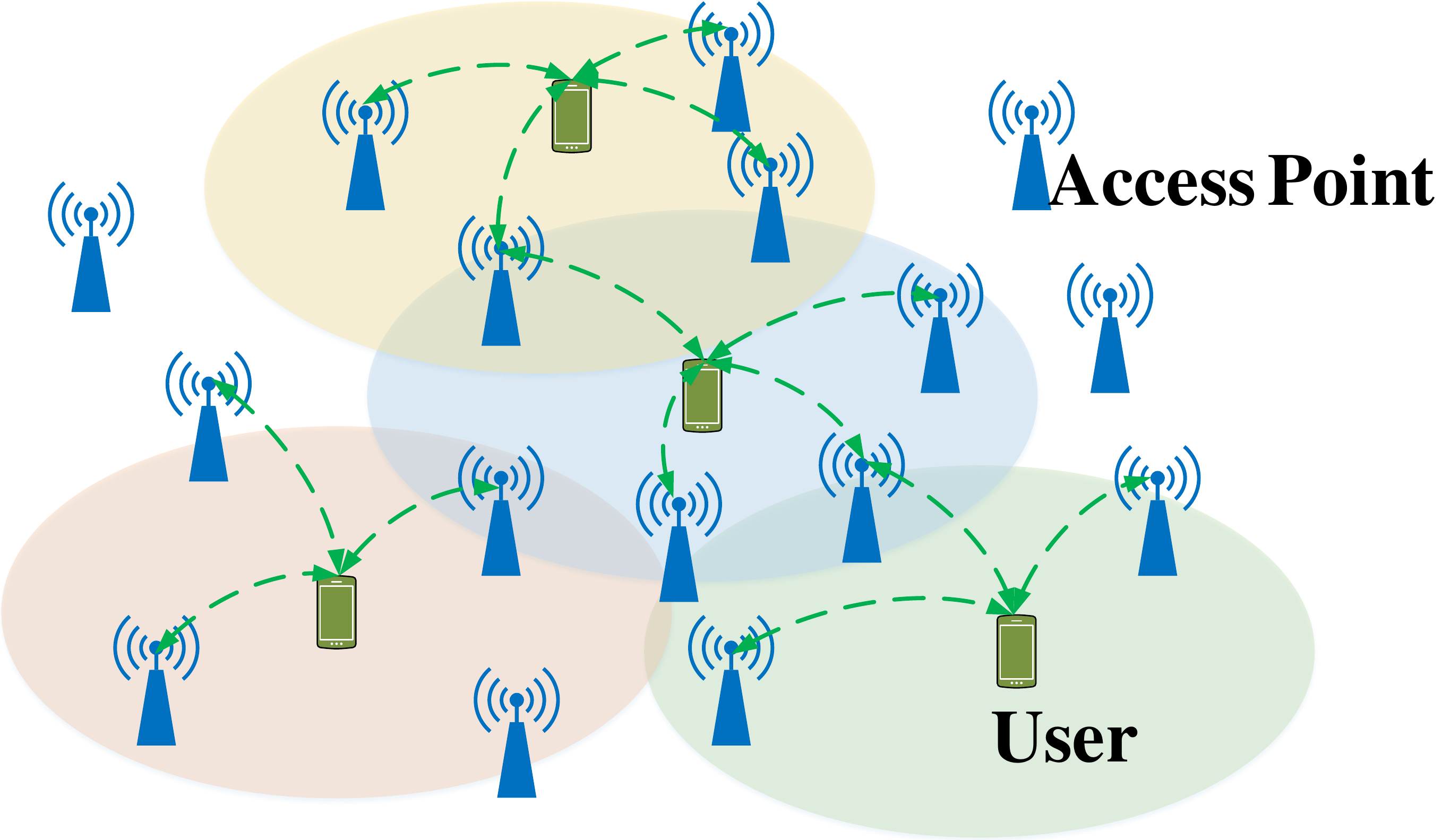}}
	\caption{The comparison between conventional cellular network and cell-free network.}
	\label{fig:cell_free}
\end{figure}

In contrast, in a cell-free network, there are a large number of distributed antennas, called APs, that serve a much smaller number of users over the same time/frequency resources~\cite{ngo2017cellfr}. ``Cell-free'' signifies that there are no cell boundaries during data downlink transmissions from the user perspective.
An AP will cooperate with different sets of APs when serving different users. The comparison between conventional cellular network and cell-free network is shown in Fig.~\ref{fig:cell_free}. It is users that select the set of APs that can provide the best service for itself, instead of the network. Namely, cell-free network is a user-centric paradigm~\cite{zhang2020prospe}.

The APs are connected via fronthaul to central processing units that are responsible for the coordination and are seen as the enabler of cell-free massive MIMO~\cite{interdonato2019ubiqui}. Each AP uses local channel estimation based on received uplink user pilot and applies conjugate beamforming (CBF) to transmit data to users~\cite{ngo2015cellfr}. Although CBF only requires local CSI, its design is based on a large-scale nonconvex problem with very high computational complexity. Hence, CBF that admits a low-scale optimization formulation for computational tractability is required~\cite{naris2021cellfr}. The proposed method can improve both the Shannon function rate and URLLC. 


Most of the traffic congestion happens at the cell edges in cellular networks so that user-experienced performance is poor. The purpose of the cell-free paradigm is not to achieve high peak performance, but to provide more uniform performance. It is proved that the cell-free massive MIMO significantly outperforms small-cell in both median and 95\%-likely performance. What is worth noting is that the cell-free massive MIMO system can provide almost 20-fold increase in 95\%-likely per-user throughput compared with small-cell system~\cite{ngo2017cellfr}. Moreover, the simulation results in~\cite{mai2018cellfe} shown that the 95\%-likely per-user throughput of cell-free system can be further improved through increasing antenna number.

One of the main challenges to design cell-free massive MIMO is how to achieve a network that is scalable in the sense of being implementable in a large network. Specifically, how to achieve the benefits of cell-free massive MIMO in a practicable way under high computational complexity and fronthaul capacity requirements should be considered. Motivated by this purpose, the framework for scalable cell-free systems should be developed and the method to make the network scalable is needed~\cite{bjornson2020scalab}. Although the proposed method in~\cite{bjornson2020scalab} is nearly optimal, power allocation for centralized and distributed operation was not considered. Power control is very important to protect users from strong interference. There have been many heuristic power allocation schemes~\cite{nayebi2017precod, interdonato2019scalab}, how to perform effective and scalable power allocation in cell-free systems still needs further study.

\subsubsection{V2X Communications}

\begin{figure}[t]
	\centering
	\includegraphics[width= \figwidth cm]{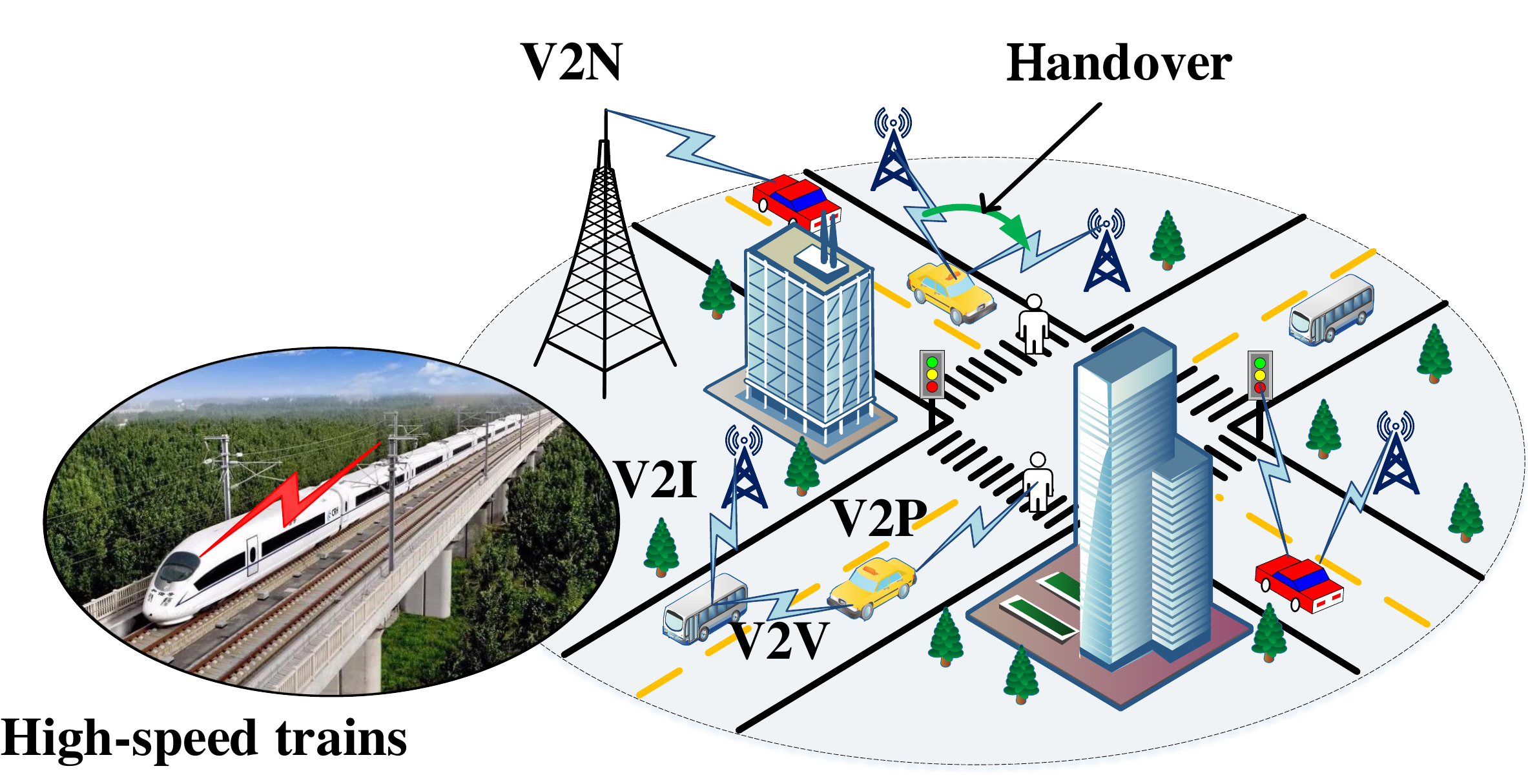}
	\caption{V2X network.}
	\label{fig:V2X_network}
\end{figure}

Autonomous driving has been an innovative technology for future intelligent transport systems, where V2X communications can enhance the safety and efficiency~\cite{gyawail2021challe}, including vehicle-to-vehicle (V2V), vehicle-to-infrastructure (V2I), vehicle-to-pedestrian (V2P) and vehicle-to-network (V2N). Future 5G cellular systems will support vehicular networks and high data transmission rates among fully connected vehicles, where vehicles will be equipped with more sensors and generate gigabit data per second. Besides, 5G is supposed to support high-speed terminals such as high-speed trains. The V2X network is shown in Fig.~\ref{fig:V2X_network}. Massive connectivity, explosive data, low latency, high-speed terminals, frequent handover and user infotainment bring great challenges for 5G vehicular communications. 

MmWave technology with large-scale antenna arrays will play a significant role in vehicular networks by providing high data transmission rates. It is proved that mmWave massive MIMO can deliver Gbps data rates for next-generation vehicular networks~\cite{busari2019millim}. Although mmWave can provide high data rates and low latency, it has high requirements for LoS. If transmission link is blocked, link quality will be greatly reduced. It is practical to model a theoretical highway communication model, where vehicles are served by mmWave BSs alongside the road and blockage was particularly considered~\cite{tassi2017modeli}. In the proposed practical application scenario, heavy vehicles in low lanes may obstruct the LoS paths between vehicles in fast lanes and BSs, through which analyzing how blockage densities impact the user achievable data rate can be studied.

One of the challenges for vehicular networks is that high-speed terminals will make widely-adopted technology used in static scenarios or low-speed scenarios inefficient duo to Doppler spread. With frequent handover and rapidly changing CSI, fast beam alignment techniques should be developed to meet vehicle’s mobility requirements. However, frequent beam sweeping will suffer large overhead and is ineffective in high mobility environment. It is necessary to design the beam alignment scheme for mmWave V2V communication between neighbor vehicles on highway with high speed without any searching steps in beam training~\cite{feng2021beamwi}. The proposed beam alignment method can provide significant throughput improvements compared to general car-following scenarios on the high-way.  


What is more remarkable is that for high speed trains with speed of over 400km/h in the future, providing services satisfying traffic demand for numerous passengers is a great challenge. Compared to highway scenarios, high-speed trains run at higher speeds, which can lead to more frequent handovers and severe inter-carrier interference due to Doppler frequency spread. Therefore many existing beam tracking methods fail to apply to high-speed trains. In terms of this issue, dynamic beam tracking strategy for mmWave high-speed railway communications that can adjust the beam direction and beam width jointly should be applied~\cite{gao2018dynami}. Doppler frequency offset can be compensated through beam alignment and data transmission can be realized through hybrid beamforming for high-speed train communications~\cite{xu2020locati}. Undoubtedly, massive antenna arrays will play an important role in such a scenario.


Much existing research on beam tracking is based on known vehicle positions and statistics CSI. Effective channel estimation is necessary for such a dynamic scenario. Wide beam design may be a feasible method against high-speed mobility. Besides, mmWave signal is sensitive to LoS path. To provide strong connection between BSs and vehicles, RISs can be intelligently applied.

{\color{black}
\subsection{Summary and Discussion}
With the demand for all-time, all-domain, and full-space network services, emerging application scenarios put forward new requirements for space/air/ground communication networks. For antenna array enabled space/air/ground communication networks,
the application of antenna arrays and the mobility of space/air/ground platforms pose substantial new characteristics and challenges. 
6G will rely on LEO satellites for providing global coverage, where proper beam coverage schemes should be employed at satellites for a variety of service scenarios. Therefore, beamforming for satellite communications  should be designed to form various beam patterns. MBA and the corresponding multi-beam forming techniques are promising solutions. Besides, special coverage schemes should be designed for the interference coordination between different beams, and between different satellite constellations~\cite{asdf}. LEO satellites travel very fast compared to the rotation of Earth, and thus result in frequent handover among satellites. However, LEO satellites have limited on-board processing capability and suffer long propagation delay. Therefore, the design of handover protocols should strive to be simple and efficient.

Airborne communications can achieve on-demand deployment and wide-area communication coverage in emergencies, which is one of the important targets for future 6G systems. The 3D mobility of airborne platforms combined with the beamforming technologies poses great DoFs to improve the communication performance, including enlarging communication capacity, saving transmit power, and enhancing physical-layer security. Mathematical optimization is a popular approach to jointly designing positioning and beamforming, where successive convex approximation  
and block coordinate descent are commonly used techniques~\cite{zeng2019access}.
On the other hand, the high mobility of aircrafts and the narrow beam generated by antenna arrays bring serious beam misalignment problem and cause large Doppler shift. The requirement for beam tracking contradicts the limited on-board computing capacity. Prior knowledge of location and mobility prediction can be exploited to assist beam tracking and reduce the overhead. To collaboratively carry out complex missions, multiple aircrafts may form an ad-hoc network, where networking issues including directional neighbor discovery, routing, and resource management should be considered. Timeliness and computational complexity remain major considerations. Prior knowledge of location or mobility prediction are helpful to facilitate the process of neighbor discovery and routing~\cite{xiao2021asurve}. It should be noted that the design of routing and resource management for aerial ad-hoc networks is a coupled decision-making process. It is interesting to develop low-complexity methods for joint routing and resource allocation problems.

Ground communications are  the main solution for providing wireless coverage for most human activities in future 6G systems. 
In order to deal with massive connectivity and improve cell-edge performance, cellular massive MIMO and cell-free massive MIMO are emerging as potential paradigm shifts in ground communication networks. Several interesting and open research problems need to be addressed to propel these emerging technologies. For example,
it is necessary to consider the inference management, FD-MIMO, uplink transmission, and  imperfect CSI  in cellular massive MIMO.  
Cell-free massive MIMO faces challenging issues including the
power control, fronthaul/backhaul provisioning, and network scalability~\cite{zhang2020prospe}. 
With the emergence of novel technologies in cellular communications, V2X communications are evolving rapidly to enable a variety of applications for road safety, traffic efficiency, and passenger infotainment~\cite{gyawail2021challe}. There are still many areas to be explored in the domain of the novel concepts to ensure efficient V2X communication networks, including the physical layer structure, high-speed terminals, synchronization, massive connectivity, resource allocations, and security issues.

%
}

\section{Future Research Directions}
To provide more insights on the design of space/air/ground communications and networking, future directions of research are pointed out as follows:	
\begin{itemize}

		\item 
		\textit{Channel estimation and new modulation methods in RISs-assisted communication systems:}
		Despite various performance gains brought by RIS, the accurate knowledge of CSI is a prerequisite for passive beamforming design or modulation schemes. However, RIS cannot send or decode pilot signals to perform channel estimation, because of its passive and RF chain-free structure. It may be possible for RIS to equip simple communication module to enable the sensing capability for channel estimation~\cite{wu2020toward}. However, it would put heavy signal processing burdens and energy consumption, which is contrary to the original intention of applying RIS as a cost and energy-effective technology. 
		Besides,  applying RIS in the wireless transceiver is a promising candidate for future innovative transceiver technologies, which greatly reduces the hardware complexity and implementation costs. However, the research of RIS-based transceiver is still in its infancy. New modulation methods including theoretical modeling, scheme design, practical measurement, and prototyping work, are worth exploring.
		Overall, practical channel estimation schemes and new modulation methods for RISs-assisted communication systems are open research directions which deserves dedicated efforts for further investigation.

	    \item \textit{Artificial intelligence (AI)-empowered space/air/ground communications and networking:} Taking a wide variety of application tasks, antenna array enabled space/air/ground communication networks are becoming increasingly complicated, decentralized, and autonomous. As a result, it may be challenging to employ mathematical model-based theories to solve problems in large-scale and dynamic cases.
	    In contrast, AI, with model-free, data-driven, adaptive, scalable, and distributed characteristics, shows great potential to achieve significant performance enhancement for space/air/ground communication networks. For example, deep reinforcement learning is promising to help beam tracking based on prior beam alignment decisions and environment information, striking a balance between the system resilience and efficiency~\cite{chiang2021machin}. Besides, AI is a potential solution to solve the complex resource scheduling problem for antenna array enabled communication systems, where not only the original decision domains of time, frequency, and power are considered, but also the beam domain is involved.
	    In summary, AI is a powerful approach to realize an antenna array enabled space/air/ground communication system having rapid response, adaptive learning, and intelligent decision.

	    \item \textit{Joint deployment and beamforming:} A distinct superiority of airborne platforms is on-demand deployment. Moreover, a distinct superiority of antenna array is beamforming. Thus, antenna array enabled airborne communication systems have a great DoF to perform joint deployment and beamforming design to improve communication performance. However, the deployment and beamforming are highly coupled. The channels are affected by different positions of airborne platforms. Moreover, when considering the dynamic scenario such as the movement of users, the design is more challenging due to the trajectory and beamforming. Besides,  practical factors such as aircraft jittering may cause beam misalignment thus deteriorating the communication link quality. As a result, the robust joint deployment and beamforming design for airborne communication networks, which concentrates on both optimization strategy and practical scenario, is an appealing future research direction.
	    
	\item \textit{Joint resource management and routing:} The directional transmission of antenna array and high dynamics of airborne platforms have brought new challenges for both resource management and routing. While in fact, the resource management in physical and media access control (MAC) layers and the routing in network layer are highly coupled. Thus, the joint design of resource management and routing for airborne ad-hoc networks is necessary and challenging. The multiple dimensional resources such as time slot, spectrum, spatial beam, and power should be carefully managed according to the communication tasks. Due to the high dynamics of aircraft, the airborne ad-hoc network's topology is rapidly changing, resulting in not only the change of routing paths but also time-varying available communication resources. To enhance the overall system performance when facing multiple concurrent tasks, it is promising to perform real-time cross layer optimization to allocate the resources in an active manner and update the routing paths according to the network state.
	
\item 	\textit{Adaptive multi-beam pattern and footprint planning:} When providing services to users in remote areas, the uneven distribution of user terminals and dynamic changes in traffic demand, and satellite network access requirements will vary with the user's access time and geographic location. Therefore, in order to meet the ubiquitous access needs of users anytime and anywhere, adaptive multi-beam patterns and footprint planning represent an important research direction. Beam patterns and fingerprints of satellite array antennas are susceptible to the uneven distribution of user and traffic requirements, channel conditions, user QoS requirements, and wireless resources. Therefore, uniform traffic load distribution, simplified radio resource management, effective load and frequency distribution need to be emphasized. The above problem is usually modeled as a compromise between unlimited resource management, load balancing, and user demand. However, this problem is generally a highly non-convex optimization problem, which is challenging to deal with.
	    
\item 	\textit{Multi-spot beam arrangement:} With the continuous expansion of the scope of people activities and the rapid growth of traffic demand, the requirement for broadband satellite capabilities has been diversified. However, the spectrum for satellite communications is becoming increasingly scarce. It is necessary to effectively use the limited spectrum resources to share resources with other communication systems. Digital beamforming has a high degree of flexibility and can be used to allocate power resources. Besides, in satellite communication systems equipped with digital beamforming technology, the theoretical relationship between multipoint beam placement and throughput is an important research direction in the future.
The internal mechanism of the distance between spot beams in the same frequency band and the distance between adjacent spot beams in different frequency bands and the overall system throughput is still unclear. To improve the overall system throughput through multi-point beam placement is usually constructed a 0-1 non-convex optimization model. Therefore, solving this problem is challenging.

\item \textit{Practical considerations for ground communications:} As the number of antenna elements arises, hardware cost will be a challenging problem for MIMO communication system. It is urgent to improve the traffic capacity and reduce the cost at the same time. Besides, CSI estimation is one of the major challenges in large-scale antenna array enabled communication system. How to design low-complexity pilot training in MIMO system to achieve channel estimation is an important research topic. As the number of access users increases, inter-cell and intra-cell interferences become much severer, especially in dense urban areas, and effective interference management methods are needed. Moreover, beam handover and beam tracking methods are supposed to be used in hot-spot areas. For the V2X network, various problems and challenges have been proposed in such highly dynamic vehicular communications. Machine learning may be a potential candidate in the handover process design.
	    
\item \textit{SAGIN:}
As the demand for communication in dense urban areas increases, it is inevitably to integrating satellite communications and airborne communications with ground communications, i.e., forming the SAGIN. As the candidate communication platforms, ground infrastructures should cooperate with aircraft and satellites to solve coverage limitations, access restrictions, and timeliness requirements, and provide users with better and more real-time services. High dynamic scenarios result in more complex and difficult routing and handover management. Future researches should pay more attention to the integrated system, such as the assignment allocation, power allocation, spectrum allocation, and equipment management. The integration of large-scale antenna arrays provides a significant technique support for SAGIN.
\end{itemize}

\section{Conclusions}
With the explosive growth in demand for information in modern society, space/air/ground communication networks are envisioned to constitute a promising architecture for building fully connected global next generation communication networks, satisfying the future network requirements of 6G. To meet the ever increasing demands of high capacity, wide coverage, low latency, and strong robustness for communications, it is encouraging to adopt antenna arrays to obtain considerable antenna gains, multiplexing gains, diversity gains, and many other benefits. This paper surveyed primary characteristics and mechanisms for the design of antenna array enabled space/air/ground communication networks. Specially, the antenna array structures and design were first introduced, where the classification, features and application scenarios of antenna array were discussed. Secondly, the antenna array enabled emerging communication technologies, mainly focusing on new beamforming technologies were considered, multi-antenna multiple access, RISs, and secure communications. Then, the distinct characteristics of antenna array enabled satellite communications, airborne communications, and ground communications were reviewed, and the unique challenges and key technologies were highlighted. Finally, future research directions and challenges were summarized. This paper offered the reader a general perspective and current research status on the design of antenna array enabled space/air/ground communication and networking, and motivated further research efforts on this topic.
%
	
	

	\bibliographystyle{IEEEtran} 
	\bibliography{IEEEabrv,bib_Choi,bib_Clerckx,bib_Han,bib_He,bib_Xiao,bib_Yi,bib_Ke,bib_ZC,bib_Le}

\end{document}